\documentclass
[superscriptaddress,secnumarabic,amssymb,amsmath,nobibnotes,aps,prd,showkeys,showpacs,nofootinbib,onecolumn]{revtex4}%
\usepackage{bigints}
\usepackage{graphicx}
\RequirePackage[colorlinks,citecolor=blue,urlcolor=blue,linkcolor=blue]{hyperref}
\usepackage{epsf}
\usepackage{bm}
\usepackage{amsmath}
\usepackage{amsfonts}
\usepackage{subfigure}
\usepackage{amssymb}%
\providecommand{\U}[1]{\protect\rule{.1in}{.1in}}

\newcommand{\be}{\begin{equation}}
\newcommand{\ee}{\end{equation}}

\newcommand{\mincir}{\raise
-3.truept\hbox{\rlap{\hbox{$\sim$}}\raise4.truept\hbox{$<$}\ }}
\newcommand{\magcir}{\raise
-3.truept\hbox{\rlap{\hbox{$\sim$}}\raise4.truept\hbox{$>$}\ }}

\begin{document}
\title{Variational symmetries and superintegrability in multifield cosmology}
\author{Alex Giacomini}
\email{alexgiacomini@uach.cl}
\affiliation{Instituto de Ciencias F\'{\i}sicas y Matem\'{a}ticas, Universidad Austral de
Chile, Valdivia, Chile}
\author{Esteban Gonz\'alez}
\email{esteban.gonzalezb@usach.cl}
\affiliation{Universidad de Santiago de Chile (USACH), Facultad de Ciencia, Departamento de F\'isica, Chile}
\author{Genly Leon}
\email{genly.leon@ucn.cl}
\affiliation{Departamento de Matem\'{a}ticas, Universidad Cat\'{o}lica del Norte, Avda.
Angamos 0610, Casilla 1280 Antofagasta, Chile}
\author{Andronikos Paliathanasis}
\email{anpaliat@phys.uoa.gr}
\affiliation{Instituto de Ciencias F\'{\i}sicas y Matem\'{a}ticas, Universidad Austral de
Chile, Valdivia, Chile}
\affiliation{Institute of Systems Science, Durban University of Technology, Durban 4000,
South Africa}

\begin{abstract}
We consider a spatially flat Friedmann--Lema\^{\i}tre--Robertson--Walker
background space with an ideal gas and a multifield Lagrangian consisting of
two minimally coupled scalar fields which evolve in a field space of constant
curvature. For this cosmological model we classify the potential function for
the scalar fields such that variational point symmetries exist. The
corresponding conservation laws are calculated. Finally, analytic solutions
are presented for specific functional forms of the scalar field potential in
which the cosmological field equations are characterized as a Liouville
integrable system by point symmetries. The free parameters of the cosmological model
are constrained in order to describe analytic solutions for an inflationary epoch. 
Finally, stability properties of exact closed-form solutions are investigated. 
These solutions are scaling solutions with important physical properties for the 
cosmological model. 

\end{abstract}
\date{\today}
\maketitle

\section{Introduction}

\label{sec1}

Scalar fields play an important role in the theoretical description of
two accelerated expansion phases of the observed universe
\cite{Teg1,Teg2,Teg3,Teg4,planck2015,su1,su2}. The early acceleration phase of
the universe known as inflation is attributed to domination for a
finite time of the inflaton \cite{Aref1,guth}. The inflaton is a single scalar
field model that drives the dynamics; providing a matter source in
the universe which can display antigravitating behavior. Because of the latter antigravitating property, scalar fields have been used in the literature as
dark energy models \cite{de1,de2,de3,de4,de5,de6} to overpass the problems of
$\Lambda$-cosmology \cite{per1}. Moreover, because of the additional degrees
of freedom provided by scalar fields, the scalar field models can be used
as unified models for the description of the dark matter and dark
energy \cite{un1,un2,un3,un4}, while scalar fields can attribute degrees
of freedom for higher-order theories of gravity \cite{lan1}  providing an
equivalent description of modified theories of gravity.

The simplest scalar field model proposed in the literature is the quintessence
model consisting of a minimally coupled single scalar field with positive
energy density and equation of state parameter $w_{Q}$ bounded in the range
$\left\vert w_{Q}\right\vert \leq1$ \cite{sf1}.\ The phantom field is an alternative to quintessence model in which the scalar fields have a
negative kinetic energy, such that the equation of state parameter $w_{P}$ is bounded as $w_{P}\leq -1$, which means that $w_{P}$ can cross the phantom
divide line $w_{P}= -1$ \cite{sf2}. Scalar fields nonminimally coupled to gravity have
been also studied in detail for example  in Brans-Dicke theory \cite{sf3}, O'Hanlon gravity
\cite{sf4}, Hordenski theory \cite{sf5} and others \cite{sf6,sf7,sf8}.
Multifield models have been widely studied  in literature;  the additional degrees of freedom that multifield models provide
overpass various problems of the single scalar field models, providing a richer cosmological evolution \cite{mf1,mf2,mf3,mf4}.

In this work we are interested in multifield cosmological models in a
spatially flat Friedmann--Lema\^{\i}tre--Robertson--Walker (FLRW) background
space consisting of two scalar fields minimally coupled to gravity which are
defined in a hyperbolic field space. Additionally, we consider the
contribution to the cosmic fluid of a perfect fluid of constant equation of state
parameter, that is, an ideal gas. This specific multifield model has been
widely studied to describe the inflationary epoch as alternative to the
inflation mechanism \cite{ch2,ch3,ch4,ch5,ch6,ch7}. In addition, it can provide
a cosmological history which explains the transition from the inflationary
epoch to the matter era and then to the late-time acceleration phase of the
universe \cite{an1,an2}. For this cosmological model we investigate the
existence of conservation laws and the integrability properties of the field
equations. The equations of motion for the scale factor and the scalar fields
are of second-order and form a Hamiltonian system described by a point-like
Lagrangian. We use that property in order to find all the functional forms
for the potential functions of the scalar fields such that variational point
symmetries exist \cite{ns1}. Specifically, we apply Noether's first and
second theorem to calculate the variational point symmetries and write the
corresponding conservation laws \cite{ns2}. Point symmetries have been widely applied
in gravitational physics and in cosmology \cite{sym1,sym2,sym3}.

The main idea of this work is to use the variational symmetries to perform a
classification scheme of the potential function. This classification of a given set
of differential equations according to the admitted symmetry vectors has
been proposed by Ovsiannikov \cite{ovs}. The requirement of the unknown potential
function to be constrained by the symmetry conditions is also a geometric
selection rule. Variational symmetries of the field equations are related to
geometric symmetries of the kinematic metric which define the point-like
Lagrangian \cite{sym4,sym5}. Consequently, the requirement that the two scalar
fields are defined in a hyperbolic field space is directly related to the
existence of variational symmetries. Previous studies for the determination of
exact solutions for the two-scalar field model are presented in
\cite{sl1,sl2,sl3,sl4,sl5}. 

The plan of the paper is as follows: in Section \ref{sec2} we present the cosmological model of our considerations and we derive the field equations. The main properties and definitions of the
variational symmetries are given in Section \ref{sec3}. The classification
problem of this work is performed in Section \ref{sec4}  where we present all
the functional forms of the scalar field potentials in which the field
equations admit variational point symmetries and conservation laws linear in
the momentum. In Section \ref{sec5} we construct analytic and exact solutions to the cosmological models that are Liouville integrable which are obtained from the classification
scheme. In Section \ref{stability} we perform a stability analysis of the scaling solutions using a similar procedure in order to obtain the scaling solutions in the previous Section. Finally, in Section \ref{sec6} we discuss
our results and  our conclusions are drawn.

\section{Field equations}

\label{sec2}

According to the cosmological principle in large scales the universe is
isotropic and homogeneous and its geometry is described by the spatially
flat Friedmann--Lema\^{\i}tre--Robertson--Walker (FLRW) spacetime%
\begin{equation}
ds^{2}=-N\left(  t\right)  ^{2}dt^{2}+a\left(  t\right)  ^{2}\left(
dx^{2}+dy^{2}+dz^{2}\right)  , \label{ss.01}%
\end{equation}
where $N\left(  t\right)  $ is the lapse function and $a\left(  t\right)  $ is
the scale factor which is the radius of a three-dimensional hypersurface.

For the gravitational model we assume the Action Integral%
\begin{equation}
S=S_{GR}+S_{\text{multifield}}+S_{m}, \label{ss.02}%
\end{equation}
where $S_{GR}$ is the action integral of General Relativity%
\begin{equation}
S=\int\sqrt{-g}dx^{4}R, \label{ss.03}%
\end{equation}
in which $R$ is the Ricci scalar of the background metric tensor $g_{\mu\nu}$.~

The Action Integral$~S_{\text{multifield}}$  is given by \cite{sl1}
\begin{equation}
S_{\text{multifield}}=-\int\sqrt{-g}\left(  \frac{1}{2}g^{\mu\nu}\nabla_{\mu}%
\phi\nabla_{\nu}\phi+\frac{\varepsilon}{2}g^{\mu\nu}e^{\kappa\phi}\nabla_{\mu
}\psi\nabla_{\nu}\psi+V\left(  \phi,\psi\right)  \right) , \label{ss.04}%
\end{equation}
where the two scalar fields are defined in a hyperbolic field space, 
$\kappa$ is the coupling parameter and it is related to the curvature
of the hyperbolic space  $R_{h}$ as $R_{h}=-\kappa^2/2 $. Parameter $\varepsilon=\pm1$ indicates if the second
field $\psi\left(  x^{\mu}\right)  $ is phantom $\left(  \varepsilon
=-1\right)  $ or not $\left(  \varepsilon=+1\right)  $. In the special case
where $\kappa=0$ and $\varepsilon=-1$, the quintom model is recovered, however
in this work we assume $\kappa\neq0.$ Finally,~$S_{m}$ corresponds to the
action integral for a perfect fluid with energy density $\rho,~$pressure $p$
and constant equation of state parameter $p=w_{m}\rho~$in which $w_{m}%
\in\left(  -1,1\right)  $ \cite{ss1,ss2}.

For the background space (\ref{ss.01}) and by assuming that scalar fields
inherit the spacetime symmetries we ended with the gravitational field equation%
\begin{align}
3H^{2} &  =\frac{1}{2N^{2}}\dot{\phi}^{2}+\frac{\varepsilon}{2N^{2}}%
e^{\kappa\phi}\dot{\psi}^{2}+V\left(  \phi\right)  +\rho,\label{ss.05}\\
-\left(  2\dot{H}+3H^{2}\right)   &  =\frac{1}{2N^{2}}\dot{\phi}^{2}%
+\frac{\varepsilon}{2N^{2}}e^{\kappa\phi}\dot{\psi}-V\left(  \phi\right)
+w_{m}\rho,\label{ss.06}%
\end{align}
where dot means derivative with respect to the time variable and $H\left(
t\right)  =\frac{1}{N}\frac{d}{dt}\left(  \ln a\right)  $ is the Hubble function.

The equations of motion for the fluid components are%
\begin{align}
\left(  \ddot{\phi}+\frac{\dot{N}}{N}\dot{\phi}+3H\dot{\phi}\right)
-\frac{\varepsilon}{2}\kappa e^{\kappa\phi}\dot{\psi}^{2}+N^{2}V_{,\phi
}=0~,\label{ss.07}%
\\
\ddot{\psi}-\varepsilon\frac{\dot{N}}{N}\dot{\psi}+3H\dot{\psi}+\kappa
\dot{\phi}\dot{\psi}+\varepsilon e^{-\kappa\phi}V_{,\psi}=0~,\label{ss.08}%
\\
\dot{\rho}+3NH(1+w_{m})\rho=0\,\label{ss.09}%
\end{align}
where from this last one, the conservation law follows $\rho\left(  t\right)
=\rho_{m0}a\left(  t\right)  ^{-3\left(  1+w_{m}\right)  }$. \newline 
By replacing in
(\ref{ss.05})-(\ref{ss.08}) and for arbitrary $N(t)$ we end with a Hamiltonian dynamical system which
is described by the singular point-like Lagrangian%
\begin{equation}
L\left(  N,a,\dot{a},\phi,\dot{\phi},\psi,\dot{\psi}\right)  =\frac
{1}{2N\left(  t\right)  }\left(  -6a\dot{a}^{2}+a^{3}\left(  \dot{\phi}%
^{2}+\varepsilon e^{\kappa\phi}\dot{\psi}^{2}\right)  \right)  -N\left(
t\right)  \left(  a^{3}V\left(  \phi,\psi\right)  +\rho_{m0}a^{-3w_{m}%
}\right)  . \label{ss.10}%
\end{equation}
\newline
 In the following we assume the lapse function $N\left(  t\right)
=a^{3w_{m}}$. Then, the point-like Lagrangian (\ref{ss.10}) describes the
trajectory $U\left(  t\right)  =U\left(  a\left(  t\right)  ,\phi\left(
t\right)  ,\psi\left(  t\right)  \right)  $ of a point particle in the three
dimensional geometry
\begin{equation}
ds_{\left(  3\right)  }^{2}=a^{-3w_{m}}\left(  -6ada^{2}+a^{3}\left(
d\phi^{2}+\varepsilon e^{\kappa\phi}d\psi^{2}\right)  \right)  , \label{ss.11}%
\end{equation}
under the action of the effective potential $V_{eff}=a^{3\left(
1+w_{m}\right)  }V\left(  \phi,\psi\right)  $. The integration constant
$\rho_{m0}$ is related to the \textquotedblleft energy~$h$\textquotedblright%
\ of the point-particle, that is, $h=-\rho_{m0}$, in which $h$ is a
conservation law for the equation of motions.

We continue with the presentation of the basic properties for the variational symmetries.

\section{Variational symmetries}

\label{sec3}

Variational symmetries can be defined for differential equations of any order 
which are deduced from a variational principle. Thus in this article we work with
second-order differential equations with a Lagrangian function $L=L\left(
t,q,\dot{q}\right)  $.

Considering the infinitesimal transformation
\begin{equation}
\bar{t}=t+\varepsilon\tau\left(  t,q\right)  ,\qquad\bar{q}=q+\varepsilon
\eta\left(  t,q\right),  \label{ss.12}%
\end{equation}
generated by the differential operator $\Gamma=\tau\partial_{t}+\eta
\partial_{q},$ where $\varepsilon$ is an infinitesimal parameter. Under this transformation, the Action Integral
$A=\int_{t_{0}}^{t_{1}}L\left(  t,q,\dot{q}\right)  \mbox{\rm d}t $ becomes
$\bar{A}=\int_{\bar{t}_{0}}^{\bar{t}_{1}}L\left(  \bar{t},\bar{q},\dot{\bar
{q}}\right)  \mbox{\rm d}\bar{t}$~,~which up to first order in the
infinitesimal  $\varepsilon$ is written as follows
\begin{align}
\bar{A}  &  =\int_{t_{0}}^{t_{1}}\left[  L+\varepsilon\left(  \tau
\displaystyle{\frac{\partial L}{\partial t}}+\eta\displaystyle{\frac{\partial
L}{\partial q}}+\zeta\displaystyle{\frac{\partial L}{\partial\dot{q}}}%
+\dot{\tau}L\right)  \right]  \mbox{\rm d}t  +\varepsilon F, \nonumber \\
& F:=  \tau{t_{1}}L(t_{1},q_{1},\dot{q}_{1})-\tau{t_{0}%
}L(t_{0},q_{0},\dot{q}_{0}),\label{4425}%
\end{align}
where now $\zeta=\dot{\eta}-\dot{q}\dot{\tau}$ and $L(t_{0},q_{0},\dot{q}%
_{0})$ and $L(t_{1},q_{1},\dot{q}_{1})$ are the values of $L$ at the endpoints
$t_{0}$ and $t_{1}$, respectively. Therefore, it follows
\begin{equation}
\bar{A}=A+\varepsilon\int_{t_{0}}^{t_{1}}\left(  \tau\displaystyle{\frac
{\partial L}{\partial t}}+\eta\displaystyle{\frac{\partial L}{\partial q}%
}+\zeta\displaystyle{\frac{\partial L}{\partial\dot{q}}}+\dot{\tau}L\right)
\mbox{\rm d}t+\varepsilon F.
\end{equation}
As
$F$ depends only upon the endpoints, we may write it as
\begin{equation}
F=-\int_{t_{0}}^{t_{1}}\dot{f}\mbox{\rm d}t.
\end{equation}
\hfill

We have to say that the generator $\Gamma$ of the infinitesimal transformation
will be a variational symmetry, that is, a Noether symmetry if $\bar{A}=A$,
i.e.
\begin{equation}
\int_{t_{0}}^{t_{1}}\left(  \tau\displaystyle{\frac{\partial L}{\partial t}%
}+\eta\displaystyle{\frac{\partial L}{\partial q}}+\zeta\displaystyle{\frac
{\partial L}{\partial\dot{q}}}+\dot{\tau}L-\dot{f}\right)  \mbox{\rm d}t=0,
\end{equation}
from which it follows the symmetry condition (Noether's first theorem): 
\begin{equation}
\dot{f}=\tau\displaystyle{\frac{\partial L}{\partial t}}+\eta
\displaystyle{\frac{\partial L}{\partial q}}+\zeta\displaystyle{\frac{\partial
L}{\partial\dot{q}}}+\dot{\tau}L. \label{44210}%
\end{equation}

Hence, according to Noether's second theorem if there exists a vector field
$\Gamma$ and a function $f\left(  t,q,\dot{q}\right)  $ such that condition
(\ref{44210}) is true, then the quantity%
\begin{equation}
I\left(  t,q,\dot{q}\right)  =f-\left[  \tau L+\left(  \eta-\dot{q}%
\tau\right)  \frac{\partial L}{\partial\dot{q}}\right]  , \label{4433}%
\end{equation}
is a conservation law for the Euler-Lagrange equations with Lagrangian
function $L\left(  t,q,\dot{q}\right)  $. \ For a recent discussion on
Noether's theorem and for extensions and generalizations we refer the reader
to \cite{ns1}.

\section{Symmetry classification}

\label{sec4}

For Lagrangian functions of the form%
\begin{equation}
L\left(  t,q,\dot{q}\right)  =\frac{1}{2}\gamma_{\alpha\beta}\left(  q\right)
\dot{q}^{\alpha}\dot{q}^{\beta}-V\left(  q\right)  , \label{ll.01}%
\end{equation}
variational symmetries are constructed by the elements of the Homothetic
algebra of the metric tensor $\gamma_{\alpha\beta}$. We omit the presentation
of the details for the derivation of the variational symmetry vectors and of
the symmetry conditions which constraint the effective potential function for
Lagrangian density of the form (\ref{ll.01}). They are presented into a
theorem in \cite{sym7} with applications which demonstrate the main theorem.

The minisuperspace line element (\ref{ss.11}) is conformally flat and admits the vector field
as elements of the Homothetic algebra
\begin{align*}
K^{1}  &  =\psi\partial_{\phi}+\left(  \frac{e^{-\kappa\phi}}{\kappa
\varepsilon}-\frac{\kappa}{4}\psi^{2}\right)  \partial_{\psi}~,~\\
K^{2}  &  =\partial_{\phi}-\frac{\kappa}{2}\psi\partial_{\psi},~\\
K^{3}  &  =\partial_{\psi},~H=\frac{2}{3\left(  1-w_{m}\right)  }a\partial_{a},%
\end{align*}
in which $K^{1},~K^{2}$ and $K^{3}$ are Killing symmetries and $H$ is the
proper Homothetic vector field. Because variational symmetries are generated
by the vector fields $\left\{  K^{1},K^{2},K^{3},H\right\}  $ or linear
combination of them, we should derive the one-dimensional optimal system of
the Lie algebra.%

\begin{table}[tbp] \centering
\caption{Commutator of the four-dimensional Homothetic algebra}%
\begin{tabular}
[c]{ccccc}\hline\hline
$\left[  ,\right]  $ & $\mathbf{K}^{1}$ & $\mathbf{K}^{2}$ & $\mathbf{K}^{3}$
& $\mathbf{H}$\\\hline
$\mathbf{K}^{1}$ & $0$ & $\frac{\kappa}{2}K^{1}$ & $-K^{2}$ & $0$\\
$\mathbf{K}^{2}$ & $\frac{\kappa}{2}K^{1}$ & $0$ & $\frac{\kappa}{2}K^{3}$ &
$0$\\
$\mathbf{K}^{3}$ & $K^{2}$ & $-\frac{\kappa}{2}K^{3}$ & $0$ & \thinspace$0$\\
$\mathbf{H}$ & $0$ & $0$ & $0$ & $0$\\\hline\hline
\end{tabular}
\label{tab1}%
\end{table}%
%

\begin{table}[tbp] \centering
\caption{Adjoint representation for the four-dimensional Homothetic Lie algebra}%
\begin{tabular}
[c]{ccccc}\hline\hline
$Ad\left(  e^{\left(  \mu\mathbf{\Gamma}_{i}\right)  }\right)  \mathbf{\Gamma
}_{j}$ & $\mathbf{K}^{1}$ & $\mathbf{K}^{2}$ & $\mathbf{K}^{3}$ & $\mathbf{H}%
$\\\hline
$\mathbf{K}^{1}$ & $K^{1}$ & $-\frac{\kappa}{2}\mu K^{1}+K^{2}$ &
$-\frac{\kappa}{4}\mu^{2}K^{1}+\mu K^{2}+K^{3}$ & $H$\\
$\mathbf{K}^{2}$ & $e^{\frac{\kappa}{2}\mu}$ & $K^{2}$ & $e^{-\frac{\kappa}%
{2}\mu}K^{3}$ & $H$\\
$\mathbf{K}^{3}$ & $K^{1}-\mu K^{2}-\frac{\kappa}{4}\mu^{2}K^{3}$ &
$K^{2}+\frac{\kappa}{2}\mu K^{3}$ & $K^{3}$ & $H$\\
$\mathbf{H}$ & $K^{1}$ & $K^{2}$ & $K^{3}$ & $H$\\\hline\hline
\end{tabular}
\label{tab2}%
\end{table}%

The one-dimensional optimal system provides all the independent dynamical
systems which do not communicate under the adjoint representation. Indeed, we
consider the two generic vector fields%
\begin{align}
\mathbf{Z} &  =\alpha_{1}K^{1}+\alpha_{2}K^{2}+\alpha_{3}K^{3}+\alpha_{0}H,~\\
\mathbf{W} &  =\beta_{1}K^{1}+\beta_{2}K^{2}+\beta_{3}K^{3}+\beta_{0}H,
\end{align}
where $\mathbf{\alpha}$ and $\beta$ are constants. The two vector fields
$\mathbf{Z,~W}$ are equivalent if and only if
\begin{equation}
\mathbf{W}=\sum_{j=i}^{n}Ad\left(  \exp\left(  \mu_{i}X^{i}\right)  \right)
\mathbf{Z}\label{sw.05}%
\end{equation}
or%
\begin{equation}
\mathbf{W}=c\mathbf{Z}~,~c=const.\label{sw.06}%
\end{equation}
where the operator $Ad\left(  \exp\left(  \mu X_{i}\right)  \right)
X_{j}=X_{j}-\mu\left[  X_{i},X_{j}\right]  +\frac{1}{2}\mu^{2}\left[
X_{i},\left[  X_{i},X_{j}\right]  \right]  +...~$is called the Adjoint
representation \cite{olver}. In Tables \ref{tab1} and \ref{tab2} the
commutators and the Adjoint representation for the Homothetic algebra
$\left\{  K^{1},K^{2},K^{3},H\right\}  $ are presented. Therefore, from the
results of the Tables we calculate the one-dimensional optimal system
given by the one-dimensional Lie algebra $\left\{  K^{1}\right\}
,~\left\{  K^{2}\right\}  ~,~\left\{  K^{3}\right\}  ~,~\left\{  H\right\}
~,~\left\{  H+\alpha K^{1}\right\}  ,~\left\{  H+aK^{2}\right\}  $ and
$\left\{  H+\alpha K^{3}\right\}  $.

However, in a special case in which $\kappa=\pm\frac{\sqrt{6}}{2}\left(
w_{m}-1\right)  $ the admitted Homothetic algebra is of higher dimension. Let
us assume $\kappa=+\frac{\sqrt{6}}{2}\left(  w_{m}-1\right)  ,$ the case with
$\kappa=-\frac{\sqrt{6}}{2}\left(  w_{m}-1\right)  $ is recovered with the
change of variables $\phi\rightarrow-\phi$. Therefore, for $\kappa=\pm\frac{\sqrt{6}}{2}\left(  w_{m}-1\right)  $, the admitted
Homothetic vector fields by the minisuperspace are%
\begin{align*}
\bar{K}^{1}  &  =\left(  a^{\frac{3}{2}}e^{\frac{\sqrt{6}}{4}\phi}\right)
^{\left(  w_{m}-1\right)  }\left(  a\psi\partial_{a}+\sqrt{6}\psi
\partial_{\phi}+\frac{4e^{-\frac{\sqrt{6}}{2}\left(  w_{m}-1\right)  \phi}%
}{\varepsilon\left(  w_{m}-1\right)  }\right)  ~,\\
\bar{K}^{2}  &  =\left(  a^{\frac{3}{2}}e^{\frac{\sqrt{6}}{4}\phi}\right)
^{\left(  w_{m}-1\right)  }\left(  a\partial_{a}+\sqrt{6}\partial_{\phi
}\right)  ~,
\end{align*}%
\begin{align*}
\bar{K}^{3}  &  =\frac{1}{8}a^{\frac{3}{2}w_{m}-\frac{1}{2}}\left(
3\varepsilon\left(  w_{m}-1\right)  ^{2}\psi^{2}e^{\frac{\sqrt{6}}{4}\left(
w_{m}-1\right)  \phi}+8e^{-\frac{\sqrt{6}}{4}\left(  w_{m}-1\right)  \phi
}\right)  \partial_{a}+\\
&  +\frac{1}{8}a^{\frac{3}{2}\left(  w_{m}-1\right)  }\left(  3\varepsilon
\left(  w_{m}-1\right)  ^{2}\psi^{2}e^{\frac{\sqrt{6}}{4}\left(
w_{m}-1\right)  \phi}-8e^{-\frac{\sqrt{6}}{4}\left(  w_{m}-1\right)  \phi
}\right)  \partial_{\phi}+\\
&  +3\left(  w_{m}-1\right)  a^{\frac{3}{2}\left(  w_{m}-1\right)  }\psi
e^{-\frac{\sqrt{6}}{4}\left(  w_{m}-1\right)  \phi}\partial_{\psi}~,~
\end{align*}%
\[
\bar{K}^{4}=\psi\partial_{\psi}+\left(  \frac{8\sqrt{6}}{\varepsilon\left(
w_m-1\right)  }e^{-\frac{\sqrt{6}}{2}\left(  w_{m}-1\right)  \phi}-3\left(
w_m-1\right)  \sqrt{6}\psi^{2}\right)  \partial_{\phi}~,~
\]%
\[
\bar{K}^{5}=\partial_{\phi}-\frac{\sqrt{6}}{4}\left(  w_{m}-1\right)
\psi\partial_{\psi}~,~\bar{K}^{6}=\partial_{\psi}~\ ,~H=\frac{2}{3\left(
1-w_{m}\right)  }a\partial_{a}~,
\]
where $H$ is the proper Homothetic vector field and $\left\{  \bar{K}^{1}%
,\bar{K}^{2},\bar{K}^{3},\bar{K}^{4},\bar{K}^{5},\bar{K}^{6}\right\}  $ form a
six dimensional Killing algebra. Consequently, when $\kappa=+\frac{\sqrt{6}%
}{2}\left(  w_m-1\right)  $ the minisuperspace is maximally symmetric. Moreover,
we observe that the vector fields $\left\{  \bar{K}^{1},\bar{K}^{2},\bar
{K}^{3}\right\}  $ are gradient from which we infer that the minisuperspace is
the flat space. Therefore, $\left\{  \bar{K}^{1},\bar{K}^{2},\bar{K}%
^{3}\right\}  $ correspond to the three translation symmetries of the three
dimensional flat space and $\left\{  \bar{K}^{4},\bar{K}^{5},\bar{K}%
^{6}\right\}  $ form the $SO\left(  3\right)  .~$The commutators and the
Adjoint representation of the seven-dimensional Homothetic algebra are
presented in Tables \ref{tab3} and \ref{tab4}.%

\begin{table}[tbp] \centering
\caption{Commutator of the seven-dimensional Homothetic algebra}%
\begin{tabular}
[c]{cccccccc}\hline\hline
$\left[  ,\right]  $ & $\mathbf{\bar{K}}^{1}$ & $\mathbf{\bar{K}}^{2}$ &
$\mathbf{\bar{K}}^{3}$ & $\mathbf{\bar{K}}^{4}$ & $\mathbf{\bar{K}}^{5}$ &
$\mathbf{\bar{K}}^{6}$ & $\mathbf{H}$\\\hline
$\mathbf{K}^{1}$ & $0$ & $0$ & $0$ & $-\frac{\sqrt{6}}{3\varepsilon\left(
w_{m}-1\right)  }\bar{K}^{3}$ & $0$ & $-\bar{K}^{2}$ & $0$\\
$\mathbf{K}^{2}$ & $0$ & $0$ & $0$ & $-\frac{\sqrt{6}\left(  w_{m}-1\right)
}{4}\bar{K}^{1}$ & $-\frac{\sqrt{6}}{4}\bar{K}^{2}$ & $0$ & $0$\\
$\mathbf{K}^{3}$ & \thinspace$0$ & \thinspace$0$ & \thinspace$0$ & $0$ &
$\frac{\sqrt{6}\left(  w_{m}-1\right)  }{4}\bar{K}^{3}$ & $-\frac{3\left(
w_{m}-1\right)  ^{2}}{4}\bar{K}^{1}$ & \thinspace$0$\\
$\mathbf{\bar{K}}^{4}$ & $\frac{\sqrt{6}}{3\varepsilon\left(  w_{m}-1\right)
}\bar{K}^{3}$ & $\frac{\sqrt{6}\left(  w_{m}-1\right)  }{4}\bar{K}^{1}$ & $0$
& $0$ & $\frac{\sqrt{6}\left(  w_{m}-1\right)  }{4}\bar{K}^{4}$ & $-\bar
{K}^{5}$ & $0$\\
$\mathbf{\bar{K}}^{5}$ & $0$ & $\frac{\sqrt{6}}{4}\bar{K}^{2}$ & $-\frac
{\sqrt{6}\left(  w_{m}-1\right)  }{4}\bar{K}^{3}$ & $-\frac{\sqrt{6}\left(
w_{m}-1\right)  }{4}\bar{K}^{4}$ & $0$ & $\frac{\sqrt{6}\left(  w_{m}%
-1\right)  }{4}\bar{K}^{6}$ & $0$\\
$\mathbf{\bar{K}}^{6}$ & $\bar{K}^{2}$ & $0$ & $\frac{3\left(  w_{m}-1\right)
^{2}}{4}\bar{K}^{1}$ & $\bar{K}^{5}$ & $-\frac{\sqrt{6}\left(  w_{m}-1\right)
}{4}\bar{K}^{6}$ & $0$ & \thinspace$0$\\
$\mathbf{H}$ & $0$ & $0$ & $0$ & $0$ & $0$ & $0$ & $0$\\\hline\hline
\end{tabular}
\label{tab3}%
\end{table}%
%

\begin{table}[tbp] \centering
\caption{Adjoint representation for the seven-dimensional Homothetic Lie algebra, in which $W=w_{m}-1$}%
\resizebox{\textwidth}{!}{\begin{tabular}
[c]{cccccccc}\hline\hline
${\small Ad}$ & $\mathbf{\bar{K}}^{1}$ & $\mathbf{\bar{K}}^{2}$ &
$\mathbf{\bar{K}}^{3}$ & $\mathbf{\bar{K}}^{4}$ & $\mathbf{\bar{K}}^{5}$ &
$\mathbf{\bar{K}}^{6}$ & $\mathbf{H}$\\\hline
$\mathbf{K}^{1}$ & ${\small \bar{K}}^{1}$ & ${\small \bar{K}}^{2}$ &
${\small \bar{K}}^{3}$ & $\frac{\sqrt{6}\varepsilon\mu}{3\left(  w-1\right)
}{\small \bar{K}}^{3}{\small +\bar{K}}^{4}$ & ${\small \bar{K}}^{5}$ &
${\small \mu\bar{K}}^{2}{\small +\bar{K}}^{6}$ & ${\small H-\mu}\bar{K}^{1}$\\
$\mathbf{K}^{2}$ & ${\small \bar{K}}^{1}$ & ${\small \bar{K}}^{2}$ &
${\small \bar{K}}^{3}$ & $\frac{\sqrt{6}\left(  w-1\right)  {\small \mu}}%
{4}{\small \bar{K}}^{1}{\small +\bar{K}}^{4}$ & $\frac{\sqrt{6}W{\small \mu}%
}{4}{\small \bar{K}}^{2}{\small +\bar{K}}^{5}$ & ${\small \bar{K}}^{6}$ &
${\small H-\mu}\bar{K}^{2}$\\
$\mathbf{K}^{3}$ & ${\small \bar{K}}^{1}$ & ${\small \bar{K}}^{2}$ &
${\small \bar{K}}^{3}$ & ${\small \bar{K}}^{4}$ & ${\small -}\frac{\sqrt
{6}W{\small \mu}}{4}{\small \bar{K}}^{3}{\small +\bar{K}}^{5}$ &
$\frac{3\varepsilon W^{2}{\small \mu}}{4}{\small \bar{K}}^{1}{\small +\bar{K}%
}^{6}$ & ${\small H-\mu}\bar{K}^{3}$\\
$\mathbf{\bar{K}}^{4}$ & ${\small \bar{K}}^{1}{\small -}\frac{\sqrt
{6}\varepsilon\mu}{3W}{\small K}^{3}$ & ${\small \bar{K}}^{2}{\small +}%
\frac{\mu^{2}}{4\varepsilon}{\small \bar{K}}^{3}{\small -}\frac{\sqrt{6}W}%
{4}{\small \mu\bar{K}}^{1}$ & ${\small \bar{K}}_{3}$ & ${\small \bar{K}}^{4}$
& ${\small -}\frac{\sqrt{6}W{\small \mu}}{4}{\small \bar{K}}^{4}%
{\small +\bar{K}}^{5}$ & ${\small \bar{K}}^{6}{\small -\frac{\sqrt{6}W}{8}%
\mu^{2}\bar{K}^{4}+\mu\bar{K}}^{5}$ & ${\small H}$\\
$\mathbf{\bar{K}}^{5}$ & ${\small \bar{K}}^{1}$ & ${\small e}^{-\frac{\sqrt
{6}W}{4}\mu}{\small \bar{K}}^{2}$ & ${\small e}^{\frac{\sqrt{6}W}{4}\mu
}{\small \bar{K}}^{3}$ & ${\small e}^{\frac{\sqrt{6}\left(  w-1\right)  }%
{4}\mu}{\small \bar{K}}^{4}$ & ${\small \bar{K}}^{5}$ & ${\small e}%
^{-\frac{\sqrt{6}W}{4}\mu}{\small \bar{K}}^{6}$ & ${\small H}$\\
$\mathbf{\bar{K}}^{6}$ & ${\small \bar{K}}^{1}{\small -\mu\bar{K}}^{2}$ &
${\small \bar{K}}^{2}$ & ${\small \bar{K}}^{3}{\small -}\frac{3\varepsilon
W^{2}{\small \mu}\left(  \bar{K}^{1}-\mu\bar{K}^{2}\right)  }{4}$ &
${\small \bar{K}}^{4}{\small -\mu\bar{K}}^{5}{\small +}\frac{\sqrt
{6}W{\small \mu}}{4}{\small \bar{K}}^{6}$ & ${\small \bar{K}}^{5}%
{\small +}\frac{\sqrt{6}W{\small \mu}}{4}{\small \bar{K}}^{6}$ &
${\small \bar{K}}^{6}$ & ${\small H}$\\
$\mathbf{H}$ & ${\small \bar{K}}^{1}$ & ${\small \bar{K}}^{2}$ &
${\small \bar{K}}^{3}$ & ${\small \bar{K}}^{4}$ & ${\small \bar{K}}^{5}$ &
${\small \bar{K}}^{6}$ & ${\small H}$\\\hline\hline
\end{tabular}}
\label{tab4}%
\end{table}%

Consequently, from Tables \ref{tab3} and \ref{tab4} we calculate the
one-dimensional system given by the one-dimensional Lie algebras $\left\{
\bar{K}^{A}\right\}  $~,~$\left\{  \bar{K}^{A}+\beta K^{B}\right\}
$~,~$\left\{  \bar{K}^{1}+\alpha\bar{K}^{2}+\beta\bar{K}^{3}\right\}$, 
$\left\{  \bar{K}^{J}\right\}$,  $\left\{  \bar{K}^{J}+\alpha\bar{K}%
^{A}\right\}$,  $\left\{  H\right\}$,  $\left\{  \bar{K}^{J}+\gamma
H\right\}$;  in which $A,B=1,2,3$ and $J=4,5,6$.

Until now we have assumed that $w_{m}-1\neq0$. The case where $w_{m}=1$ will
be studied elsewhere.

\subsection{Classification of $V\left(  \phi,\psi\right)  $}

We continue with the presentation of the special potential forms $V\left(
\phi,\psi\right)  $ where the cosmological field equations (\ref{ss.05}%
)-(\ref{ss.09}) admit conservation laws. We omit the calculations and for each
potential function which admit a variational symmetry we present the
conservation law.

\subsubsection{Arbitrary $\kappa$}

For potential $V_{A}\left(  \phi,\psi\right)  =V\left(  G_{A}\left(  \phi
,\psi\right)  \right)  ~,~G_{A}\left(  \phi,\psi\right)  =e^{-\frac{\kappa}%
{2}\phi}\left(  4+\varepsilon\kappa^{2}\psi^{2}e^{\kappa\phi}\right)  $, the
vector field $K^{1}$ is a variational symmetry with conservation law
function $I^{1}\left(  \mathbf{q,p}\right)  =K^{1}\left(  \mathbf{q}\right)
\mathbf{p}$, where $\mathbf{p}=\left(  p_{a},p_{\phi},p_{\psi}\right)  $; that
is~$I^{1}\left(  a,\phi,\psi,p_{a},p_{\phi,}p_{\psi}\right)  =\psi p_{\phi
}+\left(  \frac{e^{-\kappa\phi}}{\kappa\varepsilon}-\frac{\kappa}{4}\psi
^{2}\right)  p_{\psi}$. $\ $Functions $\mathbf{p}$ are the momentum defined as
$p_{\alpha}=\frac{\partial L}{\partial q^{\alpha}}$, i.e. $p_{a}%
=-6a^{1-3w_{m}}\dot{a}~,~p_{\phi}=a^{3\left(  1-w_{m}\right)  }\dot{\phi}$ and
$p_{\psi}=a^{3\left(  1-w_{m}\right)  }\varepsilon e^{\kappa\phi}\dot{\psi}$.

For $V_{B}\left(  \phi,\psi\right)  =V\left(  G_{B}\left(  \phi,\psi\right)
\right)  ~,~G_{B}=\psi e^{\frac{\kappa}{2}\phi}$, the vector field $K^{2}$ is
a variational symmetry and the corresponding Noetherian conservation law is
$I^{2}\left(  \mathbf{q,p}\right)  =K^{2}\left(  \mathbf{q}\right)
\mathbf{p}$.

When $V_{C}\left(  \phi,\psi\right)  =V\left(  \phi\right)  $ the field
equations admit the variational symmetry $K^{3}$ and the conservation
law~$I^{3}\left(  \mathbf{q,p}\right)  =K^{3}\left(  \mathbf{q}\right)
\mathbf{p}$.

Finally, when $V_{D}\left(  \phi,\psi\right)  =V_{0}$ the field equations admit
three variational symmetries, the fields $K^{1},~K^{2}$ and $K^{3}$ with the
corresponding conservation laws $I^{1}\left(  \mathbf{q,p}\right)
,~I^{2}\left(  \mathbf{q,p}\right)  $ and $I^{3}\left(  \mathbf{q,p}\right)  $.

\subsubsection{Case $\kappa=\frac{\sqrt{6}}{2}\left(  w_{m}-1\right)  $}

For the potential function $\bar{V}_{A}\left(  \phi,\psi\right)  =V\left(
\bar{G}_{A}\left(  \phi,\psi\right)  \right)  e^{-\frac{\sqrt{6}}{2}\left(
w_{m}+1\right)  \phi}$ with $\bar{G}_{A}\left(  \phi,\psi\right)  =\frac
{8}{3}e^{-\frac{\sqrt{6}\left(  w_{m}-1\right)  }{2}\phi}+\varepsilon\left(
w_{m}-1\right)  ^{2}\psi^{2}$ the cosmological field equations admit the
conservation law $\bar{I}^{1}\left(  \mathbf{q,p}\right)  =\bar{K}^{1}\left(
\mathbf{q}\right)  \mathbf{p}$ where the corresponding symmetry vector is the
isometry $\bar{K}^{1}$.

For $\bar{V}_{B}\left(  \phi,\psi\right)  =V\left(  \psi\right)
e^{-\frac{\sqrt{6}}{2}\left(  w_{m}+1\right)  \phi}$, vector field $\bar
{K}^{2}$ is a variational symmetry and the corresponding conservation law is
$\bar{I}^{2}\left(  \mathbf{q,p}\right)  =\bar{K}^{2}\left(  \mathbf{q}%
\right)  \mathbf{p}$.

Similarly, for $\bar{V}_{C}\left(  \phi,\psi\right)  =V\left(  \bar{G}%
_{C}\left(  \phi,\psi\right)  \right)  \,\ $with $\bar{G}_{C}\left(  \phi
,\psi\right)  =\left(  \frac{8}{3}e^{\frac{\sqrt{6}}{2}\phi}+\varepsilon
\left(  w_{m}-1\right)  ^{2}\psi^{2}\right)  e^{-\frac{\sqrt{6}}{4}\left(
w_{m}+1\right)  \phi}$ the symmetry vector is $\bar{K}^{4}$ with $\bar{I}%
^{4}\left(  \mathbf{q,p}\right)  =\bar{K}^{4}\left(  \mathbf{q}\right)
\mathbf{p\,.}$

When potential $\bar{V}_{D}\left(  \phi,\psi\right)  =V\left(  \bar{G}%
_{D}\left(  \phi,\psi\right)  \right)  ,~\bar{G}_{D}\left(  \phi,\psi\right)
=\psi e^{\frac{\sqrt{6}}{4}\left(  w_{m}+1\right)  \phi}$ the system admits
the variational symmetry $\bar{K}^{5}$ with conservation law $\bar{I}%
^{5}\left(  \mathbf{q,p}\right)  =\bar{K}^{5}\left(  \mathbf{q}\right)
\mathbf{p}$.

For $\bar{V}_{E}\left(  \phi,\psi\right)  =V\left(  \phi\right)  $ the field
equations admit the conservation law $\bar{I}^{6}\left(  \mathbf{q,p}\right)
=\bar{K}^{6}\left(  \mathbf{q}\right)  \mathbf{p}$ generated by the symmetry
vector $\bar{K}^{6}$.

Moreover, for the potential function $\bar{V}_{F}\left(  \phi,\psi\right)
=V\left(  \bar{G}_{F}\left(  \phi,\psi\right)  \right)  e^{-\frac{\sqrt{6}}%
{2}\left(  w_{m}+1\right)  \phi},~\bar{G}_{F}\left(  \phi,\psi\right)
=\frac{8}{3}e^{-\frac{\sqrt{6}\left(  w_{m}-1\right)  }{2}\phi}+\varepsilon
\left(  w_{m}-1\right)  ^{2}\psi\left(  \psi+2\beta\right)  $, the vector
field $\bar{K}^{1}+\bar{K}^{2}$ is a variational symmetry and the
corresponding Noetherian conservation law is $\bar{I}_{\beta}^{12}\left(
\mathbf{q,p}\right)  =\left(  \bar{K}^{1}\left(  \mathbf{q}\right)  +\beta
\bar{K}^{2}\left(  \mathbf{q}\right)  \right)  \mathbf{p}$.

The vector field $K^{3}$ is a variational symmetry for the field equations
with conservation law $\bar{I}^{3}\left(  \mathbf{q,p}\right)  =\bar{K}%
^{3}\left(  \mathbf{q}\right)  \mathbf{p}$ if the potential~function $\bar
{V}_{G}\left(  \phi,\psi\right)  =V\left(  \phi,\psi\right)  $ satisfies the
differential equation $F^{3}=0$ \ in which%
\begin{align}
F^{3}  &  \equiv-e^{-\frac{\sqrt{6}}{4}\left(  w_{m}+1\right)  \phi}\left(
\sqrt{6}V_{,\phi}-3\left(  w_{m}-1\right)  \psi V_{,\psi}-3\left(
w_{m}+1\right)  V\right)  +\nonumber\\
&  +3\left(  w_{m}-1\right)  ^{2}\varepsilon\psi^{2}e^{-\frac{\sqrt{6}}%
{4}\left(  w_{m}+1\right)  \phi}\left(  3\left(  w_{m}+1\right)  V+\sqrt
{6}V_{,\phi}\right)  .
\end{align}

Similarly for the linear combinations $K^{1}+\beta K^{3},~K^{2}+\beta K^{3}$
and $K^{1}+\beta K^{2}+\gamma K^{3}$ the symmetry conditions are%
\begin{equation}
0=\psi e^{\frac{\sqrt{6}}{4}\left(  w_{m}-1\right)  \phi}\varepsilon\left(
w_{m}-1\right)  \left(  3\left(  w_{m}+1\right)  V+\sqrt{6}V_{,\phi}\right)
+4e^{\frac{\sqrt{6}}{4}\left(  w_{m}-1\right)  \phi}V_{,\phi}+\varepsilon
\left(  w_{m}-1\right)  \frac{\beta}{8}F^{3}~,~
\end{equation}%
\[
0=e^{\frac{\sqrt{6}}{4}\left(  w_{m}-1\right)  \phi}\left(  3\left(
w_{m}+1\right)  V+\sqrt{6}V_{,\phi}\right)  +\frac{\beta}{8}F^{3}~,
\]%
\begin{align*}
0  &  =\psi e^{\frac{\sqrt{6}}{4}\left(  w_{m}-1\right)  \phi}\varepsilon
\left(  w_{m}-1\right)  \left(  3\left(  w_{m}+1\right)  V+\sqrt{6}V_{,\phi
}\right)  +4e^{\frac{\sqrt{6}}{4}\left(  w_{m}-1\right)  \phi}V_{,\phi}+\\
&  +\beta\varepsilon\left(  w_{m}-1\right)  e^{\frac{\sqrt{6}}{4}\left(
w_{m}-1\right)  \phi}\left(  3\left(  w_{m}+1\right)  V+\sqrt{6}V_{,\phi
}\right)  +\varepsilon\left(  w_{m}-1\right)  \frac{\gamma}{8}F^{3},
\end{align*}
respectively, while the corresponding conservation laws are $\bar{I}_{\beta
}^{13}\left(  \mathbf{q,p}\right)  =\left(  \bar{K}^{1}\left(  \mathbf{q}%
\right)  +\beta\bar{K}^{3}\left(  \mathbf{q}\right)  \right)  \mathbf{p}%
$~,~$\bar{I}_{\beta}^{23}\left(  \mathbf{q,p}\right)  =\left(  \bar{K}%
^{2}\left(  \mathbf{q}\right)  +\beta\bar{K}^{3}\left(  \mathbf{q}\right)
\right)  \mathbf{p}$~,~and $\bar{I}_{\beta\gamma}^{123}\left(  \mathbf{q,p}%
\right)  =\left(  \bar{K}^{1}\left(  \mathbf{q}\right)  +\beta\bar{K}%
^{2}\left(  \mathbf{q}\right)  +\gamma\bar{K}^{2}\left(  \mathbf{q}\right)
\right)  \mathbf{p}$.~ The rest of the linear combinations do not provide any
other potential function which satisfies a conservation law.

We have found all the possible cases for the scalar field potential where the
cosmological field equations admit at least a variational symmetry. The field
equations form a dynamical system of three dimensions, therefore in order to
infer about the integrability by point symmetries we need at least three
conservation laws. By definition an autonomous Hamiltonian system
$\mathcal{H}$ \ of dimension $n~$is characterized as Liouville integrable
\cite{arn1} if there exist at least $n-1$ independent conservation laws
$I_{J},~J=1,2,3,...,n-1,$ i.e. $\left\{  I_{J},H\right\}  =0$, which are in
involution, that is $\left\{  I_{J},I_{\bar{J}}\right\}  =0$.

For the scalar field potential
\begin{equation}
V_{I}\left(  \phi,\psi\right)  =V_{0}e^{-\frac{\sqrt{6}\left(  w_{m}+1\right)
}{2}\phi},
\end{equation}
we find that the cosmological field equations admit the conservation laws
$\bar{I}^{1}\left(  \mathbf{q,p}\right)  ,~\bar{I}^{2}\left(  \mathbf{q,p}%
\right)  $ and $\bar{I}^{6}\left(  \mathbf{q,p}\right)  =\bar{K}^{6}\left(
\mathbf{q}\right)  \mathbf{p}$.~Potential $V_{I}\left(  \phi,\psi\right)  $
admits as asymptotic solution the so-called hyperinflation model where the two
scalar fields drive the dynamics. We refer the reader to
\cite{sl5} and references therein for more details.

The potential function
\begin{equation}
V_{II}\left(  \phi,\psi\right)  =V_{0}e^{\frac{\sqrt{6}}{2}\frac{\left(
w_{m}+1\right)  ^{2}}{w_{m-1}}\phi}\left(  \frac{3}{8}\psi^{2}\left(
w_{m}-1\right)  ^{2}e^{\frac{\sqrt{6}w_{m}\phi}{2}}+8e^{\frac{\sqrt{6}\phi}%
{2}}\right)  ^{-2\frac{w_{m}+1}{w_{m}-1}}%
\end{equation}
admits the variational symmetries $\bar{K}^{1}~,~\bar{K}^{3}$ and $\bar{K}^{4}$
with \ conservation laws $\bar{I}^{1}\left(  \mathbf{q,p}\right)  ,~\bar
{I}^{3}\left(  \mathbf{q,p}\right)  $ and $\bar{I}^{6}\left(  \mathbf{q,p}%
\right)  =\bar{K}^{4}\left(  \mathbf{q}\right)  \mathbf{p}$.

For the scalar field potential%
\begin{equation}
V_{III}\left(  \phi,\psi\right)  =V_{0}e^{-\frac{\sqrt{6}\left(
w_{m}+1\right)  }{2}\phi}\psi^{-2\frac{w_{m}+1}{w_{m}-1}},
\end{equation}
the Hamiltonian system admits the conservation laws $\bar{I}^{2}\left(
\mathbf{q,p}\right)  ,~\bar{I}^{3}\left(  \mathbf{q,p}\right)  $ and $\bar
{I}^{6}\left(  \mathbf{q,p}\right)  =\bar{K}^{4}\left(  \mathbf{q}\right)
\mathbf{p}$ generated by the Noether symmetries $\bar{K}^{2},~\bar{K}^{3}$ and
$\bar{K}^{5}$.

Furthermore, when%
\begin{equation}
V_{IV}\left(  \phi,\psi\right)  =V_{0}e^{\frac{\sqrt{6}}{2}\frac{\left(
w_{m}+1\right)  ^{2}}{w_{m-1}}\phi}\left(  3\psi\left(  2\beta+\psi\right)
\left(  w_{m}-1\right)  ^{2}e^{\frac{\sqrt{6}w_{m}\phi}{2}}+8e^{\frac{\sqrt
{6}\phi}{2}}\right)  ^{-2\frac{w_{m}+1}{w_{m}-1}},
\end{equation}
the conservation laws are $\bar{I}^{12}\left(  \mathbf{q,p}\right)  ,~\bar
{I}^{3}\left(  \mathbf{q,p}\right)  $ and $\bar{I}^{45}\left(  \mathbf{q,p}%
\right)  =\left(  \bar{K}^{4}\left(  \mathbf{q}\right)  +\beta\bar{K}%
^{5}\left(  \mathbf{q}\right)  \right)  \mathbf{p}$.

The potential%
\begin{equation}
V_{V}\left(  \phi,\psi\right)  =V_{0}e^{-\frac{\sqrt{6}\left(  w_{m}+1\right)
}{2}\phi}\left(  3\beta\psi\left(  w_{m}-1\right)  ^{2}\varepsilon+4\right)
^{-2\frac{w_{m}+1}{w_{m}-1}},
\end{equation}
admits the conservation laws $\bar{I}^{13}\left(  \mathbf{q,p}\right)
,~\bar{I}^{2}\left(  \mathbf{q,p}\right)  $ and $\bar{I}^{56}\left(
\mathbf{q,p}\right)  =\left(  \bar{K}^{5}\left(  \mathbf{q}\right)
-\frac{\sqrt{6}}{3\beta\varepsilon\left(  w_{m}-1\right)  }\bar{K}^{6}\left(
\mathbf{q}\right)  \right)  \mathbf{p}$.

For the potential function%
\begin{equation}
V_{VI}\left(  \phi,\psi\right)  =V_{0}e^{\frac{\sqrt{6}}{2}\frac{\left(
w_{m}+1\right)  ^{2}}{w_{m-1}}\phi}\left(  \left(  3\gamma\psi\left(
2\beta+\psi\right)  \left(  w_{m}-1\right)  ^{2}+8\beta\right)  e^{\frac
{\sqrt{6}w_{m}\phi}{2}}+8\gamma e^{\frac{\sqrt{6}\phi}{2}}\right)  ,
\end{equation}
the Noetherian conservation laws are $\bar{I}^{12}\left(  \mathbf{q,p}\right)
=\left(  \bar{K}^{1}\left(  \mathbf{q}\right)  +\beta\bar{K}^{2}\left(
\mathbf{q}\right)  \right)  \mathbf{p}$ , $\bar{I}^{\prime12}\left(
\mathbf{q,p}\right)  =\left(  \bar{K}^{1}\left(  \mathbf{q}\right)
+\gamma\bar{K}^{2}\left(  \mathbf{q}\right)  \right)  \mathbf{p~}$and $\bar
{I}^{56}\left(  \mathbf{q,p}\right)  =\left(  \bar{K}^{5}\left(
\mathbf{q}\right)  -\frac{\sqrt{6}}{3\beta\varepsilon\left(  w_{m}-1\right)
}\bar{K}^{6}\left(  \mathbf{q}\right)  \right)  \mathbf{p}$,~with $\beta
\neq\gamma$.

For the special case where $w_{m}=0$, there exist additional scalar field
potentials which admit conservation laws. These potentials are presented in
\cite{sym10}. The field equations for the six scalar field potentials
$V_{I-VI}\left(  \phi,\psi\right)  $ are Liouville integrable and specifically
they are superintegrable because they admit more than three indepedent
conservation laws. In the following, we construct the analytic solutions for
these integrable dynamical systems.

\section{Analytic solutions}

\label{sec5}

In order to calculate the analytic solutions for the superintegrable scalar
field potentials $V_{I-VI}\left(  \phi,\psi\right)  ~$we prefer to work with
normal coordinates. We consider the change of variables%
\begin{align}
a\left(  x,y,z\right)   &  =\left(  \frac{2}{\left\vert w_{m}-1\right\vert
}\right)  ^{\frac{2}{3\left(  w_{m}-1\right)  }}\left(  4yz-\frac{3}%
{2}\varepsilon x^{2}\right)  ^{-\frac{1}{3\left(  w_{m}-1\right)  }%
}~,~\label{dc.01}\\
\phi\left(  x,y,z\right)   &  =-\frac{\sqrt{6}}{3\left(  w_{m}-1\right)  }%
\ln\left(  \frac{4yz-\frac{3}{2}\varepsilon x}{y^{2}}\left(  w_{m}-1\right)
^{2}\right)  ~,\label{dc.02}\\
\psi\left(  x,y,z\right)   &  =\frac{x}{y}, \label{dc.03}%
\end{align}
in which the point-like Lagrangian for the cosmological field equations
becomes%
\begin{equation}
L\left(  x,\dot{x},y,\dot{y},z,\dot{z}\right)  =\frac{1}{2}\varepsilon\dot
{x}^{2}-\frac{4}{3}\dot{y}\dot{z}-U_{eff}\left(  x,y,z\right)  -\rho_{m0},
\label{dc.04}%
\end{equation}
where $U_{eff}\left(  x,y,z\right)  =a^{3\left(  1+w_{m}\right)  }V\left(
\phi,\psi\right)  $. In the new variables the second-order field equations
become%
\begin{align}
\varepsilon\ddot{x}+\left(  U_{eff}\right)  _{,x}  &  =0,\label{dc.05}\\
\ddot{y}-\frac{3}{4}\left(  U_{eff}\right)  _{,z}  &  =0,\label{dc.06}\\
\ddot{z}-\frac{3}{4}\left(  U_{eff}\right)  _{,y}  &  =0, \label{dc.07}%
\end{align}
while the constraint equation becomes%
\begin{equation}
\frac{1}{2}\varepsilon\dot{x}^{2}-\frac{4}{3}\dot{y}\dot{z}+U_{eff}\left(
x,y,z\right)  +\rho_{m0}=0. \label{dc.08}%
\end{equation}

\subsection{Scalar field potential $V_{I}\left(  \phi,\psi\right)  $}

The exact solution for potential $V_{I}\left(  \phi,\psi\right)  $ was derived
 for the first time in \cite{sl5}. However, we present
the analytic solution for completeness. For the scalar field potential $V_{I}\left(  \phi
,\psi\right)  $ the effective potential in the variables $\left\{
x,y,z\right\}  $ of point-like Lagrangian (\ref{dc.04}) becomes $U_{eff}%
^{I}\left(  x,y,z\right)  =V_{0}y^{-2\frac{w_{m}+1}{w_{m}-1}}$.

Thus, the field equations (\ref{dc.05})-(\ref{dc.07}) become
\begin{equation}
\varepsilon\ddot{x}=0~,~\ddot{y}=0, \label{dc.09}%
\end{equation}%
\begin{equation}
\ddot{z}+\frac{3}{2}\frac{w_m+1}{w_{m}-1}V_{0}y^{-\frac{3w_{m}+1}{w_{m}-1}%
}=0. \label{dc.10}%
\end{equation}
Therefore, the analytic solution is
\begin{equation}
x\left(  t\right)  =x_{0}\left(  t-t_{1}\right)  ~,~y\left(  t\right)
=y_{0}\left(  t-t_{2}\right),  \label{dc.11}%
\end{equation}%
\begin{equation}
z\left(  t\right)  =-\frac{3 V_0 (w_m-1) y_0^{-\frac{3 w_m+1}{w_m-1}}
   (t-t_2)^{\frac{w_m+3}{1-w_m}}}{4 (w_m+3)}+z_{0}\left(  t-t_{3}\right) , \label{dc.12}%
\end{equation}
with constraint condition%
\begin{equation}
\rho_{m0}=\left(  \frac{4}{3}y_{0}z_{0}-\frac{\varepsilon}{2}\left(
x_{0}\right)  ^{2}\right)  . \label{dc.13}%
\end{equation}

In the special case where $w_{m}=-\frac{1}{3}$, the closed-form solution is
\begin{equation}
x\left(  t\right)  =x_{0}\left(  t-t_{1}\right)  ~,~y\left(  t\right)
=y_{0}\left(  t-t_{2}\right)  ,~ \label{dc.14}%
\end{equation}%
\begin{equation}
z\left(  t\right)  =\frac{3}{8}V_{0}(t-t_2)^{2}+z_{0}\left(  t-t_{3}\right),
\label{dc.15}%
\end{equation}
with the same constraint condition. However, when $w_{m}=-\frac{1}{3}$ the
additional matter source can play the role of the spatial curvature $k$, where
$\rho_{m0}=k$, and in such case we recover the closed-form solution for the
multifield model in a nonflat FLRW background space and spatial curvature $k$.

Thus, the scale factor is
\begin{equation}
\left(  a\left(  t\right)  \right)  ^{-3\left(  w_{m}-1\right)  }\simeq\left(
4\left(  \frac{3V_{0}\left(  1-w_{m}\right)  }{4\left(  w_{m}+3\right)
}\left(  y_{0}\left(  t-t_{2}\right)  \right)  ^{-\frac{w_{m}+3}{w_{m}-1}%
+1}+z_{0}y_{0}\left(  t-t_{2}\right)  \left(  t-t_{3}\right)  \right)
-\frac{3}{2}\varepsilon x_{0}\left(  t-t_{1}\right)  \right)  ,
\end{equation}
when the term $\left(  t-t_{2}\right)  ^{-\frac{w_{m}+3}{w_{m}-1}+1}$
dominates, \ the scale factor is approximated as $a\left(  t\right)
\simeq\left(  t-t_{2}\right)  ^{\frac{4}{3\left(  w_{m}-1\right)  ^{2}}}$.

Thus, the line element of the background FLRW space is approximated as
\begin{equation}
ds^{2}=-\left(  t-t_{2}\right)  ^{\frac{8w_{m}}{\left(  w_{m}-1\right)  ^{2}}%
}dt^{2}+\left(  t-t_{2}\right)  ^{\frac{8}{3\left(  w_{m}-1\right)  ^{2}}%
}\left(  dx^{2}+dy^{2}+dz^{2}\right),
\end{equation}
or, equivalently%
\begin{equation}
ds^{2}=-d\tau^{2}+\tau^{\frac{8}{^{3\left(  \left(  w_{m}-1\right)  ^{2}%
+w_{m}\right)  }}}\left(  dx^{2}+dy^{2}+dz^{2}\right)  .
\end{equation}

For $w_m$ the closed-form solution can be found in \cite{an1}.

\subsection{Scalar field potential $V_{II}\left(  \phi,\psi\right)  $}

For the scalar field potential $V_{II}\left(  \phi,\psi\right)  ~$we derive
$U_{eff}^{II}\left(  x,y,z\right)  =V_{0}z^{-2\frac{w_{m}+1}{w_{m}-1}}$. We
observe that changing variables $\left(  y,z\right)
\rightarrow\left(  z,y\right)  $ we end with the dynamical system of potential
$V_{I}\left(  \phi,\psi\right)  $. 
The field equations (\ref{dc.05})-(\ref{dc.07}) become
\begin{equation}
\varepsilon\ddot{x}=0~,~\ddot{z}=0, \label{dc.50}%
\end{equation}%
\begin{equation}
\ddot{y}+\frac{3}{2}\frac{w_m+1}{w_{m}-1}V_{0}z^{-\frac{3w_{m}+1}{w_{m}-1}%
}=0. \label{dc.51}%
\end{equation}

Hence, the analytic solution is
\begin{equation}
x\left(  t\right)  =x_{0}\left(  t-t_{1}\right)  ~,~z\left(  t\right)
=z_{0}\left(  t-t_{2}\right), \; 
y\left(  t\right)  =-\frac{3 V_0 (w_m-1) z_0^{-\frac{3 w_m+1}{w_m-1}}
   (t-t_2)^{\frac{w_m+3}{1-w_m}}}{4 (w_m+3)}+y_{0}\left(  t-t_{3}\right), \label{dc.52}
\end{equation}
with $\rho_{m0}=\left(  \frac{4}{3}y_{0}z_{0}-\frac{\varepsilon}{2}\left(
x_{0}\right)  ^{2}\right)  $. At the limit where $t\rightarrow t_{2}$ we have
a similar behaviour as for the previous potential.

\subsection{Scalar field potential $V_{III}\left(  \phi,\psi\right)  $}

For the third potential of our analysis $V_{III}\left(  \phi,\psi\right)  $
in the new variables it follows~$U_{eff}^{III}\left(  x,y,z\right)
=V_{0}x^{-2\frac{w_{m}+1}{w_{m}-1}}$. 

The field equations (\ref{dc.05})-(\ref{dc.07}) become
\begin{align}
&    \ddot{x}-2V_{0}\varepsilon\frac{w_{m}+1}{w_{m}-1}x^{-\frac{3w_{m}+1}{w_{m}-1}%
}=0, \label{dc.17}\\
& \ddot{y}=0,  \quad \ddot{z}=0. \label{yz}
\end{align}
Furthermore, from the Friedmann equation it follows %
\begin{equation}
\frac{1}{2}\varepsilon\dot{x}^{2}+ V_{0}x^{-2\frac{w_{m}+1}{w_{m}-1}}-\frac
{4}{3}y_{0}z_{0}+\rho_{m0}=0, \label{dc.18}%
\end{equation}
that is
\begin{equation}
  \bigints \frac{dx}{\sqrt{ 2 \varepsilon\left(\frac{4}{3}%
y_{0}z_{0}-\rho_{m0}-V_{0}x^{-2\frac{w_{m}+1}{w_{m}-1}}\right)}} =t-t_{0}.  
\end{equation}
Firstly, we solve \eqref{dc.17}. 
The general solution is given in terms of a hypergeometric function $_2F_1(a, b; c; u)$: 
\begin{align}
    \frac{x ^2 \left(1-\frac{2 V_0 \varepsilon  x ^{-\frac{2
   (w_m+1)}{w_m-1}}}{c_1}\right) \,
   _2F_1\left(\frac{1}{2},\frac{1-w_m}{2
   w_m+2};\frac{w_m+3}{2 w_m+2};\frac{2 V_0
   \varepsilon  x ^{-\frac{2
   (w_m+1)}{w_m-1}}}{c_1}\right){}^2}{c_1-2 V_0 \varepsilon
    x ^{-\frac{2 (w_m+1)}{w_m-1}}}=\left(c_2+t\right){}^2.
\end{align}
Given the complexity in solving this implicit equation, we investigate powerlaw solutions of the system.

In the special case where $\frac{4}{3}y_{0}z_{0}-\rho_{m0}=0$, the closed form
solution is $x\left(  t\right)  \simeq\left(  t-t_{0}\right)  ^{\frac{w_{m}%
-1}{2w_{m}}}\,$, where for $w_{m}=-\frac{1}{3}$, $x\left(  t\right)
\simeq\left(  t-t_{0}\right)  ^{2}$. 

For $w_m\neq 0$ we can obtain a power-law solution as follows. We propose the ansatz $x(t)= x_0 (t-t_1)^p$.
Then  we obtain the equations: 
\begin{align}
    & (p-1) p x_0 (t-t_1)^{p-2}-\frac{2 V_0 (w_m+1) \varepsilon  x_0^{-\frac{4}{w_m-1}-3} (t-t_1)^{\frac{3 p w_m+p}{1-w_m}}}{w_m-1}=0, \\
    & \frac{1}{2} p^2 x_0^2 \varepsilon  (t-t_1)^{2
   p-2}+V_0 x_0^{-\frac{2 (w_m+1)}{w_m-1}} (t-t_1)^{-\frac{2 p
   (w_m+1)}{w_m-1}}+\rho_{m0}-\frac{4 y_0 z_0}{3}=0. 
\end{align}
The first equation must be valid for all $t$. Then, balancing the powers we obtain $p=\frac{w_m-1}{2 w_m}$. Substituting back in the second equation we obtain
\begin{equation}
    \rho_{m0}+\frac{x_0^2 (t-t_1)^{-\frac{w_m+1}{w_m}} \left(8
   V_0 w_m^2 x_0^{-\frac{4 w_m}{w_m-1}}+(w_m-1)^2 \varepsilon \right)}{8 w_m^2}-\frac{4
   y_0 z_0}{3}=0.
\end{equation}
This expression must be valid for all $t$. Therefore, we have the additional restrictions in the parameters:
\begin{equation}
   V_0= -\frac{(w_m-1)^2 \varepsilon  x_0^{\frac{4 w_m}{w_m-1}}}{8 w_m^2}, \quad \rho_{m0} -\frac{4
   y_0 z_0}{3}=0.
\end{equation}
Finally, we have the powerlaw analytic solution 
\begin{equation}
x(t)= x_0 (t-t_1)^{\frac{w_m-1}{2 w_m}}, \; y\left(  t\right)  =y_{0}\left(  t-t_{2}\right)  ~,~z\left(  t\right)
=z_{0}\left(  t-t_{3}\right).  \label{dc.16}%
\end{equation}

Consider now the limit where $x\left(  t\right)  \simeq\left(  t-t_1\right)
^{\frac{w_{m}-1}{2w_{m}}}$ then for $w_{m}>0$, because $\frac{w_{m}-1}{2w_{m}%
}<0$, the scale factor at the limit $t\rightarrow t_1$ is approximated as
$a\left(  t\right)  \simeq\left(  t-t_1\right)  ^{-\frac{1}{6w_{m}}}$.
Therefore, the line element for the background space becomes%
\begin{equation}
ds^{2}=-\left(  t-t_1\right)  ^{-1}dt^{2}+\left(  t-t_1\right)
^{-\frac{1}{3w_{m}}}\left(  dx^{2}+dy^{2}+dz^{2}\right)  ,
\end{equation}
or
\begin{equation}
ds^{2}=-d\tau^{2}+e^{-\frac{1}{3w_{m}}\tau}\left(  dx^{2}+dy^{2}%
+dz^{2}\right)  ,
\end{equation}
which describes a de Sitter universe.

On the other hand when $w_{m}<0$, when $t\rightarrow t_1$ the scale factor
is approximated as $a\left(  t\right)  \simeq t^{-\frac{2}{3\left(
w_{m}-1\right)  }}$. Therefore, the line element for background space is
simplified as
\begin{equation}
ds^{2}=-t^{-\frac{4w_{m}}{\left(  w_{m}-1\right)  }}dt^{2}+t^{-\frac
{4}{3\left(  w_{m}-1\right)  }}\left(  dx^{2}+dy^{2}+dz^{2}\right),
\label{ss1}%
\end{equation}
that is,%
\begin{equation}
ds^{2}=d\tau+\tau^{\frac{2}{3\left(  1+w_{m}\right)  }}\left(  dx^{2}%
+dy^{2}+dz^{2}\right),  \label{ss2}%
\end{equation}
which describes an accelerated universe for $w_{m}<-\frac{1}{3}$.

\subsection{Scalar field potential $V_{IV}\left(  \phi,\psi\right)  $}

The effective potential which corresponds to $V_{IV}\left(  \phi,\psi\right)  $
is derived $U^{IV}\left(  x,y,z\right)  =\bar{V}_{0}\left(  3\beta\varepsilon
x+4z\right)  ^{-2\frac{w_{m}+1}{w_{m}-1}}$ with $\bar{V}_{0}=V_{0}\left(
2\left(  w-1\right)  ^{2}\right)  ^{-2\frac{w_{m}+1}{w_{m}-1}}$. The field
equations become%
\begin{align}
\ddot{x}-6\bar{V}_{0}\frac{w_{m}+1}{w_{m}-1}\beta \varepsilon Z^{-\frac{3w_{m}+1}{w_{m}%
-1}}  &  =0,\label{eqIVa}\\
\ddot{y}+6\bar{V}_{0}\frac{w_{m}+1}{w_{m}-1}Z^{-\frac{3w_{m}+1}{w_{m}-1}}  &
=0, \label{eqIVb}\\
\ddot{Z}-18\bar{V}_{0}\frac{w_{m}+1}{w_{m}-1}\beta^{2}\varepsilon
Z^{-\frac{3w_{m}+1}{w_{m}-1}}  &  =0, \label{eqIVc}
\end{align}
where $Z=3\beta\varepsilon x+4z$. The analytic solution of \eqref{eqIVc} is expressed in terms
of a hypergeometric function
\begin{align}
    \frac{Z ^2 \left(1-\frac{18 \beta ^3 V_0 \varepsilon  Z ^{-\frac{2
   (w_m+1)}{w_m-1}}}{c_1}\right) \,
   _2F_1\left(\frac{1}{2},\frac{1-w_m}{2
   w_m+2};\frac{w_m+3}{2 w_m+2};\frac{18 \bar{V}_0 \beta
   ^3 \varepsilon  Z ^{-\frac{2
   (w_m+1)}{w_m-1}}}{c_1}\right){}^2}{c_1-18 \beta ^3
   \bar{V}_0 \varepsilon  Z ^{-\frac{2
   (w_m+1)}{w_m-1}}}=\left(c_2+t\right){}^2.
\end{align}

For $w_m\neq 0$ there exists the exact solution $Z\left(
t\right)  =Z_{0}\left(  t-t_{1}\right)  ^{\frac{w_{m}-1}{2w_{m}}}$, where
$Z_{0}$ is a solution of $\left(  w_{m}-1\right)  ^{2}+72Z_{0}^{-\frac
{4}{w_{m}-1}-1}\bar{V}_0 w^{2}\beta^{2}\varepsilon=0$. Hence, we have the powerlaw solution
\begin{align}
& x\left(  t\right)
=\frac{Z_0 (t-t_1)^{\frac{w_m-1}{2 w_m}}}{3 \beta}+x_{0}\left(
t-t_{2}\right) , \label{solIVa} \\
& y\left(  t\right) =-\frac{Z_0 (t-t_1)^{\frac{w_m-1}{2
   w_m}}}{3 \beta ^2} \varepsilon+y_{0}\left(  t-t_{3}\right), \label{solIVb} \\
& Z\left(
t\right)  =Z_{0}\left(  t-t_{1}\right)  ^{\frac{w_{m}-1}{2w_{m}}}. \label{solIVc} 
\end{align}

The restriction $\rho_{m0}+\bar{V}_0 Z^{-\frac{2 (w_m+1)}{w_m-1}}-\frac{1}{3} \dot{y} \left(\dot{Z}-3 \beta  \dot{x}\right)+\frac{1}{2} \varepsilon  {\dot{x}}^2=0$ becomes
$\rho_{m0}+\frac{x_0^2 \varepsilon }{2}+\beta  x_0 y_0=0$.

Thus for $w_{m}>0$ and in the limit $t\rightarrow t_{1}$ the scale factor
becomes $a\left(  t\right)  \simeq\left(  t-t_{1}\right)  ^{-\frac{3w_{m}%
-1}{6w_{m}\left(  w_{m}-1\right)  }}$ for $\frac{w_{m}-1}{2w_{m}}+1<0\,$.
Hence the background space becomes%
\begin{equation}
ds^{2}=-\left(  t-t_{1}\right)  ^{-\frac{3w_{m}-1}{\left(  w_{m}-1\right)  }%
}dt^{2}+\left(  t-t_{1}\right)  ^{-\frac{3w_{m}-1}{3w_{m}\left(
w_{m}-1\right)  }}\left(  dx^{2}+dy^{2}+dz^{2}\right)  ,
\end{equation}
or equivalently%
\begin{equation}
ds^{2}=-d\tau^{2}+\tau^{\frac{1}{3w_{m}}}\left(  dx^{2}+dy^{2}+dz^{2}\right)
,
\end{equation}
which describes an accelerated universe for $w_{m}\in\left(  0,\frac{1}%
{3}\right)  $.

On the other hand, for $w_{m}<0$ for large values of $t$, the scale factor is
approximated as~$a\left(  t\right)  \simeq t^{-\frac{2}{3\left(
w_{m}-1\right)  }}$, which leads to the line element (\ref{ss1}).

\subsection{Scalar field potential $V_{V}\left(  \phi,\psi\right)  $}

From the potential function $V_{V}\left(  \phi,\psi\right)  $ we calculate
$U_{eff}^{V}\left(  x,y,z\right)  =V_{0}\left(  \left(  w-1\right)  ^{2}%
\beta\varepsilon x+4y\right)  ^{-2\frac{w_{m}+1}{w_{m}-1}}$. We define the new
variable $Y=\left(  w-1\right)  ^{2}\beta\varepsilon x+4y$ where the field
equations are written as
\begin{align}
\ddot{Y}-2V_{0}\left( w_{m}-1\right)^3 (w_m+1)  \varepsilon\beta^{2}%
Y^{-\frac{3w_{m}+1}{w_{m}-1}}~  &  =0, \label{eqVa}\\
\ddot{x}-2V_{0}\left(  w_{m}^{2}-1\right)  \beta Y^{-\frac{3w_{m}+1}{w_{m}%
-1}}  &  =0, \label{eqVb}\\
\ddot{z}+\frac{6V_{0}\left(  w_{m}+1\right)  }{w_{m}-1}Y^{-\frac{3w_{m}%
+1}{w_{m}-1}}  &  =0, \label{eqVc}
\end{align}
with constraint%
\begin{align}
\rho_{m0}+V_0 Y^{-\frac{2 (w_{m}+1)}{w_{m}-1}}-\frac{1}{3} \dot{z} \left(\dot{Y}-\beta 
   (w_{m}-1)^2 \varepsilon  \dot{x}\right)+\frac{1}{2} \varepsilon 
   {\dot{x}}^2=0.
\end{align}
The closed-form solution of equation \eqref{eqVa} is expressed in terms of hypergeometric function 
\begin{align}
 \frac{Y ^2 \left(1-\frac{2 \beta ^2 V_0 ( w_m-1)^4 \varepsilon 
   Y ^{-\frac{2 ( w_m+1)}{ w_m-1}}}{c_1}\right) \,
   _2F_1\left(\frac{1}{2},\frac{1- w_m}{2
    w_m+2};\frac{ w_m+3}{2  w_m+2};\frac{2 V_0
   ( w_m-1)^4 \beta ^2 \varepsilon  Y ^{-\frac{2
   ( w_m+1)}{ w_m-1}}}{c_1}\right){}^2}{c_1-2 \beta ^2 V_0
   ( w_m-1)^4 \varepsilon  Y ^{-\frac{2
   ( w_m+1)}{ w_m-1}}}=\left(c_2+t\right){}^2.
\end{align}
For $w_{m}=-\frac{1}{3}$ we recover the closed-form solution%
\begin{align}
x\left(  t\right)     =-\frac{8}{9} \beta  V_0 (t-t_1)^2+ x_0 (t-t_2), \; 
Y\left(  t\right)     =Y_0 (t-t_1)^2, \; 
z\left(  t\right)    =\frac{3}{2} V_0 (t-t_1)^2+z_0 (t-t_3),
\end{align}
with $V_0= -\frac{81 Y_0}{128 \beta ^2 \varepsilon }, \quad 6 \rho_{m0}+x_0 \varepsilon  \left(3 x_0+\frac{32 \beta  z_0}{9}\right)=0$. 

The special exact solutions for arbitrary value of $w_{m} \neq 0$ exist and for this
potential in a similar way as we calculated them for potential $V_{IV}\left(
\phi,\psi\right)  $ where we have to perform the change of variables $\left(
y,z\right)  \rightarrow\left(  z,y\right)$.
We can obtain a power-law solution 
\begin{align}
& Y(t) = Y_{0}(t-t_1)^{\frac{w_{m}-1}{2w_{m}}}, \label{solVa}\\
& x(t)=\frac{\varepsilon Y_0 (t-t_1)^{\frac{w_m-1}{2 w_m}}}{\beta  (w_m-1)^2}+x_0 (t-t_2), \label{solVb} \\
& z(t)=-\frac{3 \varepsilon Y_0 (t-t_1)^{\frac{w_m-1}{2 w_m}}}{\beta ^2 (w_m-1)^4}+z_0 (t-t_3), \label{solVc}
\end{align}
where $8 \beta ^2 V_0 (w_m-1)^3 w_m^2
   (w_m+1) \varepsilon  Y_0^{-\frac{4 w_m}{w_m-1}}+w_m^2-1=0$ gives  $V_0$ and $6 \rho_{m0}+x_0 \varepsilon  \left(3 x_0+2 \beta  (w_m-1)^2 z_0\right)=0$. 

\subsection{Scalar field potential $V_{VI}\left(  \phi,\psi\right)  $}

Finally, from $V_{VI}\left(  \phi,\psi\right)  $ it follows $U_{eff}%
^{VI}\left(  x,y,z\right)  =V_{0}\left(  6\left(  w_{m}-1\right)  ^{2}%
\beta\varepsilon x+8\left(  \beta y+\left(  w-1\right)  ^{2}\right)  z\right)^{-2\frac{w_{m}+1}{w_{m}-1}}
$. With the use of the new variable $U=6\left(  w_{m}-1\right)  ^{2}%
\beta\varepsilon x\gamma+8\left(  \beta y+\left(  w-1\right)  ^{2}\gamma
z\right)  $, the field equations become%
\begin{align}
\label{dc.82}
\ddot{U}-24\gamma V_{0}\left(  w_{m}^{2}-1\right)  \beta\left(  \varepsilon
\beta\left(  w_{m}-1\right)  ^{2}-8\right)  U^{-\frac{3w_{m}+1}{w_{m}-1}}~  &
=0,\\
\label{dc.83}
\ddot{x}-12V_{0}\gamma\left(  w_{m}^{2}-1\right)  \beta U^{-\frac{3w_{m}%
+1}{w_{m}-1}}  &  =0,\\
\label{dc.84}
\ddot{z}+\frac{12V_{0}\left(  w_{m}+1\right)  }{w_{m}-1}\beta U^{-\frac
{3w_{m}+1}{w_{m}-1}}  &  =0.
\end{align}
with constraint%
\begin{align}
 \rho _{m0}-\frac{\dot{U} \dot{z}}{6 \beta }+V_0 U^{-\frac{2
   (w+1)}{w-1}}+\gamma  (w-1)^2 \varepsilon  \dot{x} \dot{z}+\frac{4 \gamma 
   (w-1)^2 \dot{z}^2}{3 \beta }+\frac{1}{2} \varepsilon  \dot{x}^2=0.
\end{align}

The results from the previous two potentials follow: the behaviour of
the solution at the limits has similar properties as before.

Firstly, we solve \eqref{dc.82}. 
The general solution is given in terms of a hypergeometric function:
\begin{align}
& \frac{U^2 \left(1-\frac{24 \beta  \gamma  V_0 (w_m-1)^2
   U^{-\frac{2 (w_m+1)}{w_m-1}} \left(\beta 
   (w_m-1)^2 \varepsilon -8\right)}{c_1}\right) \,
   _2F_1\left(\frac{1}{2},\frac{1-w_m}{2
   w_m+2};\frac{w_m+3}{2 w_m+2};\frac{24 V_0
   (w_m-1)^2 \beta  \gamma  \left((w_m-1)^2 \beta  \varepsilon
   -8\right) U^{-\frac{2
   (w_m+1)}{w_m-1}}}{c_1}\right){}^2}{c_1-24 \beta  \gamma 
   V_0 (w_m-1)^2 U^{-\frac{2 (w_m+1)}{w_m-1}}
   \left(\beta  (w_m-1)^2 \varepsilon -8\right)}\nonumber \\
   & =\left(c_2+t\right){}^2. \label{eq.89}
\end{align}

For $w_m\neq 0$ the system \eqref{dc.82}, \eqref{dc.83}, \eqref{dc.84} admits the following power-law solution 
\begin{align}
\label{dc.90}
& U(t)=U_0 (t-t_1)^{\frac{w_m-1}{2 w_m}}, \\
\label{dc.91}
& x(t)= \frac{U_0 (t-t_1)^{\frac{w_m-1}{2 w_m}}}{6 \beta  \gamma  (w_m-1)^2
   \varepsilon -16}+x_0 (t-t_2),\\
   \label{dc.92}
& z(t)=-\frac{U_0 (t-t_1)^{\frac{w_m-1}{2 w_m}}}{2
   \gamma  (w_m-1)^2 \left(3 \beta  \gamma  (w_m-1)^2 \varepsilon -8\right)}+ z_0 (t-t_3),
\end{align}
where $U_0^{\frac{4 w_m}{w_m-1}}+96 \beta  \gamma  V_0 w_m^2 \left(3 \beta 
   \gamma  (w_m-1)^2 \varepsilon -8\right)=0$ gives $V_0$ and $6 \beta   \rho_{m0}+2 \gamma  (w-1)^2 z_{0} (3 \beta  x_{0}
   \varepsilon +4 z_0)+3 \beta  x_{0}^2 \varepsilon =0$ gives $\rho_{m0}$. 

\section{Stability analysis of the scaling solutions}
\label{stability}

According to the methods in \cite{Ratra:1987rm,Liddle:1998xm,Uzan:1999ch} let be
\begin{equation}
    F(\ddot\phi, \dot\phi, \phi)=0,
\end{equation}
a second order differential equation in one dimension which admits a singular powerlaw solution 
\begin{equation}
    \phi_c(t)= \phi_0 t^\beta.
\end{equation}
To examine the stability of the solution $\phi_c$, the logarithmic time $\tau$ through  
$t= e^{\tau}$ is introduced, such that $t\rightarrow 0$ as  $\tau \rightarrow -\infty$ and $t\rightarrow +\infty$ as $\tau\rightarrow +\infty$. We use $\phi'\equiv \frac{d \phi}{d\tau}$ in the following discussion. 

The following dimensionless function is introduced\begin{equation}
    u(\tau)= \frac{\phi(\tau)}{\phi_c(\tau)},
\end{equation}
and the stability analysis in translated into the analysis of the stability of $u=1$ of a transformed dynamical system. 
To construct the aforementioned system the following relations are useful: 
\begin{equation}
  \dot\phi= e^{-\tau}\phi',  \quad  \ddot\phi= e^{-2 \tau} (\phi''-\phi'), \quad \text{and}
 \quad   \frac{\phi_c'}{\phi_c}=\beta \quad \text{if} \quad \phi_c(t)= \phi_0 t^\beta.
\end{equation}
In this section we use a similar procedure for analyzing  stability of the scaling solutions that are obtained in Section \ref{sec5}.
\newline 
Due to the complexity of the dynamics, we have supported our analytical results of this section in numerical simulations by using initial conditions within the constraint  surfaces: $(\phi_{1}-1)^{2}+(\phi_{2}-1)^{2}+(\phi_{3}-1)^{2}=1$ and $\Phi_{1}^{2}+\Phi_{2}^{2}+\Phi_{3}^{2}=1$. The auxiliary variables $\phi_i, \Phi_i$ are dummy variables that are defined for each potential, and they are used to classify the stability of the powerlaw solutions obtained for each potential. The forward integration is based in two simple basis: i) if the point is an attractor, the boundary surfaces act as trapping surfaces due to all the orbits are attracted by the basing of attraction of the (1,0,1,0,1,0); ii) if the point (1,0,1,0,1,0) have saddle behavior, some orbits would abandon this region passing through the boundary surfaces to higher radii, say, having to the past and to the future that $(\phi_{1}-1)^{2}+(\phi_{2}-1)^{2}+(\phi_{3}-1)^{2}>1$ and $\Phi_{1}^{2}+\Phi_{2}^{2}+\Phi_{3}^{2}>1$. 

\subsection{Scalar field potential $V_{I}\left(  \phi,\psi\right)  $}

In this section we analyze the stability of the analytic solution \eqref{dc.11}, \eqref{dc.12} of equations \eqref{dc.09} \eqref{dc.10}. We  set for simplicity the integration constants $t_1, t_2, t_3$ to zero  because they are not relevant as $t\rightarrow \infty$. 

With the time variable $\tau=\ln (t)$, and defining the new variables $Y= V_0 e^{2 \tau} y^{-\frac{3w_{m}+1}{w_{m}-1}}$ and $Z=z(\tau)-z_0 e^{\tau}$
the equations \eqref{dc.09} and \eqref{dc.10} become \begin{align*}
& x''(\tau)-x'(\tau)=0, \quad Z''(\tau) - Z'(\tau) +\frac{3}{2}\frac{w_m+1}{w_{m}-1}Y(\tau)=0, \\
& -4 w_m Y'(\tau )^2+Y(\tau ) \left((3 w_m+1) Y''(\tau )+(w_m-5) Y'(\tau
   )\right)+2 (w_m+3) Y(\tau )^2=0. 
\end{align*}
The analytic solution \eqref{dc.11}, \eqref{dc.12} becomes:
\begin{equation*}
x_c\left( \tau\right)  =x_{0} e^{\tau}, ~Y_c\left(  \tau\right)
=Y_{0} e^{-\frac{\tau  (w_m+3)}{w_m-1}}, \quad Y_0=V_0 y_0^{-\frac{3 w_m+1}{w_m-1}} \; 
Z_c\left(  \tau \right)  = z_1  e^{-\frac{\tau  (w_m+3)}{w_m-1}}, \quad z_1=-\frac{3 (w-1) Y_0}{4 (w+3)}.
\end{equation*}
Defining the dimensionless variables 
\begin{equation}
    \phi_1= \frac{x}{x_c}, \quad \phi_2= \frac{Y}{Y_c}, \quad \phi_3= \frac{Z}{Z_c},
\end{equation}
we obtain the dynamical system
\begin{align}
& \phi_1'= \Phi_1, \quad  \Phi_1'=  -\Phi_1, \label{systemphi1a} \\
& \phi_2'= \Phi_2, \quad  \Phi_2'= -\Phi_2 +\frac{4 w_m \Phi_2^2}{\phi_2(1+3 w_m)}, \label{systemphi1b} \\
& \phi_3' =\Phi_3, \quad  \Phi_3'=  \frac{(3 w_m+5)  \Phi_{3}}{w_m-1}+\frac{2 (w_m+1) (w_m+3) ( \phi_{2} - \phi_{3})}{(w_m-1)^2}. \label{systemphi1c}
\end{align}
Now we analyze the stability of the solution 
$P:=(\phi_1, \Phi_1, \phi_2, \Phi_2, \phi_3, \Phi_3)= (1,0,1,0,1,0)$.
The subsystems for $(\phi_1,\Phi_1)$, $(\phi_2, \Phi_2)$ and $(\phi_2, \Phi_2, \phi_3, \Phi_3)$ are  decoupled. 
The Jacobian matrix of the full system is 
\begin{equation}
  J:= \left(
\begin{array}{cccccc}
 0 & 1 & 0 & 0 & 0 & 0 \\
 0 & -1 & 0 & 0 & 0 & 0 \\
 0 & 0 & 0 & 1 & 0 & 0 \\
 0 & 0 & -\frac{4 w_m  \Phi_{2}^2}{(3 w_m+1) \phi_{2}^2} & \frac{8 w_m  \Phi_{2}}{3 w_m  \phi_{2}+ \phi_{2}}-1 & 0 & 0 \\
 0 & 0 & 0 & 0 & 0 & 1 \\
 0 & 0 & \frac{2 (w_m+1) (w_m+3)}{(w_m-1)^2} & 0 & -\frac{2 (w_m+1) (w_m+3)}{(w_m-1)^2} & 3+\frac{8}{w_m-1} \\
\end{array}
\right).
\end{equation}
Evaluating $J$  at the fixed point $P$  the eigenvalues 
$\left\{0,0,-1,-1,-\frac{2 (w_m+1)}{1-w_m},-\frac{w_m+3}{1-w_m}\right\}$ are obtained. The stable manifold is 4D if $-1<w_m<1$. 
If the analysis is restricted to the subspace $(\phi_2, \Phi_2, \phi_3, \Phi_3)$ the eigenvalues are
$\left\{0,-1,-\frac{2 (w_m+1)}{1-w_m},-\frac{w_m+3}{1-w_m}\right\}$.  The stable manifold in this subspace is 3D if $-1<w_m<1$. 

Introducing the new variables 
\begin{align}
& u_1= \Phi_{2}+ \phi_{2}-1, \; u_2= \Phi_1+ \phi_{1}-1, \; v_1= \frac{ \Phi_{2} (w_m+3)}{3 w_m+1}, \; v_2= \Phi_1, \label{var1a} \\
& v_3= \frac{2 (w_m+1) ( \Phi_{3} (w_m-1) (3 w_m+1)+(w_m+3) ( \Phi _{2} (w_m-1)+3 w_m  \phi_{2}-(3 w_m+1)
    \phi_{3}+ \phi_{2}))}{(w_m-1)^2 (3 w_m+1)}, \label{var1b}\\
&v_4= -\frac{(w_m+3) ( \Phi_{2} (w_m-1)+\Phi_{3} (w_m-1)+2 (w_m+1)  \phi_{2}-2 (w_m+1)  \phi _{3})}{(w_m-1)^2}, \label{var1c}
\end{align}

\begin{figure*}
    \centering
    \subfigure[Projections in the space ($\phi_{1}$, $\phi_{2}$, $\phi_{3}$) (left) and ($\Phi_{1}$, $\Phi_{2}$, $\Phi_{3}$) (right). We have represented the spheres $(\phi_{1}-1)^{2}+(\phi_{2}-1)^{2}+(\phi_{3}-1)^{2}=r^2$ and $\Phi_{1}^{2}+\Phi_{2}^{2}+\Phi_{3}^{2}=r^2$,  $r\in\{1,\sqrt{2}\}$.]{\includegraphics[scale = 0.29]{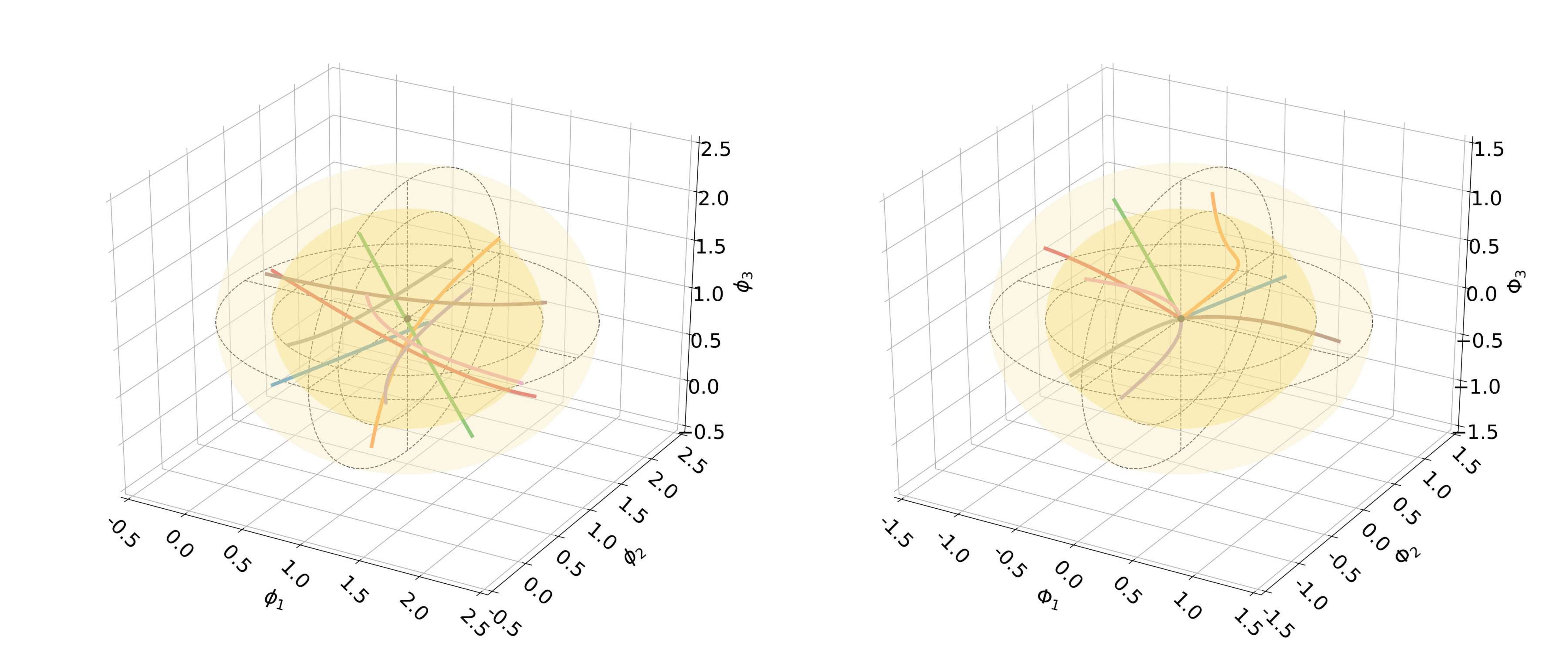}}
    \subfigure[From left to right, from top to bottom:  Projections in the planes ($u_{1},u_
    {2}$), ($v_{1},v_{2}$), ($v_{1},v_{3}$), ($v_{1},v_{4}$), ($v_{2},v_{3}$), ($v_{2},v_{4}$), ($v_{3},v_{4}$), ($u_{1},v_{1}$) and ($u_{2},v_{2}$). The red line in fig 8 (resp. in fig 9) of the bottom array denotes the invariant set $v_1=0$ (resp. $v_2=0$) in the projection $u_1$ vs $v_1$ (resp.  $u_2$ vs $v_2$).]{\includegraphics[scale = 0.46]{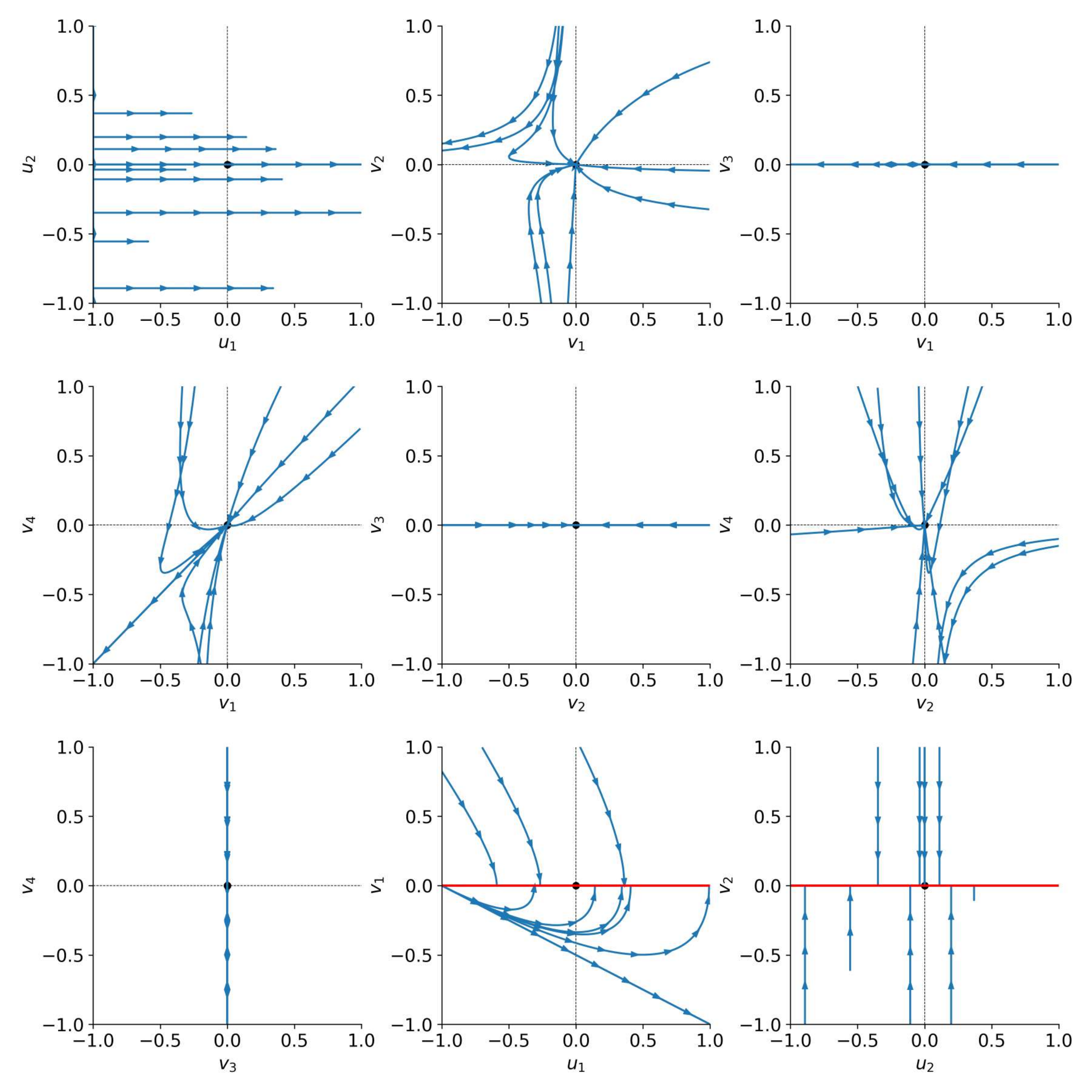}}
    \caption{Some solutions of (a) the system \eqref{systemphi1a}-\eqref{systemphi1c}  and (b) the system \eqref{system1}  for the potential $V_{I}(\phi,\psi)$ when $\omega_{m}=-1$.  Notice in the projection $u_1$ vs $u_2$ (which contains the center manifold)  the origin behaves as a saddle point. The orbits for the left along the $u_1$-axis tend to the origin, but for the right the orbits depart from the origin.}
    \label{fig:PIwm-1}
\end{figure*}

\begin{figure*}
    \centering
    \subfigure[Projections in the space ($\phi_{1}$, $\phi_{2}$, $\phi_{3}$) (left) and ($\Phi_{1}$, $\Phi_{2}$, $\Phi_{3}$) (right). We have represented the spheres $(\phi_{1}-1)^{2}+(\phi_{2}-1)^{2}+(\phi_{3}-1)^{2}=r^2$ and $\Phi_{1}^{2}+\Phi_{2}^{2}+\Phi_{3}^{2}=r^2$,  $r\in\{1,\sqrt{2}\}$.]{\includegraphics[scale = 0.29]{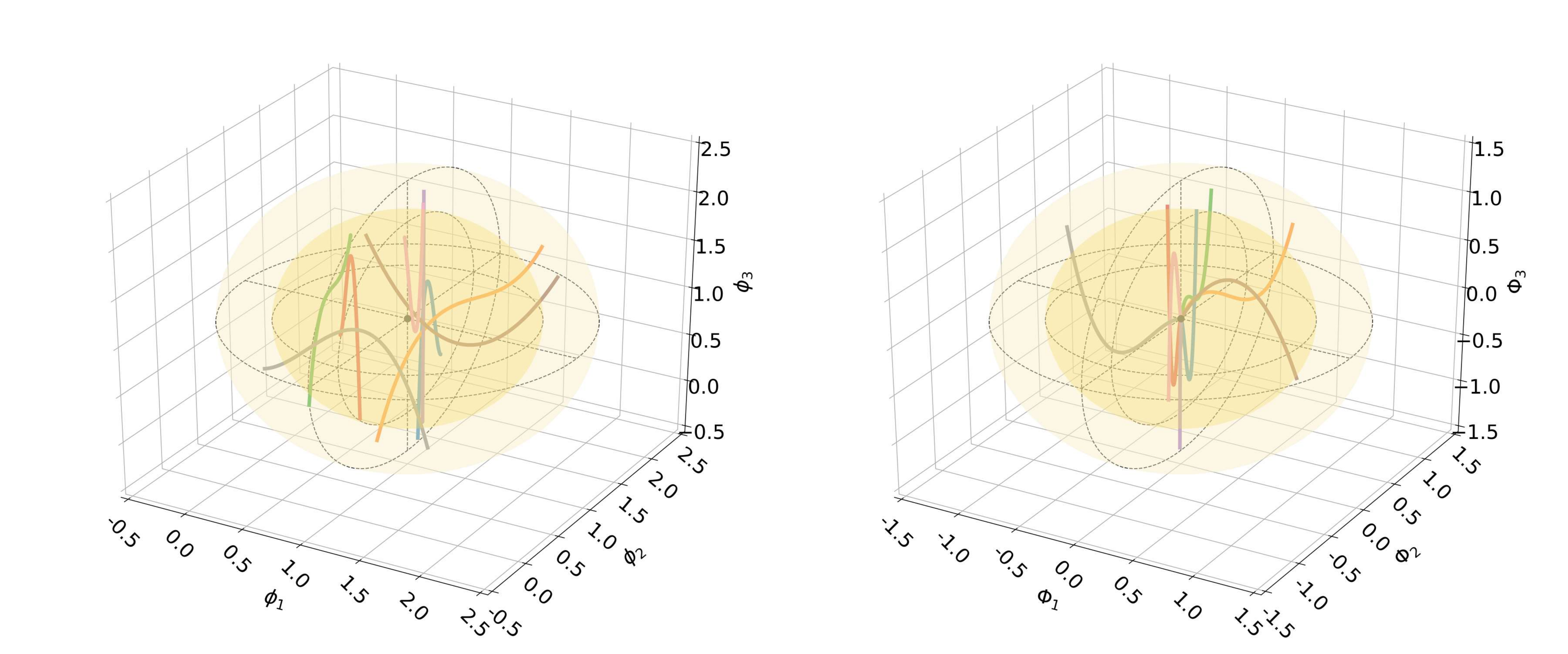}}
    \subfigure[From left to right, from top to bottom: Projections in the planes ($u_{1},u_
    {2}$), ($v_{1},v_{2}$), ($v_{1},v_{3}$), ($v_{1},v_{4}$), ($v_{2},v_{3}$), ($v_{2},v_{4}$), ($v_{3},v_{4}$), ($u_{1},v_{1}$) and ($u_{2},v_{2}$). The red line in fig 8 (resp. in fig 9) of the bottom array denotes the invariant set $v_1=0$ (resp. $v_2=0$) in the projection $u_1$ vs $v_1$ (resp.  $u_2$ vs $v_2$).]{\includegraphics[scale = 0.46]{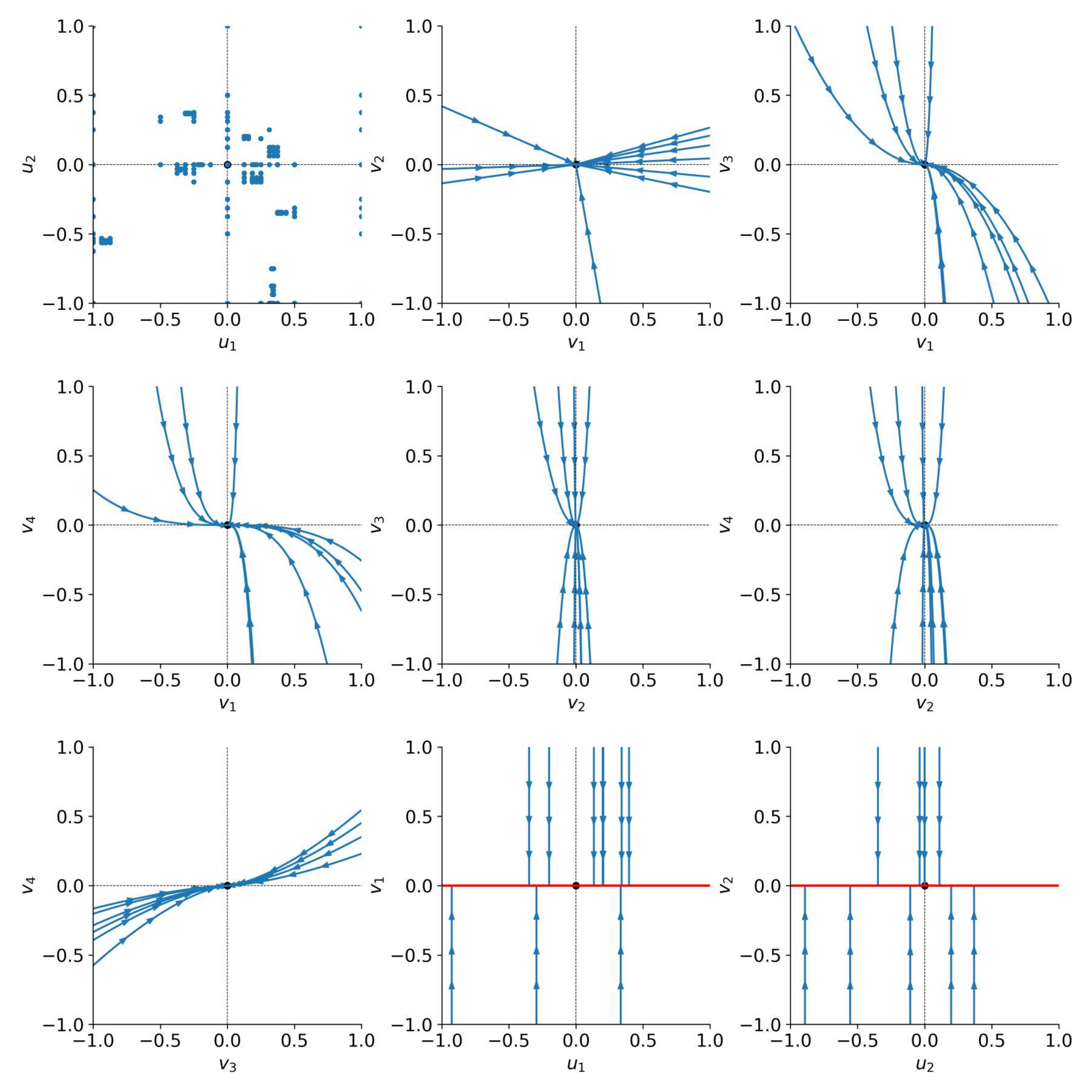}}
    \caption{Some solutions of (a) the system \eqref{systemphi1a}-\eqref{systemphi1c}  and (b) the system \eqref{system1}  for the potential $V_{I}(\phi,\psi)$ when $\omega_{m}=0$.  Notice in the projection $u_1$ vs $u_2$ (which contains the center manifold)  the origin is stable (but not asymptotically stable) since  any $\epsilon$-neighborhood of the origin  will contain a $\delta$-neighborhood of origin with other points apart of the origin with $(u_1', u_2')|_{u_1=u_1^*, u_2=u_2^*}=(0,0)$. Therefore, they remain  in $\delta$-neighborhood of origin.}
    \label{fig:PIwm0}
\end{figure*}

\begin{figure*}
    \centering
    \subfigure[Projections in the space ($\phi_{1}$, $\phi_{2}$, $\phi_{3}$) (left) and ($\Phi_{1}$, $\Phi_{2}$, $\Phi_{3}$) (right). We have represented the spheres $(\phi_{1}-1)^{2}+(\phi_{2}-1)^{2}+(\phi_{3}-1)^{2}=r^2$ and $\Phi_{1}^{2}+\Phi_{2}^{2}+\Phi_{3}^{2}=r^2$,  $r\in\{1,\sqrt{2}\}$.]{\includegraphics[scale = 0.29]{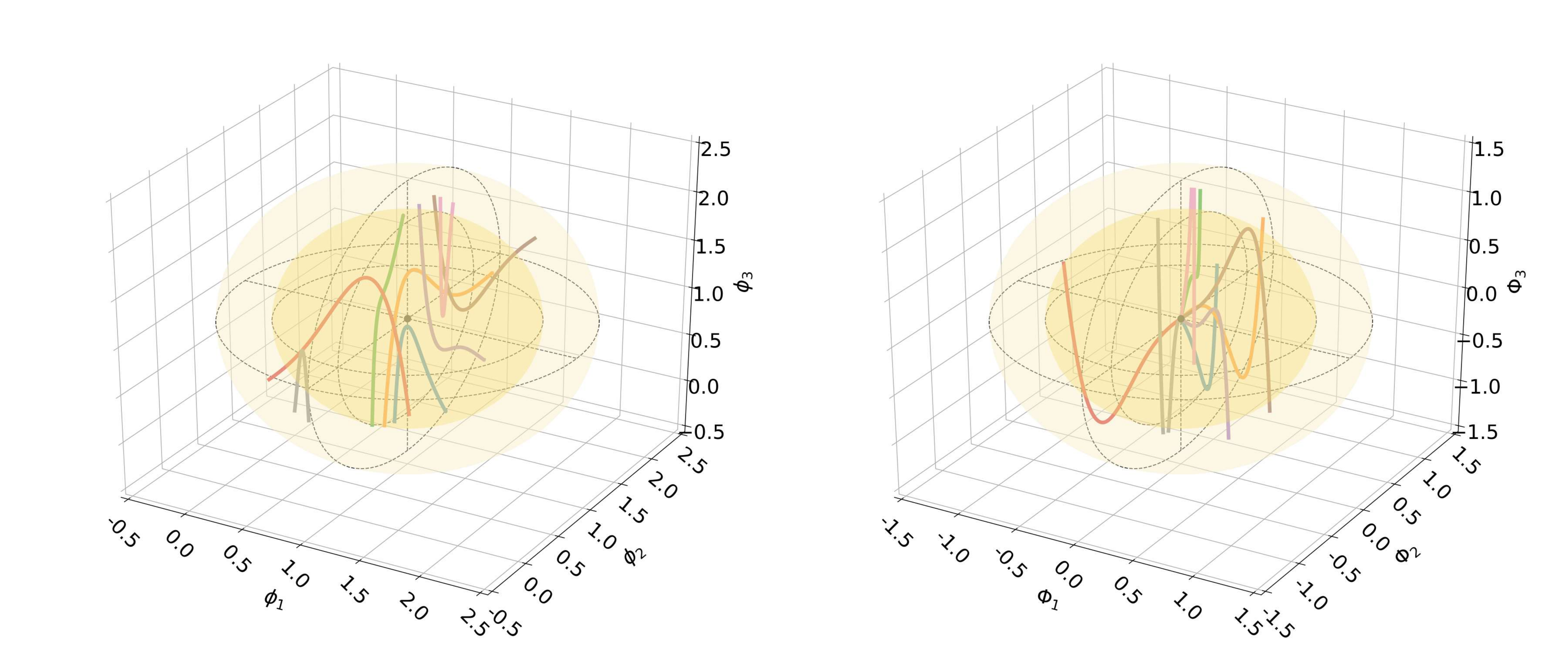}}
    \subfigure[From left to right, from top to bottom:  Projections in the planes ($u_{1},u_
    {2}$), ($v_{1},v_{2}$), ($v_{1},v_{3}$), ($v_{1},v_{4}$), ($v_{2},v_{3}$), ($v_{2},v_{4}$), ($v_{3},v_{4}$), ($u_{1},v_{1}$) and ($u_{2},v_{2}$). The red line in fig 8 (resp. in fig 9) of the bottom array denotes the invariant set $v_1=0$ (resp. $v_2=0$) in the projection $u_1$ vs $v_1$ (resp.  $u_2$ vs $v_2$).]{\includegraphics[scale = 0.46]{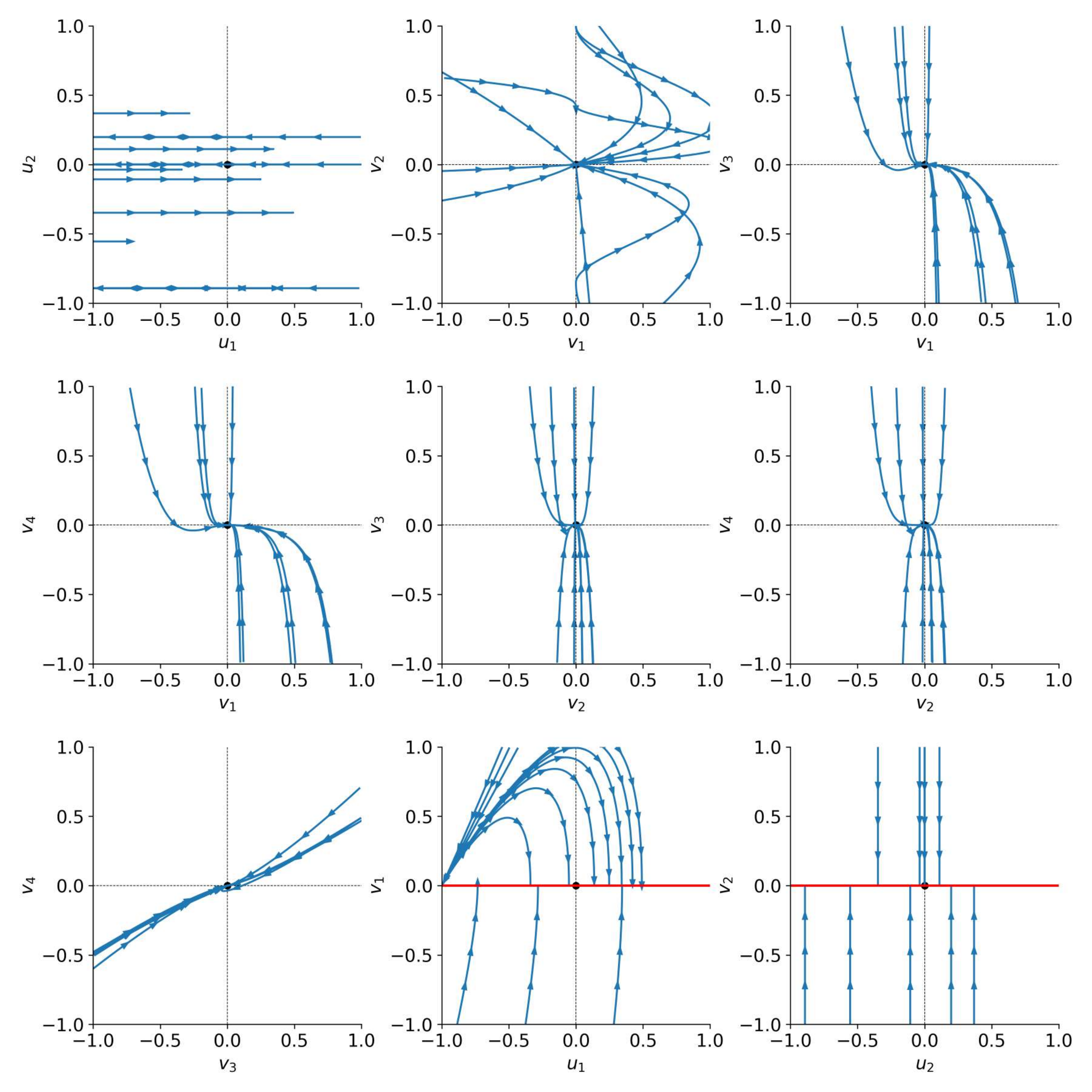}}
    \caption{Some solutions of (a) the system \eqref{systemphi1a}-\eqref{systemphi1c}  and (b) the system \eqref{system1}  for the potential $V_{I}(\phi,\psi)$ when $\omega_{m}=\frac{1}{3}$.  Notice in the projection $u_1$ vs $u_2$ (which contains the center manifold)  the origin behaves as a saddle point. The orbits for the left along the $u_1$-axis tend to the origin, but for the right the orbits depart from the origin. }
    \label{fig:PIwm1f3}
\end{figure*}
we obtain the system 
\begin{subequations}
\label{system1}
\begin{align}
 & u_1'= \frac{4 v_1^2 w_m (3 w_m+1)}{(w_m+3) (u_1 (w_m+3)-v_1 (3 w_m+1)+w_m+3)},\\
 & u_2'= 0, \\
 & v_1'= \frac{v_1 (-u_1 (w_m+3)+7 v_1 w_m+v_1-w_m-3)}{u_1 (w_m+3)-v_1 (3 w_m+1)+w_m+3},\\
 & v_2'= -v_2,\\
 & v_3'= \frac{2 (w_m+1)
   \left(v_3 (u_1 (w_m+3)-3 v_1 w_m-v_1+w_m+3)+4 v_1^2 w_m\right)}{(w_m-1) (u_1 (w_m+3)-v_1 (3 w_m+1)+w_m+3)},\\
 & v_4'= \frac{v_4 (w_m+3) (u_1 (w_m+3)-3 v_1 w_m-v_1+w_m+3)-4 v_1^2 w_m (3
   w_m+1)}{(w_m-1) (u_1 (w_m+3)-v_1 (3 w_m+1)+w_m+3)}.
\end{align}
\end{subequations}
The center manifold of the origin is given by 
\begin{equation}
    \Bigg\{(u_1, u_2,v_1, v_2, v_3, v_4)\in \mathbb{R}^6: v_i=h_i(u_1, u_2), \frac{\partial h_i}{\partial u_1}(0,0)=0, \frac{\partial h_i}{\partial u_2}(0,0)=0, h_i(0, 0)=0, i=1 \ldots 4 \Bigg\}.
\end{equation}
where $h_i(u_1,u_2)$ satisfies the system of quasi-linear partial differential equations
\begin{align}
  & h_1 \left(h_1 \left(4 w_m (3 w_m+1) \frac{\partial h_1}{\partial u_1}+(-w_m-3) (7 w_m+1)\right)+(u_1+1) (w_m+3)^2\right)=0,\\
  & 4 w_m (3 w_m+1) h_1^2
   \frac{\partial h_2}{\partial u_1}+(w_m+3) h_2 ((u_1+1) (w_m+3)-(3 w_m+1) h_1)=0,\\
  & 2 w_m ((2-3 w_m) w_m+1) h_1^2
   \frac{\partial h_3}{\partial u_1}+(w_m+1) (w_m+3) \left(h_3 ((u_1+1) (w_m+3)-(3 w_m+1) h_1)+4 w_m h_1^2\right)=0,\\
   & -(u_1+1) (w_m+3)^3
   h_4 + (3 w_m+1) h_1 \left(4 w_m h_1 \left((w_m-1) \frac{\partial h_4}{\partial u_1}+w_m+3\right)+(w_m+3)^2 h_4\right)=0.
\end{align}
Using the expansion in series 
\begin{equation}
    h_i(u_1, u_2)= \sum_{n=2}^{N}\sum_{k=0}^{n} a^{[i]}_{n k} u_1^{n-k} u_2^{k} + \mathcal{O}(\|(u_1,u_2)\|^{N+1}), i=1, \ldots 4  
\end{equation}
the zero solution for any given accuracy is found.  

In figures \ref{fig:PIwm-1}-\ref{fig:PIwm1f3} some solutions of the systems \eqref{systemphi1a}-\eqref{systemphi1c} and \eqref{system1} for the potential $V_{I}(\phi,\psi)$ when $\omega_{m}=-1$, $0$ and $\frac{1}{3}$ are represented. More specific, in the upper panel the solutions are projected in the space ($\phi_{1}$, $\phi_{2}$, $\phi_{3}$) and ($\Phi_{1}$, $\Phi_{2}$, $\Phi_{3}$), where we have represented the spheres $(\phi_{1}-1)^{2}+(\phi_{2}-1)^{2}+(\phi_{3}-1)^{2}=r^2$ and $\Phi_{1}^{2}+\Phi_{2}^{2}+\Phi_{3}^{2}=r^2$ with $r\in\{1,\sqrt{2}\}$. In the lower panel projections in the spaces ($u_{1},u_{2}$), ($v_{1},v_{2}$), ($v_{1},v_{3}$), ($v_{1},v_{4}$), ($v_{2},v_{3}$), ($v_{2},v_{4}$), ($v_{3},v_{4}$), ($u_{1},v_{1}$) and ($u_{2},v_{2}$) are represented. In these figures we have depicted a red line in the projections ($u_{1},v_{1}$) and ($u_{2},v_{2}$) which denotes the invariant set $v_{1}=0$ and $v_{2}=0$, respectively. Both lines are stable in these projections. Notice that in figures \ref{fig:PIwm-1} and \ref{fig:PIwm1f3}, the projection $u_1$ vs $u_2$, which contains the center manifold; the origin behaves as a saddle point. The orbits from the left along the $u_1$-axis tend to the origin, but from the right the orbits depart from the origin. Then, the solution is unstable (saddle behavior). This behavior is also represented in the 3D  projection $(\phi_1, \phi_2,\phi_3)$ where some orbits abandon the inner spheres backward and forward in time. On the other hand, the projection $u_1$ vs $u_2$ in figure \ref{fig:PIwm0}; the origin is stable (but not asymptotically stable) since  any $\epsilon$-neighborhood of the origin  will contain a $\delta$-neighborhood of origin with other points apart of the origin with $(u_1', u_2')|_{u_1=u_1^*, u_2=u_2^*}=(0,0)$. Therefore, they remain in $\delta$-neighborhood of origin. 

\subsection{Scalar field potential $V_{II}\left(  \phi,\psi\right)$}

In this section we study the stability of the solution \eqref{dc.52} of the system \eqref{dc.50}, \eqref{dc.51}. We  set for simplicity the integration constants $t_1, t_2, t_3$ to zero  because they are not relevant as $t\rightarrow \infty$.  Noticing for the scalar field potential $V_{II}\left(  \phi,\psi\right)  ~$we derive
$U_{eff}^{II}\left(  x,y,z\right)  =V_{0}z^{-2\frac{w_{m}+1}{w_{m}-1}}$. We
observe that doing the change of variables $\left(  y,z\right)
\rightarrow\left(  z,y\right)  $ we end with the dynamical system of potential
$V_{I}\left(  \phi,\psi\right)  $. Hence, with the time variable $\tau=\ln (t)$, and defining the new variables $Z= V_0 e^{2 \tau} z^{-\frac{3w_{m}+1}{w_{m}-1}}$ and $Y=y(\tau)-y_0 e^{\tau}$
the field equations \eqref{dc.50}, \eqref{dc.51} become 
\begin{align*}
& x''(\tau)-x'(\tau)=0,  \quad  Y''(\tau) - Y'(\tau) +\frac{3}{2}\frac{w_m+1}{w_{m}-1}Z(\tau)=0,\\
& -4 w_m Z'(\tau )^2+Z(\tau ) \left((3 w_m+1) Z''(\tau )+(w_m-5) Z'(\tau)\right)+2 (w_m+3) Z(\tau )^2=0.
\end{align*}
The analytical solution of the original system becomes
\begin{equation*}
x_c\left( \tau\right)  =x_{0} e^{\tau}, ~Y_c\left(  \tau \right)  = y_1  e^{-\frac{\tau  (w_m+3)}{w_m-1}}, \; y_1=-\frac{3 (w-1) Z_0}{4 (w+3)}, \; 
Z_c\left(  \tau\right)
=Z_{0} e^{-\frac{\tau  (w_m+3)}{w_m-1}}, \quad Z_0=V_0 z_0^{-\frac{3 w_m+1}{w_m-1}},
\end{equation*}
Defining the dimensionless variables 
\begin{equation*}
    \phi_1= \frac{x}{x_c}, \quad \phi_2= \frac{Y}{Y_c}, \quad \phi_3= \frac{Z}{Z_c},
\end{equation*}
we obtain the dynamical system 
\begin{align}
& \phi_1'= \Phi_1, \quad  \Phi_1'=  -\Phi_1, \label{systemphi2a}\\
& \phi_2' =\Phi_2, \quad  \Phi_2'=  \frac{(3 w_m+5)  \Phi_{2}}{w_m-1}+\frac{2 (w_m+1) (w_m+3) ( \phi_{3} - \phi_{2})}{(w_m-1)^2}, \label{systemphi2b}\\
& \phi_3'= \Phi_3, \quad  \Phi_3'= -\Phi_3 +\frac{4 w_m \Phi_3^2}{\phi_3(1+3 w_m)}.\label{systemphi2c} 
\end{align}
Now we analyze the stability of the fixed point 
$P:= (\phi_1, \Phi_1, \phi_2, \Phi_2, \phi_3, \Phi_3)= (1,0,1,0,1,0)$.
The subsystems for $(\phi_1,\Phi_1)$, $(\phi_3,\Phi_3)$ and $(\phi_2, \Phi_2, \phi_3, \Phi_3)$ are  decoupled.

The Jacobian matrix of the full system is 
\begin{equation}
J:=    \left(
\begin{array}{cccccc}
 0 & 1 & 0 & 0 & 0 & 0 \\
 0 & -1 & 0 & 0 & 0 & 0 \\
 0 & 0 & 0 & 1 & 0 & 0 \\
 0 & 0 & -\frac{2 (w_m+1) (w_m+3)}{(w_m-1)^2} & 3+\frac{8}{w_m-1} & \frac{2 (w_m+1)
   (w_m+3)}{(w_m-1)^2} & 0 \\
 0 & 0 & 0 & 0 & 0 & 1 \\
 0 & 0 & 0 & 0 & -\frac{4 w_m  \Phi_{3}^2}{(3 w_m+1)  \phi_{3}^2} &
   \frac{8 w_m \Phi_{3}}{(3 w_m +1) \phi_{3}}-1 \\
\end{array}
\right).
\end{equation}
Evaluating $J$  at the fixed point $P$  the eigenvalues 
$\left\{0,0,-1,-1,-\frac{2 (w_m+1)}{1-w_m},-\frac{w_m+3}{1-w_m}\right\}$ are found.
The stable manifold is 4D if $-1<w_m<1$. 
If the analysis is restricted to the subspace $(\phi_2, \Phi_2, \phi_3, \Phi_3)$ the eigenvalues are
$\left\{0,-1,-\frac{2 (w_m+1)}{1-w_m},-\frac{w_m+3}{1-w_m}\right\}$.  The stable manifold in this subspace is 3D if $-1<w_m<1$. 

Introducing the new variables 
\begin{align*}
   & u_1=  \Phi_{3}+ \phi_{3}-1, \; u_2= \Phi_{1}+ \phi_{1}-1, \;  v_1= \Phi_{3}, \;  v_2= \Phi_{1},\\
   & v_3= \frac{2 (w_m+1) ( \Phi_{3} (w_m-1) (w_m+3)+(-3 w_m-1) ( \Phi _{2}-w_m ( \Phi_{2}+ \phi_{3})+(w_m+3) \phi_{2}-3  \phi_{3}))}{(w_m-1)^2 (3 w_m+1)},\\
   & v_4= \frac{(w_m+3) ( \Phi_{2}+ \Phi_{3}-w_m ( \Phi_{2}+ \Phi_{3}+2  \phi_{3})+2 (w_m+1) \phi_{2}-2  \phi_{3})}{(w_m-1)^2},
\end{align*}

\begin{figure*}
    \centering
    \subfigure[Projections in the space ($\phi_{1}$, $\phi_{2}$, $\phi_{3}$) (left) and ($\Phi_{1}$, $\Phi_{2}$, $\Phi_{3}$) (right). We have represented the spheres $(\phi_{1}-1)^{2}+(\phi_{2}-1)^{2}+(\phi_{3}-1)^{2}=r^2$ and $\Phi_{1}^{2}+\Phi_{2}^{2}+\Phi_{3}^{2}=r^2$,  $r\in\{1,\sqrt{2}\}$.]{\includegraphics[scale = 0.29]{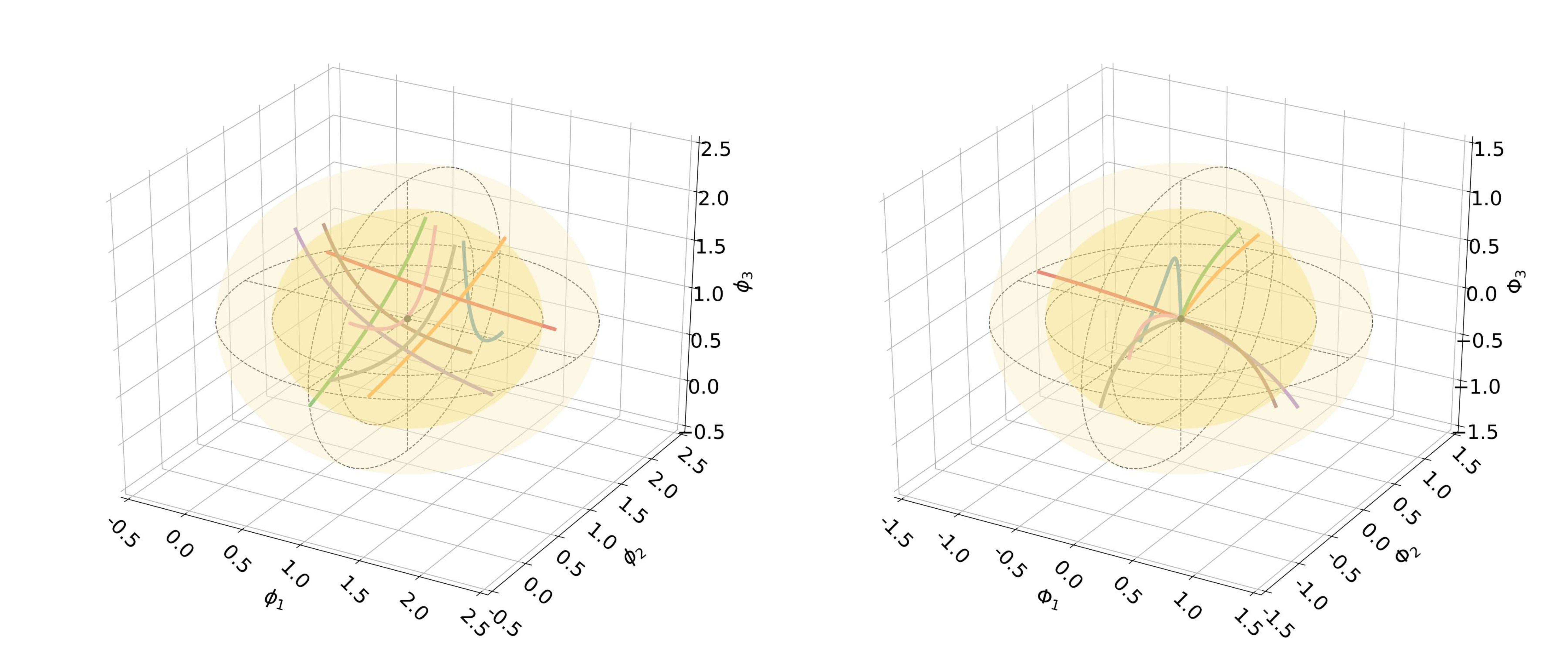}}
    \subfigure[From left to right, from top to bottom: Projections in the planes ($u_{1},u_
    {2}$), ($v_{1},v_{2}$), ($v_{1},v_{3}$), ($v_{1},v_{4}$), ($v_{2},v_{3}$), ($v_{2},v_{4}$), ($v_{3},v_{4}$), ($u_{1},v_{1}$) and ($u_{2},v_{2}$). The red line in fig 8 (resp. fig 9) of the bottom array denotes the invariant set $v_1=0$ (resp. $v_{2}=0$) in the projection $u_1$ vs $v_1$ (resp. $u_{2}$ vs $v_{2}$).]{\includegraphics[scale = 0.46]{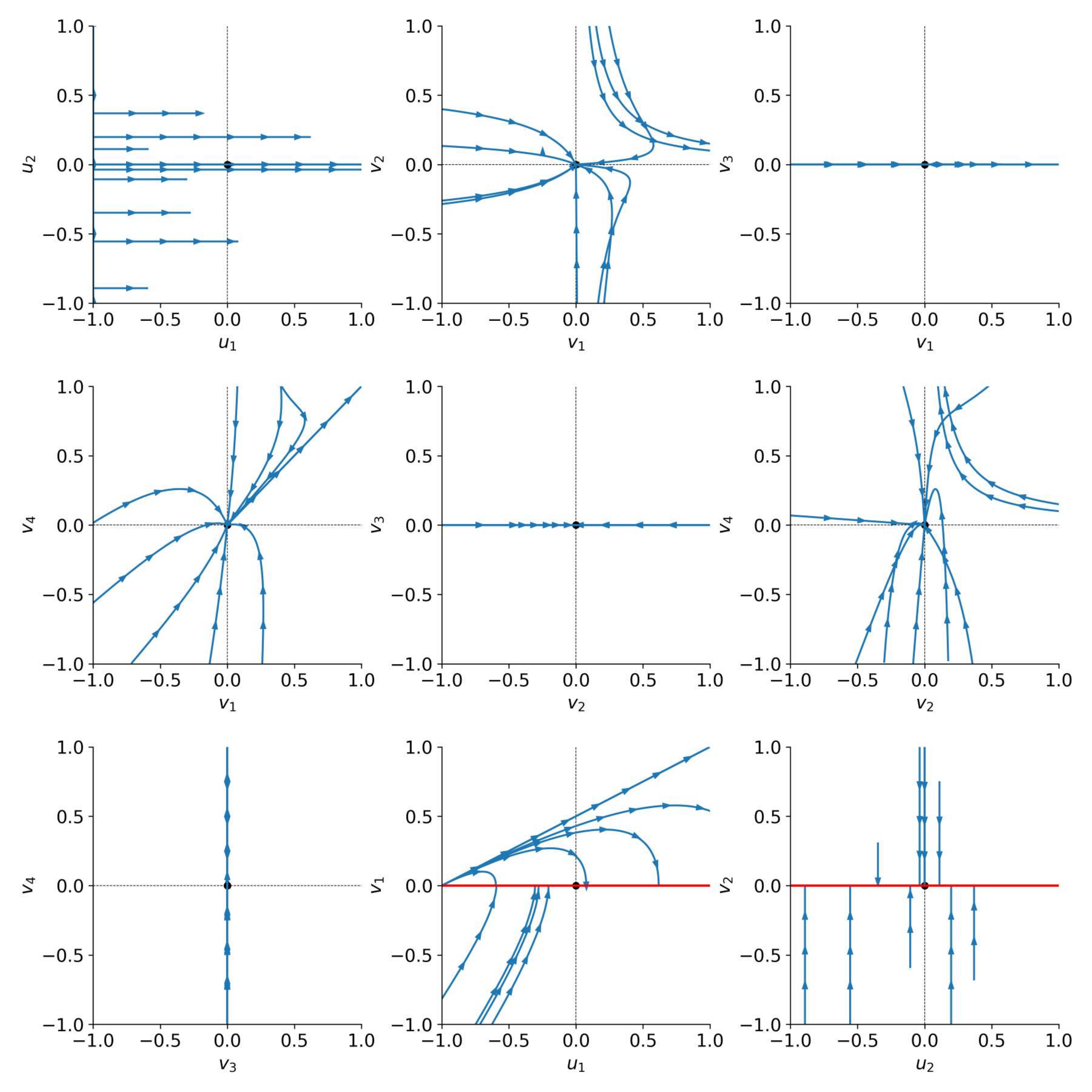}}
    \caption{Some solutions of (a) the system \eqref{systemphi2a}-\eqref{systemphi2c}  and (b) the system \eqref{system2}  for the potential $V_{II}(\phi,\psi)$ when $\omega_{m}=-1$.  Notice in the projection $u_1$ vs $u_2$ (which contains the center manifold)  the origin behaves as a saddle point. The orbits for the left along the $u_1$-axis tend to the origin, but for the right the orbits depart from the origin.}
    \label{fig:PIIwm-1}
\end{figure*}

\begin{figure*}
    \centering
    \subfigure[Projections in the space ($\phi_{1}$, $\phi_{2}$, $\phi_{3}$) (left) and ($\Phi_{1}$, $\Phi_{2}$, $\Phi_{3}$) (right). We have represented the spheres $(\phi_{1}-1)^{2}+(\phi_{2}-1)^{2}+(\phi_{3}-1)^{2}=r^2$ and $\Phi_{1}^{2}+\Phi_{2}^{2}+\Phi_{3}^{2}=r^2$,  $r\in\{1,\sqrt{2}\}$.]{\includegraphics[scale = 0.29]{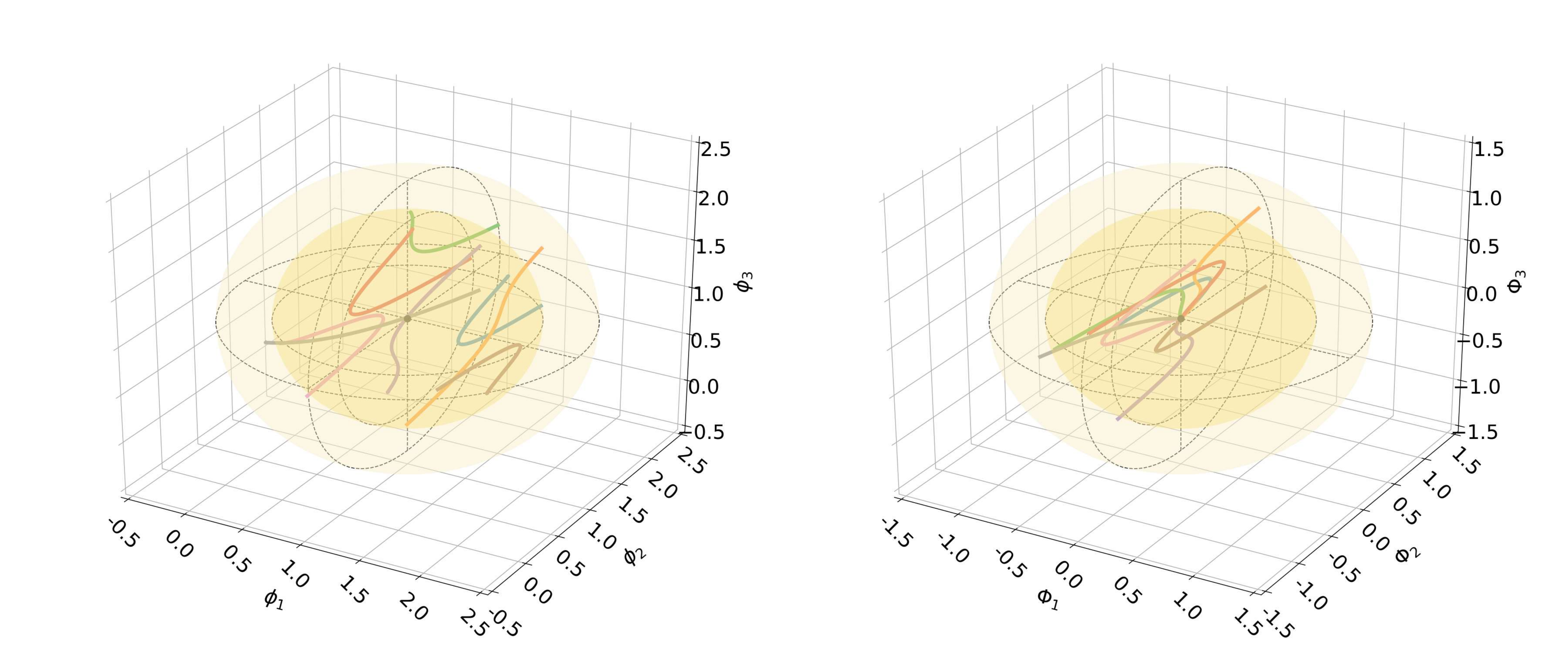}}
    \subfigure[From left to right, from top to bottom: Projections in the planes ($u_{1},u_
    {2}$), ($v_{1},v_{2}$), ($v_{1},v_{3}$), ($v_{1},v_{4}$), ($v_{2},v_{3}$), ($v_{2},v_{4}$), ($v_{3},v_{4}$), ($u_{1},v_{1}$) and ($u_{2},v_{2}$). The red line in fig 8 (resp. fig 9) of the bottom array denotes the invariant set $v_1=0$ (resp. $v_{2}=0$) in the projection $u_1$ vs $v_1$ (resp. $u_{2}$ vs $v_{2}$).]{\includegraphics[scale = 0.46]{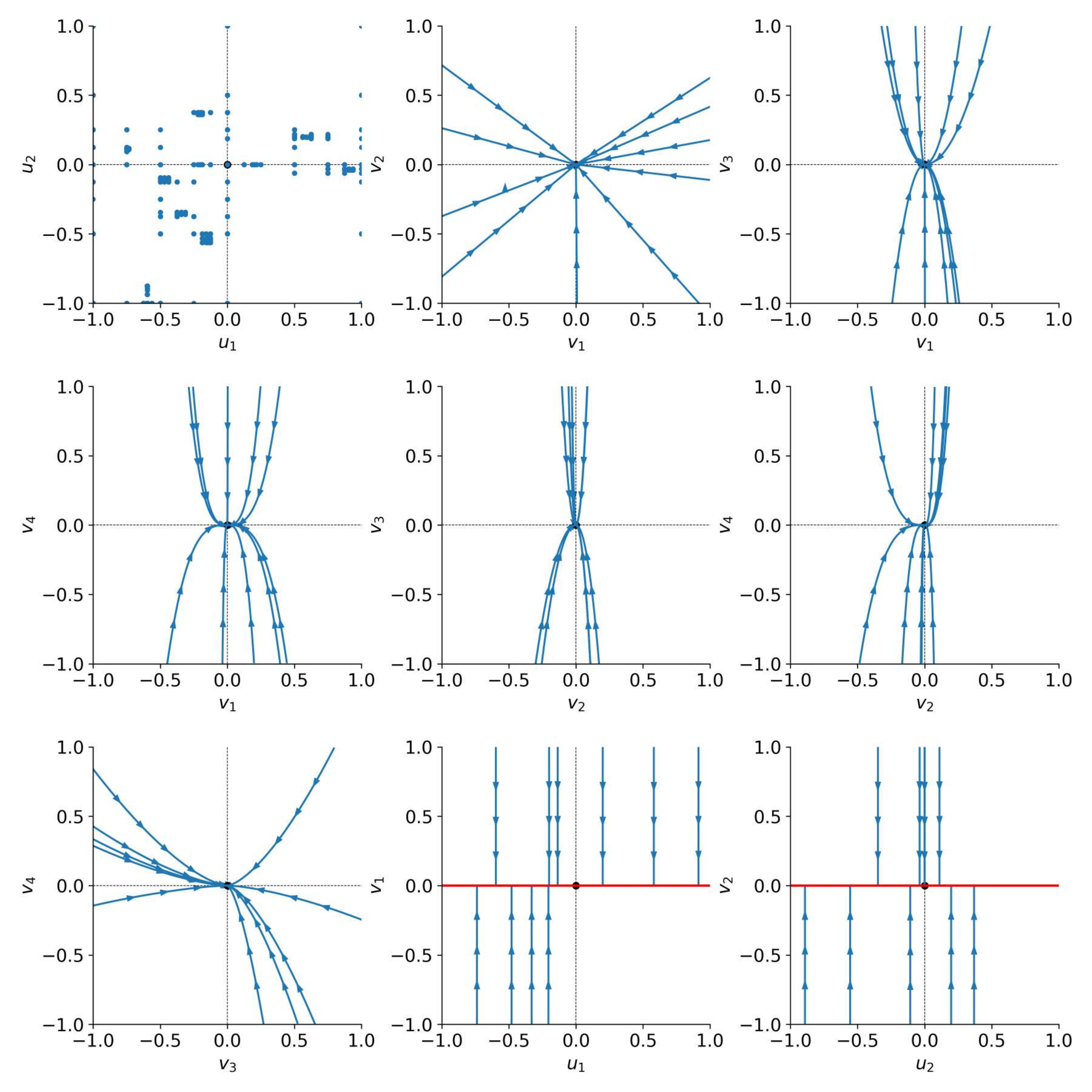}}
    \caption{Some solutions of (a) the system \eqref{systemphi2a}-\eqref{systemphi2c} and (b) the system \eqref{system2}  for the potential $V_{II}(\phi,\psi)$ when $\omega_{m}=0$. Notice in the projection $u_1$ vs $u_2$ (which contains the center manifold)  the origin is stable (but not asymptotically stable) since  any $\epsilon$-neighborhood of the origin  will contain a $\delta$-neighborhood of origin with other points apart of the origin with $(u_1', u_2')|_{u_1=u_1^*, u_2=u_2^*}=(0,0)$. Therefore, they remain  in $\delta$-neighborhood of origin.}
    \label{fig:PIIwm0}
\end{figure*}

\begin{figure*}
    \centering
    \subfigure[Projections in the space ($\phi_{1}$, $\phi_{2}$, $\phi_{3}$) (left) and ($\Phi_{1}$, $\Phi_{2}$, $\Phi_{3}$) (right). We have represented the spheres $(\phi_{1}-1)^{2}+(\phi_{2}-1)^{2}+(\phi_{3}-1)^{2}=r^2$ and $\Phi_{1}^{2}+\Phi_{2}^{2}+\Phi_{3}^{2}=r^2$,  $r\in\{1,\sqrt{2}\}$.]{\includegraphics[scale = 0.29]{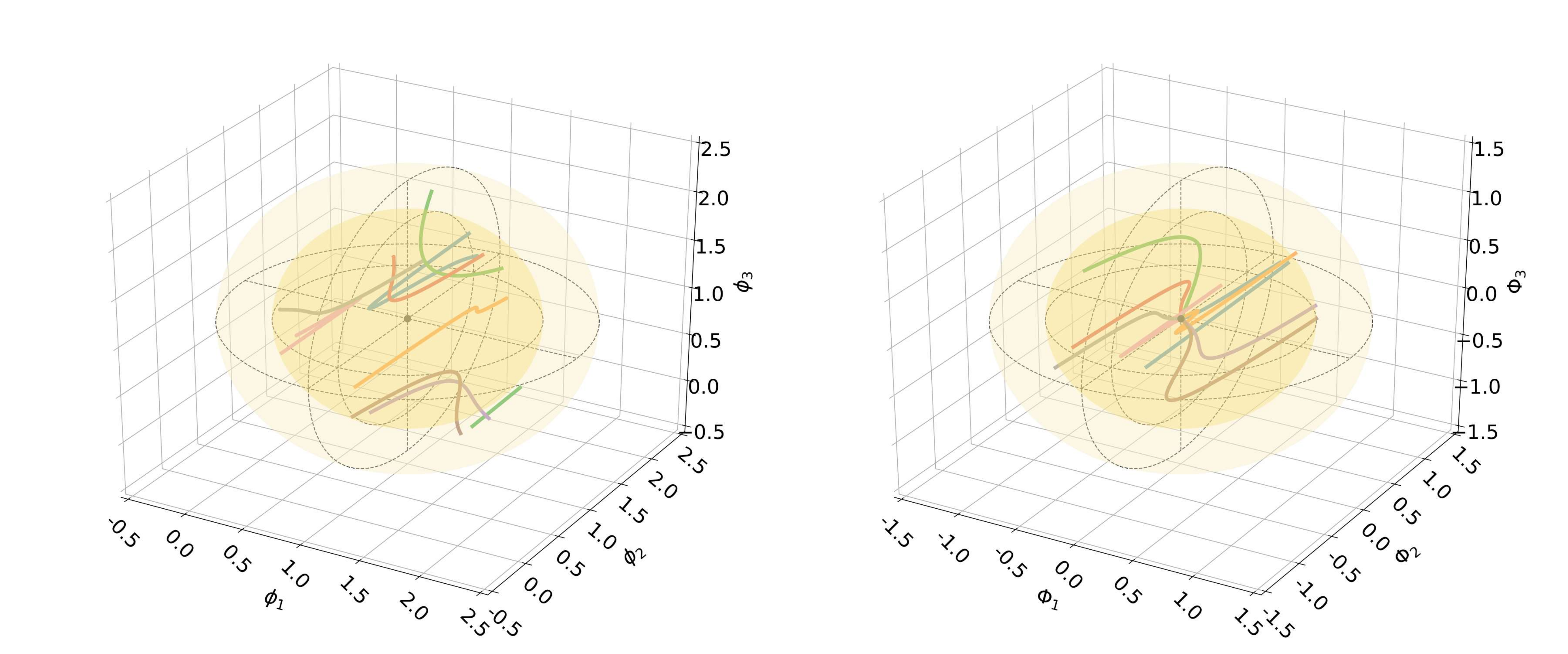}}
    \subfigure[From left to right, from top to bottom: Projections in the planes ($u_{1},u_
    {2}$), ($v_{1},v_{2}$), ($v_{1},v_{3}$), ($v_{1},v_{4}$), ($v_{2},v_{3}$), ($v_{2},v_{4}$), ($v_{3},v_{4}$), ($u_{1},v_{1}$) and ($u_{2},v_{2}$). The red line in fig 8 (resp. fig 9) of the bottom array denotes the invariant set $v_1=0$ (resp. $v_{2}=0$) in the projection $u_1$ vs $v_1$ (resp. $u_{2}$ vs $v_{2}$).]{\includegraphics[scale = 0.46]{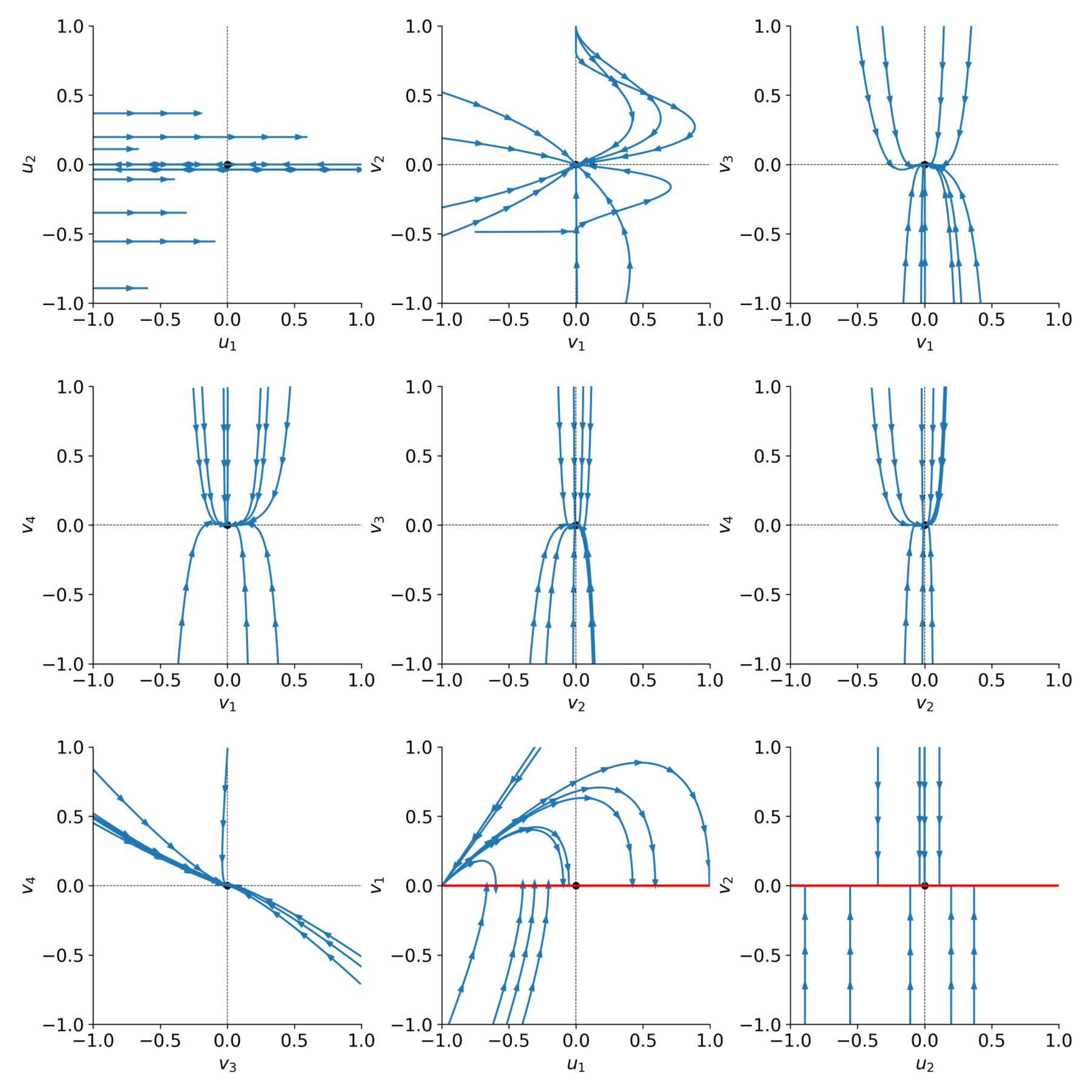}}
    \caption{Some solutions of (a) the system \eqref{systemphi2a}-\eqref{systemphi2c}  and (b) the system \eqref{system2}  for the potential $V_{II}(\phi,\psi)$ when $\omega_{m}=\frac{1}{3}$.  Notice in the projection $u_1$ vs $u_2$ (which contains the center manifold)  the origin behaves as a saddle point. The orbits for the left along the $u_1$-axis tend to the origin, but for the right the orbits depart from the origin.}
    \label{fig:PIIwm1f3}
\end{figure*}
we obtain 
\begin{subequations}
\label{system2}
\begin{align}
  & u_1'=  \frac{4 v_1^2 w_m}{(3 w_m+1) (u_1-v_1+1)}, \\
  & u_2'= 0,\\
  & v_1'= v_1 \left(\frac{4 v_1 w_m}{(3 w_m+1) (u_1-v_1+1)}-1\right), \\
  & v_2'= -v_2, \\
  & v_3'= \frac{2 (w_m+1) \left(v_3 (3 w_m+1)^2 (u_1-v_1+1)+4
   v_1^2 w_m (w_m+3)\right)}{(w_m-1) (3 w_m+1)^2 (u_1-v_1+1)},\\
   & v_4'= \frac{(w_m+3) \left(v_4 (3 w_m+1) (u_1-v_1+1)-4 v_1^2 w_m\right)}{(w_m-1) (3 w_m+1) (u_1-v_1+1)}.
\end{align}
\end{subequations}
The center manifold of the origin is given by 
\begin{equation}
    \Bigg\{(u_1, u_2,v_1, v_2, v_3, v_4)\in \mathbb{R}^6: v_i=h_i(u_1, u_2), \frac{\partial h_i}{\partial u_1}(0,0)=0, \frac{\partial h_i}{\partial u_2}(0,0)=0, h_i(0, 0)=0, i=1 \ldots 4 \Bigg\}.
\end{equation}
where $h_i(u_1,u_2)$ satisfies the system of quasi-linear partial differential equations
\begin{align}
& h_1  ((7 w_m +1) h_1 +(-u_1-1) (3 w_m +1))-4 w_m  h_1 ^2 \frac{\partial h_1}{\partial u_1}=0,\\
& 4 w_m  h_1 ^2
   \frac{\partial h_2}{\partial u_1} +(3 w_m +1) (-h_1 +u_1+1) h_2=0,\\
   & 2 w_m  ((2-3 w_m ) w_m +1) h_1 ^2 \frac{\partial h_3}{\partial u_1} +(w_m +1)
   \left((3 w_m +1)^2 (-h_1 +u_1+1) h_3 +4 w_m  (w_m +3) h_1 ^2\right)=0,\\
   & (w_m +3) \left((3 w_m +1) (-h_1 +u_1+1)
   h_4 -4 w_m  h_1 ^2\right)-4 (w_m -1) w_m  h_1 ^2 \frac{\partial h_4}{\partial u_1}=0.
   \end{align}
   Using the expansion in series 
\begin{equation}
    h_i(u_1, u_2)= \sum_{n=2}^{N}\sum_{k=0}^{n} a^{[i]}_{n k} u_1^{n-k} u_2^{k} + \mathcal{O}(\|(u_1,u_2)\|^{N+1}), i=1, \ldots 4  
\end{equation}
the zero solution is found for any given accuracy. 

In figures \ref{fig:PIIwm-1}-\ref{fig:PIIwm1f3} some solutions of the systems \eqref{systemphi2a}-\eqref{systemphi2c} and \eqref{system2} for the potential $V_{II}(\phi,\psi)$ when $\omega_{m}=-1$, $0$ and $\frac{1}{3}$ are represented. More specific, in the upper panel the solutions are projected in the space ($\phi_{1}$, $\phi_{2}$, $\phi_{3}$) and ($\Phi_{1}$, $\Phi_{2}$, $\Phi_{3}$), where we have represented the spheres $(\phi_{1}-1)^{2}+(\phi_{2}-1)^{2}+(\phi_{3}-1)^{2}=r^2$ and $\Phi_{1}^{2}+\Phi_{2}^{2}+\Phi_{3}^{2}=r^2$ with $r\in\{1,\sqrt{2}\}$. In the lower panel projections in the spaces ($u_{1},u_{2}$), ($v_{1},v_{2}$), ($v_{1},v_{3}$), ($v_{1},v_{4}$), ($v_{2},v_{3}$), ($v_{2},v_{4}$), ($v_{3},v_{4}$), ($u_{1},v_{1}$) and ($u_{2},v_{2}$) are represented. In these figures we have depicted a red line in the projections ($u_{1},v_{1}$) and ($u_{2},v_{2}$) which denotes the invariant set $v_{1}=0$ and $v_{2}=0$, respectively. Both lines are stable in these projections. Notice that in figures \ref{fig:PIIwm-1} and \ref{fig:PIIwm1f3}, the projection $u_1$ vs $u_2$ the origin behaves as a saddle point. The orbits from the left along the $u_1$-axis tend to the origin, but from the right the orbits depart from the origin. Then, the solution is unstable (saddle behavior). This behavior is also represented in the 3D  projection $(\phi_1, \phi_2,\phi_3)$ where some orbits abandon the inner spheres backward and forward in time. On the other hand, the projection $u_1$ vs $u_2$ in figure \ref{fig:PIIwm0} the origin is stable (but not asymptotically stable)
since  any $\epsilon$-neighborhood of the origin  will contain a $\delta$-neighborhood of origin with other points apart of the origin with $(u_1', u_2')|_{u_1=u_1^*, u_2=u_2^*}=(0,0)$. Therefore, they remain in $\delta$-neighborhood of origin. 

\subsection{Scalar field potential $V_{III}\left(  \phi,\psi\right)  $}

In this section we analyze the stability of the analytic solution \eqref{dc.16}  of equations \eqref{dc.17}, \eqref{yz}. We  set for simplicity the integration constants $t_1, t_2, t_3$ to zero  because they are not relevant as $t\rightarrow \infty$ and we assume  $w_m\neq 0, 1$.  

With the time variable $\tau=\ln (t)$  and defining the new variable $X= V_0 \varepsilon e^{2 \tau} x^{r}$, with $r=-\frac{4 w_m}{w_m-1}$ to balance the powers of $t$, 
the equations \eqref{dc.17}, \eqref{yz} become
\begin{align*}
&-32 w_m^2 (w_m+1) X(\tau )^3-(w_m-1)^2 \left((1-5 w_m) X'(\tau )^2-4 X(\tau ) \left(X'(\tau
   )-w_m X''(\tau )\right)+4 (w_m+1) X(\tau )^2\right)=0, \\
& y''(\tau)-y'(\tau)=0, \quad  z''(\tau)-z'(\tau)=0. 
\end{align*}
The analytical solution of the original system becomes
\begin{equation*}
    X_c(t)= -\frac{(w_m-1)^2}{8 w_m^2}, \; y_c(\tau)= y_0 e^{\tau}, \; z_c(\tau)= z_0 e^{\tau}, 
\end{equation*}
Defining the dimensionless variables 
\begin{equation*}
    \phi_1= \frac{X}{X_c}, \quad \phi_2= \frac{y}{y_c}, \quad \phi_3= \frac{z}{z_c},
\end{equation*}
we obtain for $w_m\neq 0, 1$  the dynamical system 
\begin{align}
& \phi_1'= \Phi_1, \quad  \Phi_1'=  \frac{ \Phi_{1}}{w_{m}}+\frac{ \Phi_{1}^2 (5 w_{m}-1)}{4 w_{m}
    \phi_{1}}+\frac{(w_{m}+1) ( \phi_{1}-1)  \phi_{1}}{w_{m}}, \label{systemphi3a}\\
& \phi_2' =\Phi_2, \quad  \Phi_2'=  -\Phi_2, \label{systemphi3b}\\
& \phi_3'= \Phi_3, \quad  \Phi_3'= -\Phi_3 \label{systemphi3c}. 
\end{align}
Now we analyze the stability of the fixed point 
$P:= (\phi_1, \Phi_1, \phi_2, \Phi_2, \phi_3, \Phi_3)= (1,0,1,0,1,0)$.
The subsystems for $(\phi_1,\Phi_1)$, $(\phi_2,\Phi_2)$ and $(\phi_3, \Phi_3)$ are  decoupled.

The Jacobian matrix of the full system is 
\begin{equation}
J:=    \left(
\begin{array}{cccccc}
 0 & 1 & 0 & 0 & 0 & 0 \\
 \frac{4 (w_m+1) (2  \phi_{1}-1) \phi_{1}^2+(1-5 w_m)  \Phi_{1}^2}{4 w_m
    \phi_{1}^2} & \frac{\frac{(5 w_m-1)  \Phi_{1}}{ \phi_{1}}+2}{2 w_m}
   & 0 & 0 & 0 & 0 \\
 0 & 0 & 0 & 1 & 0 & 0 \\
 0 & 0 & 0 & -1 & 0 & 0 \\
 0 & 0 & 0 & 0 & 0 & 1 \\
 0 & 0 & 0 & 0 & 0 & -1 \\
\end{array}
\right).
\end{equation}
Evaluating $J$  at the fixed point $P$  the eigenvalues  $\left\{0,0,-1,-1,-1,\frac{1}{w_m}+1\right\}$ are obtained. 
Therefore, if $-1<w_m<0$ the stable manifold of $P$ for the full system is 4D. If the analysis is restricted to the subspace $(\phi_2, \Phi_2, \phi_3, \Phi_3)$ the eigenvalues are
$\left\{0,0,-1,-1\right\}$.  The stable manifold in this subspace is 2D for all $w_m \neq 0, 1$.

Defining the new variables 
\begin{align*}
& u_1=  \Phi_{3}+ \phi_{3}-1, \;  u_2=  \Phi_{2}+ \phi_{2}-1, \;  v_1=  \Phi_{3}, \;  v_2= \Phi_{2},\\
& v_3=  \frac{\Phi_{1} w_m+(-w_m-1) (\phi_{1}-1)}{2
   w_m+1}, \;  v_4=  \frac{(w_m+1) ( \Phi_{1}+ \phi_{1}-1)}{2w_m+1}.
\end{align*}

\begin{figure*}
    \centering
    \includegraphics[scale = 0.60]{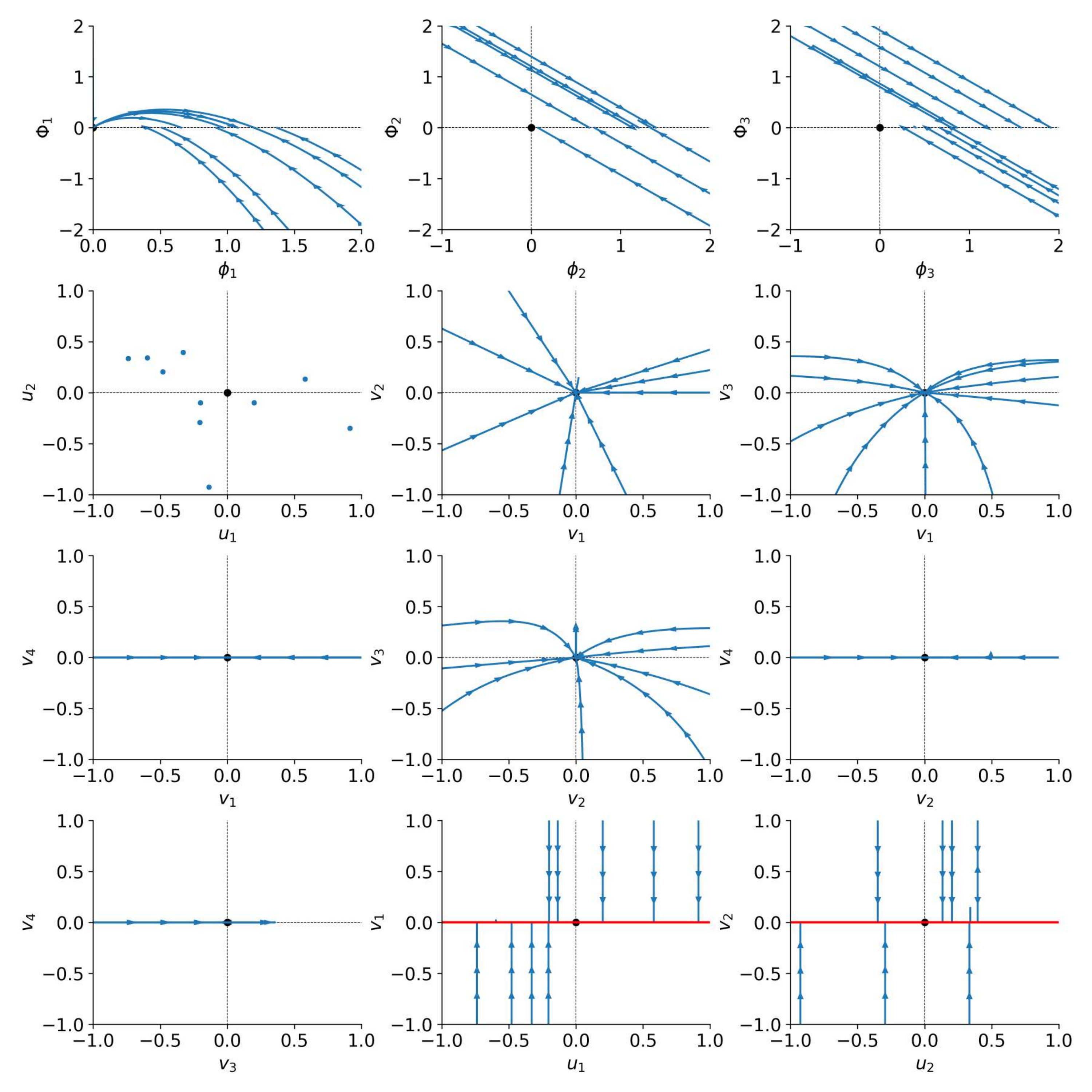}
    \caption{Some solutions of (a) the system \eqref{systemphi3a}-\eqref{systemphi3c} and (b) the system \eqref{system3} for the potential $V_{III}(\phi,\psi)$ when $\omega_{m}=-1$. From left to right, from top to bottom: Projections in the planes $(\phi_{1},\Phi_{1})$, $(\phi_{2},\Phi_{2})$, $(\phi_{3},\Phi_{3})$, $(u_{1}, u_{2})$, ($v_{1},v_{2}$), ($v_{1},v_{3}$), ($v_{1},v_{4}$), ($v_{2},v_{3}$), ($v_{2},v_{4}$), ($v_{3},v_{4}$), ($u_{1},v_{1}$) and ($u_{2},v_{2}$). The red line in fig 11 (resp. fig 12) of the bottom array denotes the invariant set $v_{1}=0$ (resp. $v_{2}=0$) in the projections $u_{1}$ vs $v_{1}$ (resp. $u_{2}$ vs $v_{2}$). Notice in the projection $u_1$ vs $u_2$ (which contains the center manifold)  the origin is stable (but not asymptotically stable) since  any $\epsilon$-neighborhood of the origin  will contain a $\delta$-neighborhood of origin with other points apart of the origin with $(u_1', u_2')|_{u_1=u_1^*, u_2=u_2^*}=(0,0)$. Therefore, they remain  in $\delta$-neighborhood of origin.}
    \label{fig:PIIIwm-1}
\end{figure*}

\begin{figure*}
    \centering
    \includegraphics[scale = 0.60]{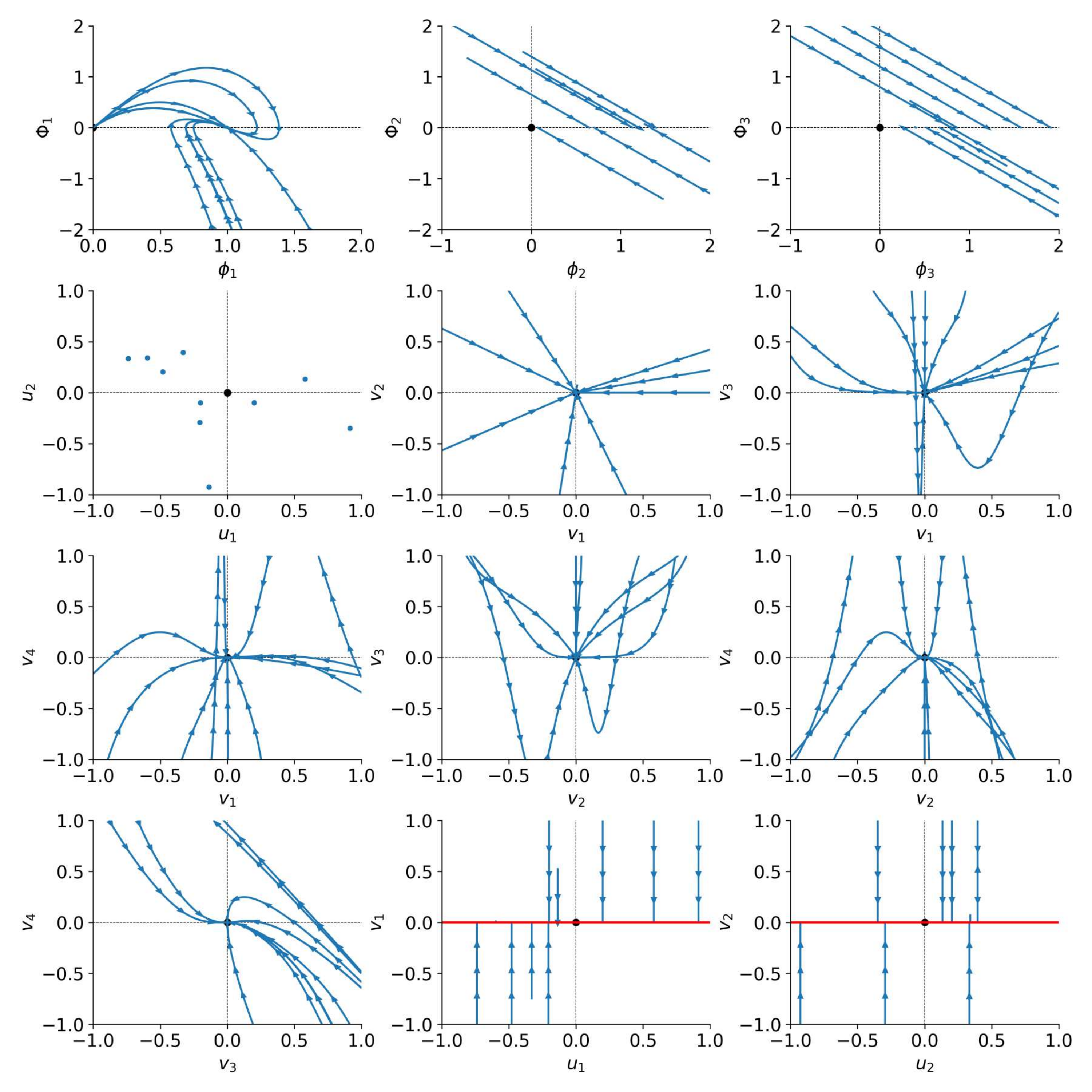}
    \caption{Some solutions of (a) the system \eqref{systemphi3a}-\eqref{systemphi3c} and (b) the system \eqref{system3} for the potential $V_{III}(\phi,\psi)$ when $\omega_{m}=-\frac{1}{3}$. From left to right, from top to bottom: Projections in the planes $(\phi_{1},\Phi_{1})$, $(\phi_{2},\Phi_{2})$, $(\phi_{3},\Phi_{3})$, $(u_{1}, u_{2})$, ($v_{1},v_{2}$), ($v_{1},v_{3}$), ($v_{1},v_{4}$), ($v_{2},v_{3}$), ($v_{2},v_{4}$), ($v_{3},v_{4}$), ($u_{1},v_{1}$) and ($u_{2},v_{2}$). The red line in fig 11 (resp. fig 12) of the bottom array denotes the invariant set $v_{1}=0$ (resp. $v_{2}=0$) in the projections $u_{1}$ vs $v_{1}$ (resp. $u_{2}$ vs $v_{2}$). Notice in the projection $u_1$ vs $u_2$ (which contains the center manifold)  the origin is stable (but not asymptotically stable) since  any $\epsilon$-neighborhood of the origin  will contain a $\delta$-neighborhood of origin with other points apart of the origin with $(u_1', u_2')|_{u_1=u_1^*, u_2=u_2^*}=(0,0)$. Therefore, they remain  in $\delta$-neighborhood of origin.}
    \label{fig:PIIIwm-1f3}
\end{figure*}

\begin{figure*}
    \centering
    \includegraphics[height=1.0\textwidth]{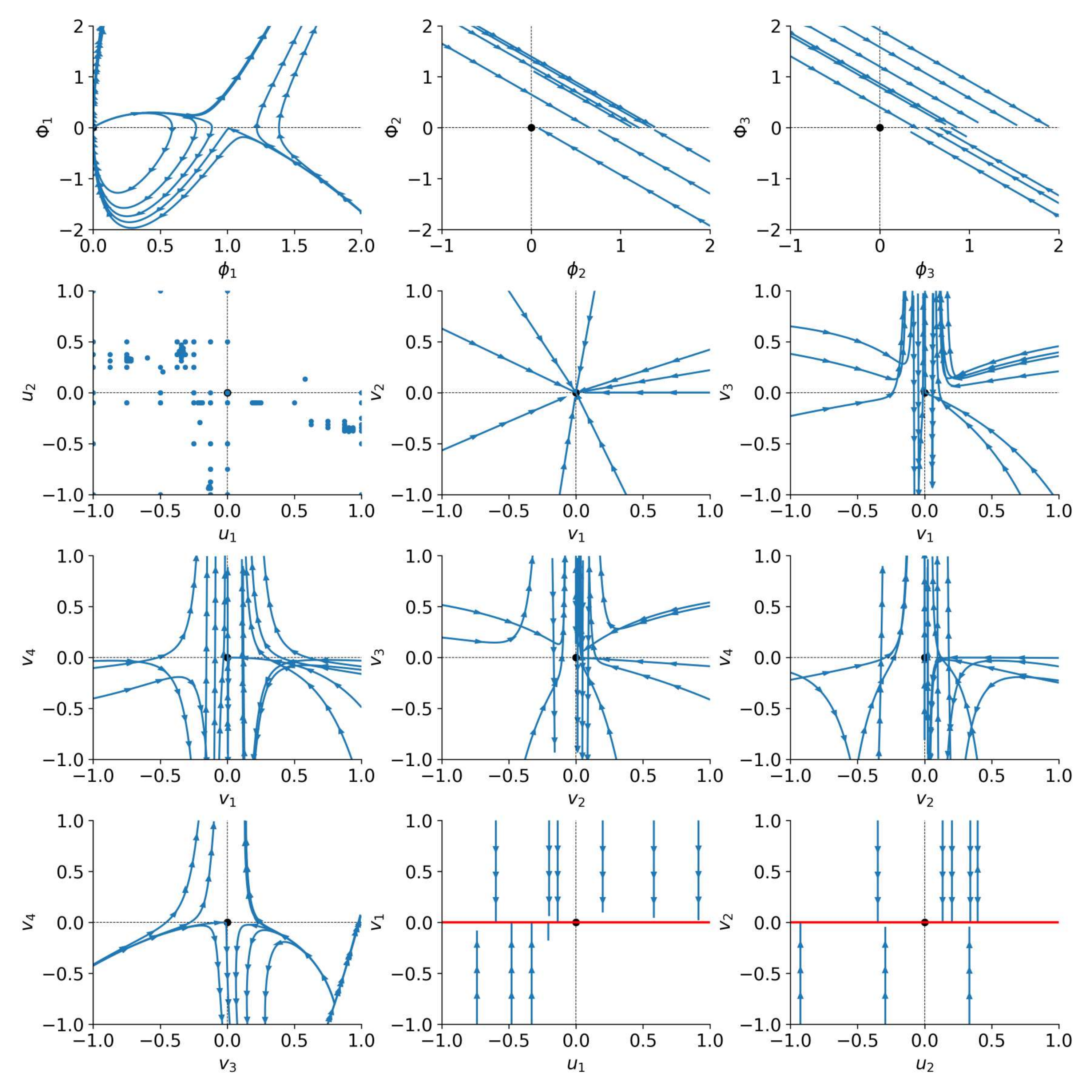}
    \caption{Some solutions of (a) the system \eqref{systemphi3a}-\eqref{systemphi3c} and (b) the system \eqref{system3} for the potential $V_{III}(\phi,\psi)$ when $\omega_{m}=\frac{1}{3}$. From left to right, from top to bottom: Projections in the planes $(\phi_{1},\Phi_{1})$, $(\phi_{2},\Phi_{2})$, $(\phi_{3},\Phi_{3})$, $(u_{1}, u_{2})$, ($v_{1},v_{2}$), ($v_{1},v_{3}$), ($v_{1},v_{4}$), ($v_{2},v_{3}$), ($v_{2},v_{4}$), ($v_{3},v_{4}$), ($u_{1},v_{1}$) and ($u_{2},v_{2}$). The red line in fig 11 (resp. fig 12) of the bottom array denotes the invariant set $v_{1}=0$ (resp. $v_{2}=0$) in the projections $u_{1}$ vs $v_{1}$ (resp. $u_{2}$ vs $v_{2}$). This plot shows that the equilibrium point is a saddle for $w_m>0$.}
    \label{fig:PIIIwm1f3}
\end{figure*}
Using the previous variables we obtain the decoupled equations: 
\begin{subequations}
\label{system3}
\begin{align}
    & u_1'= 0,\\
    & u_2'= 0,\\
    & v_1'= -v_1,\\
    & v_2'= -v_2,\\
    & v_3'= \frac{\frac{(5 w_m-1)
   (v_3+v_4)^2}{-v_3-\frac{v_4}{w_m+1}+v_4+1}-4 (w_m+1)
   (v_3+v_4)+\frac{4 (w_m (v_3-v_4-1)+v_3-1) (v_3
   w_m+v_3-v_4 w_m)}{w_m+1}+4 (v_3+v_4)}{4 (2 w_m+1)},\\
   & v_4'= \frac{(w_m+1)
   \left(\frac{(5 w_m-1) (v_3+v_4)^2}{4 w_m
   \left(-v_3-\frac{v_4}{w_m+1}+v_4+1\right)}+\frac{v_3+v_4}{w_m}+\frac{(w_m (v_3-v_4-1)+v_3-1) (v_3
   w_m+v_3-v_4 w_m)}{w_m (w_m+1)}+v_3+v_4\right)}{2 w_m+1}.
\end{align}
\end{subequations}
In figures \ref{fig:PIIIwm-1}-\ref{fig:PIIIwm1f3} some solutions of the systems \eqref{systemphi3a}-\eqref{systemphi3c} and \eqref{system3}   for the potential $V_{III}(\phi,\psi)$ when $\omega_{m}=-1$, $-\frac{1}{3}$ and $\frac{1}{3}$ are represented. More specific,  projections in the spaces ($\phi_{1}$, $\Phi_{1}$), ($\phi_{2}$, $\Phi_{2}$), ($\phi_{3}$, $\Phi_{3}$), ($u_{1},u_{2}$), ($v_{1},v_{2}$), ($v_{1},v_{3}$), ($v_{1},v_{4}$), ($v_{2},v_{3}$), ($v_{2},v_{4}$), ($v_{3},v_{4}$), ($u_{1},v_{1}$) and ($u_{2},v_{2}$) are represented. In these figures we have depicted a red line in the projections ($u_{1},v_{1}$) and ($u_{2},v_{2}$) that denotes the invariant set $v_{1}=0$ and $v_{2}=0$, respectively. Both lines are stable in these projections. Notice that in figures \ref{fig:PIIIwm-1} and \ref{fig:PIIIwm-1f3} in the projection $u_1$ vs $u_2$ (which contains the center manifold)  the origin is stable (but not asymptotically stable) since  any $\epsilon$-neighborhood of the origin  will contain a $\delta$-neighborhood of origin with other points apart of the origin with $(u_1', u_2')|_{u_1=u_1^*, u_2=u_2^*}=(0,0)$. Therefore, they remain in $\delta$-neighborhood of origin if $-1<w_m<0$. On the other hand, in figure \ref{fig:PIIIwm1f3} the origin behaves as a saddle point since $w_m>0$.

\subsection{Scalar field potential $V_{IV}\left(  \phi,\psi\right)  $}
In this section we study the stability of the solution 
\eqref{solIVa}, \eqref{solIVb}, \eqref{solIVc} of \eqref{eqIVa}, \eqref{eqIVb}, \eqref{eqIVc}. 
We  set for simplicity the integration constants $t_1, t_2, t_3$ to zero  because they are not relevant as $t\rightarrow \infty$.

With the time variable $\tau=\ln (t)$, and defining the new variables $X(\tau)=x(\tau)-x_0 e^{\tau}$, $Y(\tau)=y(\tau)-y_0 e^{\tau}$ and $\mathcal{Z}(\tau)= 6 \bar{V}_0 e^{2 \tau} Z^{r}$ with $r=-\frac{3 w_m+1}{w_m-1}$  the equations \eqref{eqIVa}, \eqref{eqIVb}, \eqref{eqIVc} become
\begin{align*}
&X''(\tau)- X'(\tau)-\frac{w_{m}+1}{w_{m}-1}\beta \mathcal{Z}(\tau)   =0, \quad Y''(\tau)-Y'(\tau)+\frac{w_{m}+1}{w_{m}-1}\mathcal{Z}(\tau)   
=0,\\
& \mathcal{Z}''(\tau )=\frac{-(w_m-5) \mathcal{Z}(\tau ) \mathcal{Z}'(\tau )+4 w_m \mathcal{Z}'(\tau )^2-2 (w_m+3) \mathcal{Z}(\tau )^2}{(3 w_m+1) \mathcal{Z}(\tau )}-\frac{(w_m+1) (3 w_m+1) Z_1
   \varepsilon  e^{-\frac{2 \tau  (w_m-1)}{3 w_m+1}} \mathcal{Z}(\tau )^{\frac{7 w_m+1}{3 w_m+1}}}{(w_m-1)^2},
\end{align*}
where $Z_1=\beta ^2 2^{\frac{1-w_m}{3 w_m+1}} 3^{\frac{2 (w_m+1)}{3 w_m+1}} {\bar{V}_0}^{\frac{1-w_m}{3 w_m+1}}$.
The solution 
\eqref{solIVa}, \eqref{solIVb}, \eqref{solIVc} becomes 
\begin{align*}
& X_c\left(\tau\right)
=\frac{Z_0 e^{\frac{\tau  (w_m-1)}{2 w_m}}}{3 \beta},  \;  Y_c\left(  \tau\right) =-\frac{Z_0 e^{\frac{\tau  (w_m-1)}{2 w_m}}}{3 \beta ^2} \varepsilon,  \;  \mathcal{Z}_c\left(
\tau\right)=-\frac{(w_m-1)^2 Z_0 e^{\frac{\tau  (w_m-1)}{2 w_m}}}{12 \beta ^2 w_m^2  } \varepsilon. 
\end{align*}
Defining the dimensionless variables 
\begin{equation}
    \phi_1= \frac{X}{X_c}, \quad \phi_2= \frac{Y}{Y_c}, \quad \phi_3= \frac{\mathcal{Z}}{\mathcal{Z}_c},
\end{equation}
we obtain for $w_m\neq 0, -1/3$ the dynamical system 
\begin{align}
& \phi_1'=  \Phi_{1}, \;  \Phi_1'= \frac{ \Phi_{1}}{w_m}+\frac{\left(w_m^2-1\right)  \phi_{1}}{4 w_m^2}+\frac{1}{4} \left(\frac{1}{w_m^2}-1\right) \phi_{3}, \label{systemphi4a} \\
& \phi_2'=  \Phi_{2}, \;  \Phi_2'= \frac{ \Phi_{2}}{w_m}+\frac{\left(w_m^2-1\right) \ \phi_{2}}{4 w_m^2}+\frac{1}{4} \left(\frac{1}{w_m^2}-1\right)  \phi_{3}, \label{systemphi4b}  \\
& \phi_3'=  \Phi_{3}, \;  \Phi_3'= \frac{ \Phi_{3}}{w_m} +\frac{4
   \Phi_{3}^2 w_m}{(3 w_m+1)  \phi_{3}}+\frac{(w_m+1) (3 w_m+1)  \phi_{3}^{\frac{7 w_m+1}{3 w_m+1}}}{4 w_m^2}-\frac{(w_m+1) (3 w_m+1)  \phi_{3}}{4 w_m^2}. \label{systemphi4c} 
\end{align}
Now we analyze the stability of the fixed point 
$P:= (\phi_1, \Phi_1, \phi_2, \Phi_2, \phi_3, \Phi_3)= (1,0,1,0,1,0)$.
The Jacobian matrix of the full system is 
\begin{equation}
J:= \left(
\begin{array}{cccccc}
 0 & 1 & 0 & 0 & 0 & 0 \\
 \frac{1}{4}-\frac{1}{4 w_m^2} & \frac{1}{w_m} & 0 & 0 & \frac{1}{4} \left(\frac{1}{w_m^2}-1\right) & 0 \\
 0 & 0 & 0 & 1 & 0 & 0 \\
 0 & 0 & \frac{1}{4}-\frac{1}{4 w_m^2} & \frac{1}{w_m} & \frac{1}{4} \left(\frac{1}{w_m^2}-1\right) & 0 \\
 0 & 0 & 0 & 0 & 0 & 1 \\
 0 & 0 & 0 & 0 & \frac{(w_m+1) (7 w_m+1)  \phi_{3}^{\frac{4 w_m}{3 w_m+1}}-(w_m+1) (3 w_m+1)-\frac{16  \Phi_{3}^2 w_m^3}{(3 w_m+1)  \phi_{3}^2}}{4 w_m^2} & \frac{8  \Phi_{3} w_m}{3 w_m  \phi_{3}+ \phi_{3}}+\frac{1}{w_m} \\
\end{array}
\right).
\end{equation}
Evaluating $J$  at the fixed point $P$  the eigenvalues  $\left\{-1,-\frac{w_m-1}{2 w_m},-\frac{w_m-1}{2 w_m},\frac{w_m+1}{2 w_m},\frac{w_m+1}{2 w_m},\frac{ w_m+1}{w_m}\right\}$ are obtained. For $-1<w_m <0$ $P$ is a sink, or a saddle for $w_m>0$. 

Defining the variables
\begin{align*}
 & v_1= \frac{(w_m+1) \left(w_m^2 (6  \Phi _{2}+2  \Phi_{3}+3  \phi_{2}+ \phi_{3}-4)-2 w_m (- \Phi_{2}+ \Phi_{3}+\phi_{2}+ \phi_{3}-2)- \phi_{2}+ \phi_{3}\right)}{4 w_m^2 (3 w_m+1)},\\
& v_2= \frac{(w_m-1) \left(2 w_m (w_m (3 \Phi_{2}+\Phi_{3}+2)+ \Phi_{2}- \Phi_{3}+2)-(w_m+1) (3 w_m+1)  \phi_{2}+w_m^2 (- \phi_{3})+ \phi_{3}\right)}{4 w_m^2 (3 w_m+1)},\\
& v_3= \frac{ \Phi_{3} w_m+w_m-(w_m+1)  \phi_{3}+1}{2 w_m+1},\\
& v_4= \frac{(w_m+1) \left(w_m^2 (6  \Phi_{1}+2 \Phi_{3}+3  \phi_{1}+ \phi_{3}-4)-2 w_m (- \Phi_{1}+ \Phi_{3}+ \phi_{1}+ \phi _{3}-2)- \phi_{1}+ \phi_{3}\right)}{4 w_m^2 (3 w_m+1)},\\
& v_5= \frac{(w_m-1) \left(2 w_m (w_m (3  \Phi_{1}+ \Phi_{3}+2)+ \Phi _{1}- \Phi_{3}+2)-(w_m+1) (3 w_m+1)  \phi_{1}+w_m^2 (- \phi _{3})+ \phi_{3}\right)}{4 w_m^2 (3 w_m+1)},\\
& v_6= \frac{(w_m+1) ( \Phi_{3}+ \phi_{3}-1)}{2 w_m+1},
\end{align*}

\begin{figure*}
    \centering
    \subfigure[Projections in the space ($\phi_{1}$, $\phi_{2}$, $\phi_{3}$) (left) and ($\Phi_{1}$, $\Phi_{2}$, $\Phi_{3}$) (right). We have represented the spheres $(\phi_{1}-1)^{2}+(\phi_{2}-1)^{2}+(\phi_{3}-1)^{2}=r^2$ and $\Phi_{1}^{2}+\Phi_{2}^{2}+\Phi_{3}^{2}=r^2$,  $r\in\{1,\sqrt{2}\}$.]{\includegraphics[scale = 0.29]{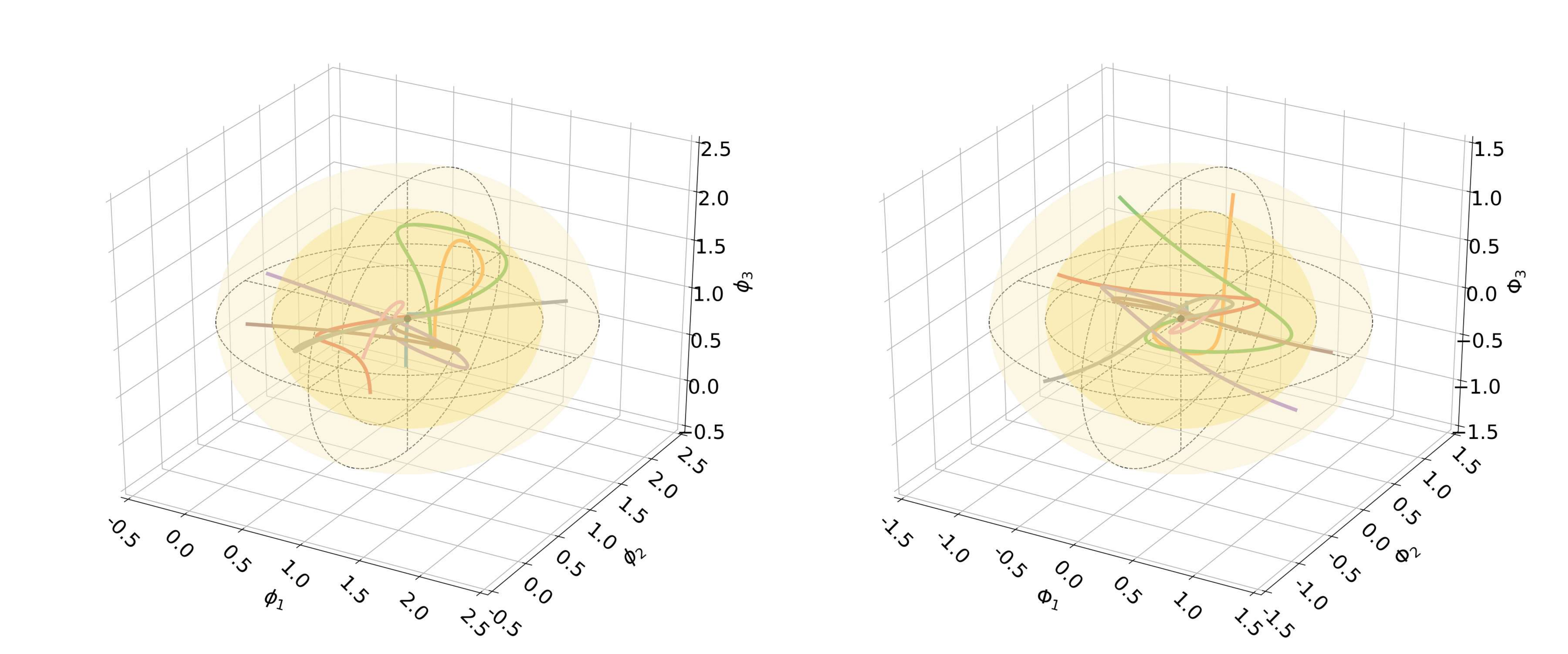}}
   \subfigure[From left to right, from top to bottom: Projections in the planes ($v_{1},v_
    {2}$), ($v_{1},v_{4}$), ($v_{1},v_{5}$), ($v_{2},v_{3}$), ($v_{2},v_{4}$), ($v_{2},v_{5}$), ($v_{3},v_{6}$), ($v_{4},v_{5}$) and ($v_{5},v_{6}$).]{\includegraphics[scale = 0.46]{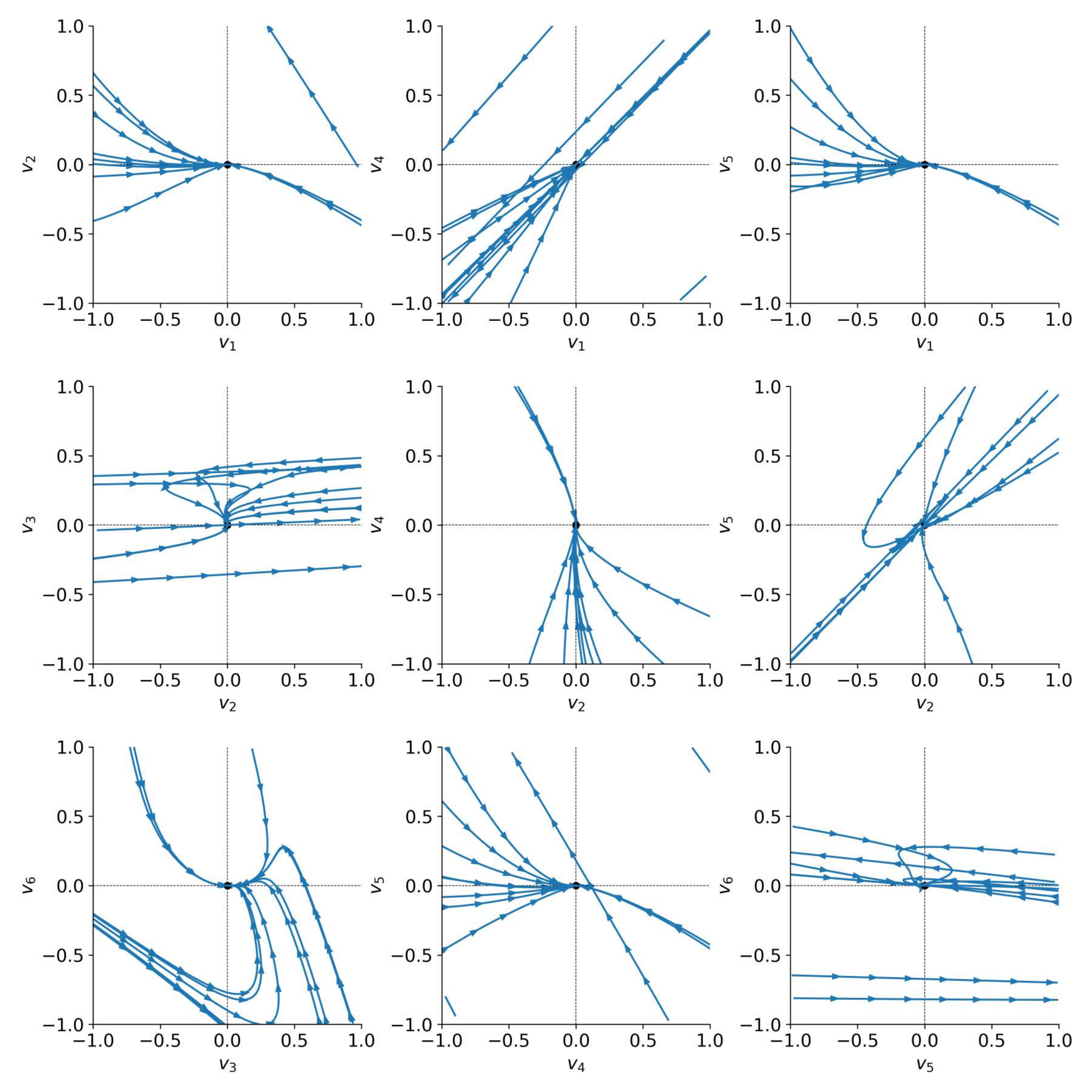}}
    \caption{Some solutions of (a) 
    the system \eqref{systemphi4a}-\eqref{systemphi4c} and (b) the system \eqref{system4} for the potential $V_{IV}(\phi,\psi)$ when $\omega_{m}=-\frac{1}{4}$. The point $P$ is asymptotically  stable.}
    \label{fig:PIVwm-1f4}
\end{figure*}

\begin{figure*}
    \centering
    \subfigure[Projections in the space ($\phi_{1}$, $\phi_{2}$, $\phi_{3}$) (left) and ($\Phi_{1}$, $\Phi_{2}$, $\Phi_{3}$) (right). We have represented the spheres $(\phi_{1}-1)^{2}+(\phi_{2}-1)^{2}+(\phi_{3}-1)^{2}=r^2$ and $\Phi_{1}^{2}+\Phi_{2}^{2}+\Phi_{3}^{2}=r^2$,  $r\in\{1,\sqrt{2}\}$.]{\includegraphics[scale = 0.29]{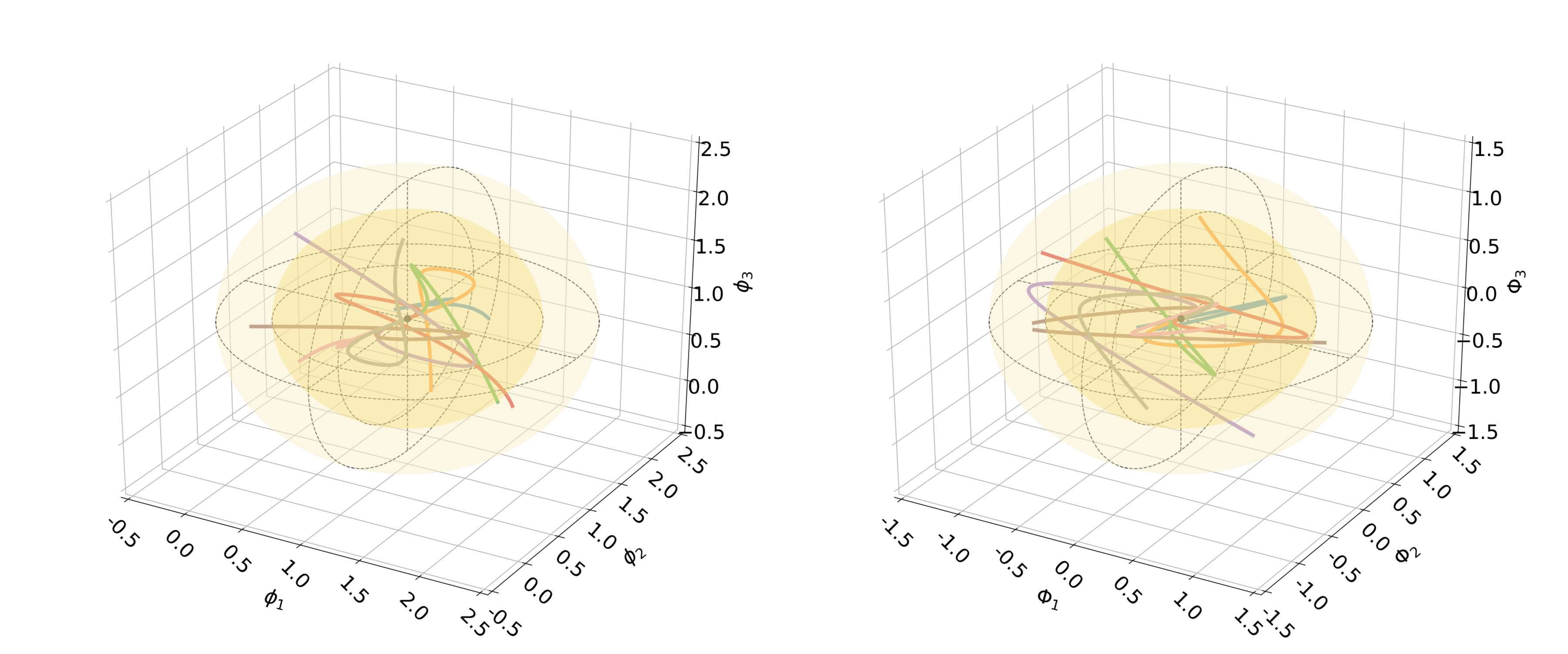}}
    \subfigure[From left to right, from top to bottom: Projections in the planes ($v_{1},v_
    {2}$), ($v_{1},v_{4}$), ($v_{1},v_{5}$), ($v_{2},v_{5}$), ($v_{2},v_{6}$), ($v_{3},v_{6}$), ($v_{4},v_{5}$), ($v_{4},v_{6}$) and ($v_{5},v_{6}$).]{\includegraphics[scale = 0.46]{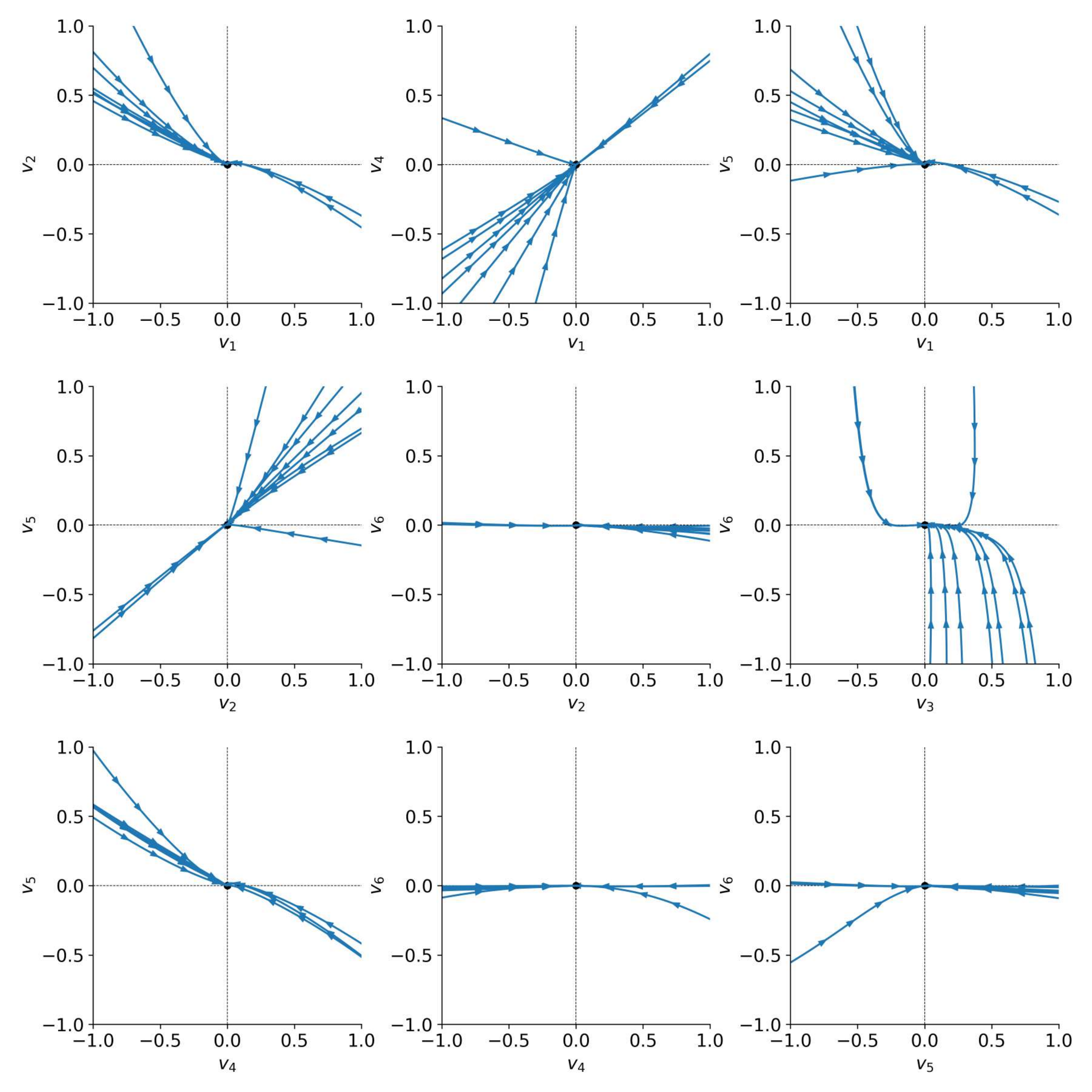}}
    \caption{Some solutions of (a) the system \eqref{systemphi4a}-\eqref{systemphi4c}  and (b) the system \eqref{system4} for the potential $V_{IV}(\phi,\psi)$ when $\omega_{m}=-\frac{1}{7}$. The point $P$ is asymptotically  stable.}
    \label{fig:PIVwm-1f7}
\end{figure*}

\begin{figure*}
    \centering
    \subfigure[Projections in the space ($\phi_{1}$, $\phi_{2}$, $\phi_{3}$) (left) and ($\Phi_{1}$, $\Phi_{2}$, $\Phi_{3}$) (right). We have represented the spheres $(\phi_{1}-1)^{2}+(\phi_{2}-1)^{2}+(\phi_{3}-1)^{2}=r^2$ and $\Phi_{1}^{2}+\Phi_{2}^{2}+\Phi_{3}^{2}=r^2$,  $r\in\{1,\sqrt{2}\}$.]{\includegraphics[scale = 0.29]{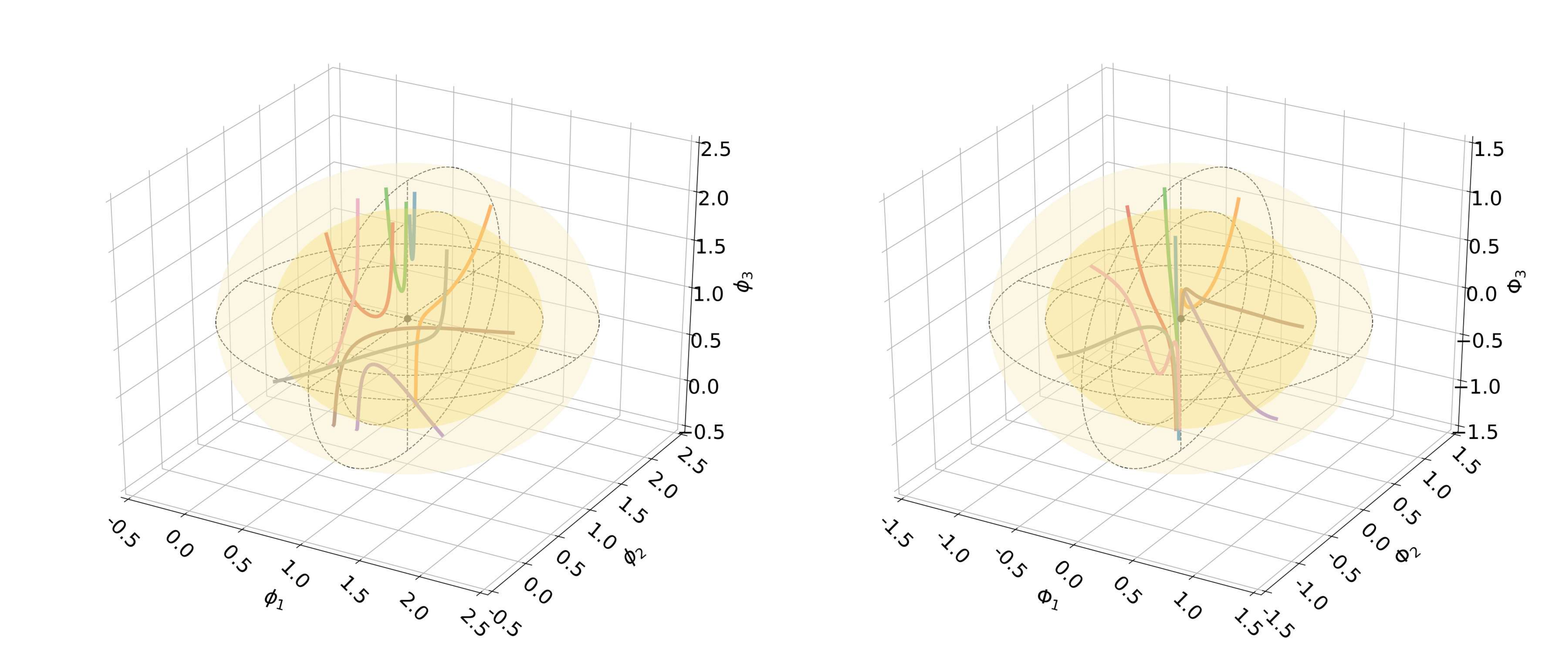}}
    \subfigure[From left to right, from top to bottom: Projections in the planes ($v_{1},v_
    {2}$), ($v_{1},v_{3}$), ($v_{1},v_{4}$), ($v_{1},v_{5}$), ($v_{1},v_{6}$), ($v_{3},v_{4}$), ($v_{3},v_{6}$), ($v_{4},v_{6}$) and ($v_{5},v_{6}$).]{\includegraphics[scale = 0.46]{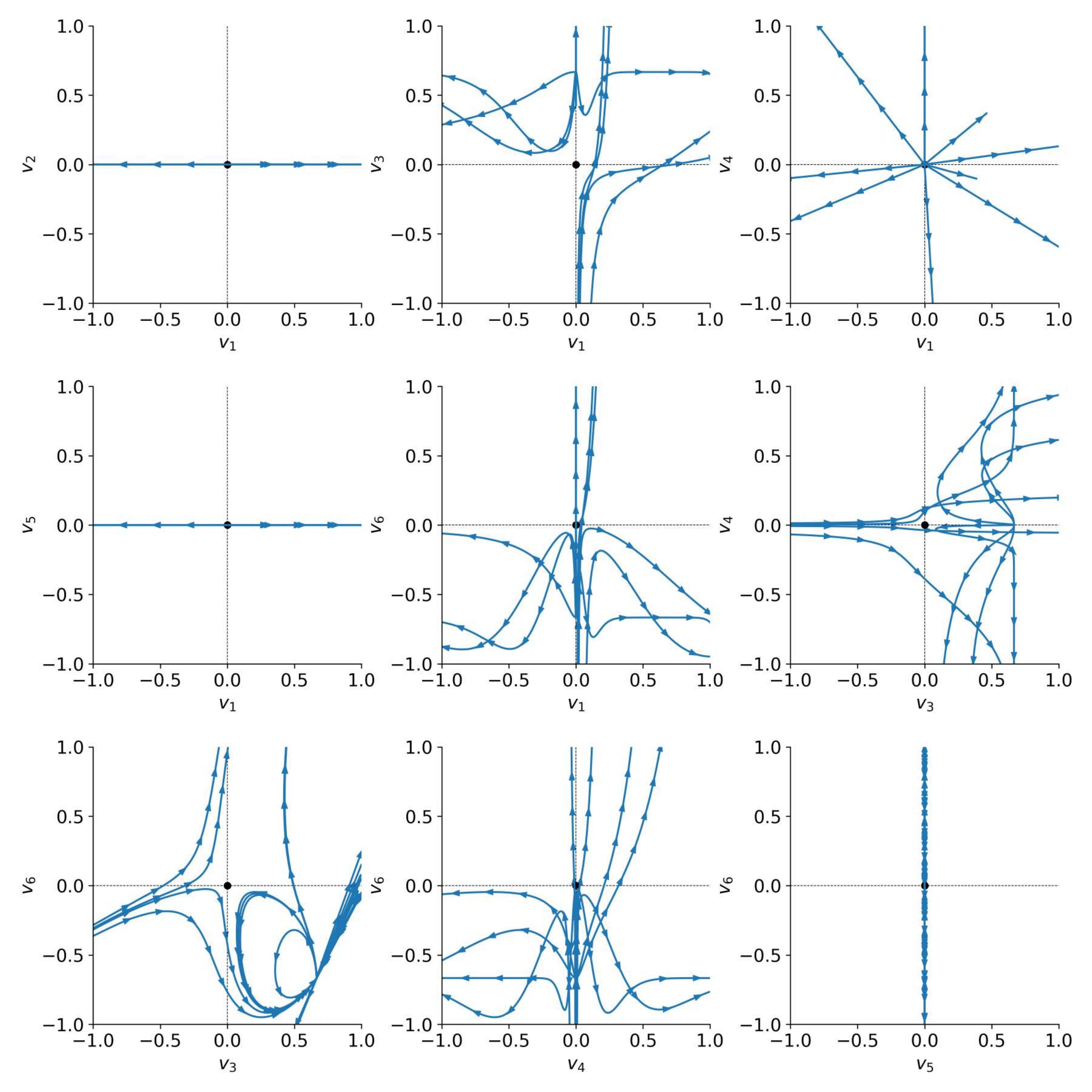}}
    \caption{Some solutions of (a) the system \eqref{systemphi4a}-\eqref{systemphi4c}  and (b) the system \eqref{system4} for the potential $V_{IV}(\phi,\psi)$ when $\omega_{m}=1$. The point $P$ is a saddle.}
    \label{fig:PIVwm1}
\end{figure*}
we obtain the system 
\begin{small}
\begin{subequations}
\label{system4}
\begin{align}
& v_1'= \frac{(w_m-1) (w_m+1)^2 \left(-v_3+v_6-\frac{v_6}{w_m+1}+1\right)^{\frac{7 w_m+1}{3 w_m+1}}}{8 w_m^3} \nonumber \\
& +\frac{1}{8 w_m^3 (3 w_m+1)^2 (v_3+(v_3-v_6-1) w_m-1)} \Bigg(v_3^2
   (w_m-1) (5 w_m+1) (w_m (w_m+6)+1) (w_m+1)^2 \nonumber \\
   & -2 v_3 \left(w_m \left(w_m \left(-2 v_1 (3 w_m+1)^2+w_m (15 w_m+8)-14\right)+v_6 (w_m-1) (w_m (37 w_m+10)+1)-8\right)-1\right) (w_m+1)^2 \nonumber \\
   & +w_m \Big(v_6^2
   w_m (w_m+1) (5 w_m-1) (w_m-1)^2-w_m \left(4 v_1 (w_m+1)^2 (3 w_m+1)^2-w_m (w_m (3
   w_m (3 w_m+8)+13)-16)+21\right) \nonumber \\
   & -2 v_6 (w_m+1) (3 w_m+1) (w_m (w_m (-5 w_m+v_1 (6 w_m+2)-1)+5)+1)-8\Big)-1\Bigg),
\\
  & v_2'= \frac{(w_m-1)^2 (w_m+1)
   \left(-v_3+v_6-\frac{v_6}{w_m+1}+1\right)^{\frac{7 w_m+1}{3 w_m+1}}}{8 w_m^3} \nonumber \\
   & +\frac{1}{8 w_m^3 (3
   w_m+1)^2 (v_3+(v_3-v_6-1) w_m-1)}\Bigg(v_3^2 (w_m+1) (5 w_m+1)
   (w_m (w_m+6)+1) (w_m-1)^2 \nonumber \\
   & +w_m \Big(v_6^2 w_m (5 w_m-1) (w_m-1)^3+2 v_6 (3 w_m+1) \left(w_m \left((6 v_2+5) w_m^2+2 v_2 w_m+w_m-5\right)-1\right) (w_m-1)\nonumber \\
   & +w_m \left(4 v_2 \left(w_m^2-1\right) (3 w_m+1)^2+w_m (3 w_m+4) (w_m (3 w_m-2)-3)+7\right)+6\Big)\nonumber \\
   & -2 v_3 \left(w_m^2-1\right) \left(w_m \left(w_m \left(2
   v_2 (3 w_m+1)^2+w_m (15 w_m+8)-14\right)+v_6 (w_m-1) (w_m (37 w_m+10)+1)-8\right)-1\right)+1\Bigg),
\\
   & v_3'=  \frac{\left(3 w_m^2+4 w_m+1\right) \left(-v_3+v_6-\frac{v_6}{w_m+1}+1\right)^{\frac{7
   w_m+1}{3 w_m+1}}}{8 w_m^2+4 w_m} \nonumber \\
   & +\frac{1}{4 w_m (2 w_m+1) (3 w_m+1) (v_3+(v_3-v_6-1) w_m-1)}\Bigg(-(w_m-1) (w_m+1) (w_m (19 w_m+8)+1) v_3^2 \nonumber \\
   & -2 (w_m
   (w_m (w_m+2) (3 w_m+10)+v_6 (w_m (w_m (25 w_m+37)+9)+1)+8)+1) v_3 \nonumber \\
   & +w_m \left((5 v_6 (v_6+6)+9)
   w_m^3+2 ((23-3 v_6) v_6+12) w_m^2+v_6 (v_6+18) w_m+22 w_m+2 v_6+8\right)+1\Bigg),
 \\
   & v_4'= \frac{(w_m-1) (w_m+1)^2 \left(-v_3+v_6-\frac{v_6}{w_m+1}+1\right)^{\frac{7 w_m+1}{3
   w_m+1}}}{8 w_m^3}\nonumber \\
   & +\frac{1}{8 w_m^3 (3 w_m+1)^2 (v_3+(v_3-v_6-1) w_m-1)}\Bigg(v_3^2 (w_m-1) (5 w_m+1) (w_m (w_m+6)+1) (w_m+1)^2 \nonumber \\
   & -2 v_3 \left(w_m
   \left(w_m \left(-2 v_1 (3 w_m+1)^2+w_m (15 w_m+8)-14\right)+v_6 (w_m-1) (w_m (37 w_m+10)+1)-8\right)-1\right)
   (w_m+1)^2 \nonumber \\
   & +w_m \Big(v_6^2 w_m (w_m+1) (5 w_m-1) (w_m-1)^2-w_m \left(4 v_1 (w_m+1)^2 (3 w_m+1)^2-w_m (w_m (3 w_m (3 w_m+8)+13)-16)+21\right) \nonumber \\
   & -2 v_6 (w_m+1) (3 w_m+1) (w_m (w_m (-5 w_m+v_1 (6 w_m+2)-1)+5)+1)-8\Big)-1\Bigg),
\\
   & v_5'= \frac{(w_m-1)^2 (w_m+1)
   \left(-v_3+v_6-\frac{v_6}{w_m+1}+1\right)^{\frac{7 w_m+1}{3 w_m+1}}}{8 w_m^3} \nonumber \\
   & +\frac{1}{8 w_m^3 (3
   w_m+1)^2 (v_3+(v_3-v_6-1) w_m-1)}\Bigg(v_3^2 (w_m+1) (5 w_m+1)
   (w_m (w_m+6)+1) (w_m-1)^2 \nonumber \\
   & +w_m \Big(v_6^2 w_m (5 w_m-1) (w_m-1)^3+2 v_6 (3 w_m+1) \left(w_m \left((6 v_5+5) w_m^2+2 v_5 w_m+w_m-5\right)-1\right) (w_m-1) \nonumber \\
   & +w_m \left(4 v_5 \left(w_m^2-1\right) (3 w_m+1)^2+w_m (3 w_m+4) (w_m (3 w_m-2)-3)+7\right)+6\Big) \nonumber \\
   & -2 v_3 \left(w_m^2-1\right) \left(w_m \left(w_m \left(2
   v_5 (3 w_m+1)^2+w_m (15 w_m+8)-14\right)+v_6 (w_m-1) (w_m (37 w_m+10)+1)-8\right)-1\right)+1\Bigg),
  \end{align}
   \begin{align}
   & v_6'= \frac{(w_m+1)^2 (3 w_m+1) \left(-v_3+v_6-\frac{v_6}{w_m+1}+1\right)^{\frac{7
   w_m+1}{3 w_m+1}}}{4 w_m^2 (2 w_m+1)} \nonumber \\
   & +\frac{(w_m+1)}{4 w_m^2
   (2 w_m+1) (3 w_m+1) (v_3+(v_3-v_6-1) w_m-1)}\Bigg( (w_m+1) (5 w_m+1) (w_m (w_m+6)+1)
   v_3^2 \nonumber \\
   & -2 (w_m+1) (5 w_m+1) (w_m (3 w_m+v_6 (5 w_m-1)+4)+1) v_3 \nonumber \\
   & +w_m \left(-w_m (w_m (19
   w_m+26)+3) v_6^2+2 (w_m-1) (w_m+1) (3 w_m+1) v_6+(3 w_m+4) (w_m (3 w_m+4)+2)\right)+1\Bigg).
\end{align}
\end{subequations}
\end{small}

In figures \ref{fig:PIVwm-1f4}-\ref{fig:PIVwm1} some solutions of the system \eqref{systemphi4a}-\eqref{systemphi4c} and of the  system  \eqref{system4}  for the potential $V_{IV}(\phi,\psi)$ when $\omega_{m}=-\frac{1}{4}$, $-\frac{1}{7}$ and $1$ are represented. More specific, in then upper panel the solutions are projected in the space ($\phi_{1}$, $\phi_{2}$, $\phi_{3}$) and ($\Phi_{1}$, $\Phi_{2}$, $\Phi_{3}$), where we have represented the spheres $(\phi_{1}-1)^{2}+(\phi_{2}-1)^{2}+(\phi_{3}-1)^{2}=r^2$ and $\Phi_{1}^{2}+\Phi_{2}^{2}+\Phi_{3}^{2}=r^2$ with $r\in\{1,\sqrt{2}\}$. In the lower panel projections  in the spaces ($v_{1},v_{2}$), ($v_{1},v_{4}$), ($v_{1},v_{5}$), ($v_{2},v_{3}$), ($v_{2},v_{4}$), ($v_{2},v_{5}$), ($v_{3},v_{6}$), ($v_{4},v_{5}$) and ($v_{5},v_{6}$) are represented. From these figures, it is confirmed that the point $(1,0,1,0,1,0)$ is a saddle when $\omega_{m}>0$ while for $-1<\omega_{m}<0$ is an attractor of the system, i. e., it  is asymptotically stable.

\subsection{Scalar field potential $V_{V}\left(  \phi,\psi\right)  $}
In this section we study the stability of the solution \eqref{solVa}, \eqref{solVb}, \eqref{solVc} of system \eqref{eqVa}, \eqref{eqVb}, \eqref{eqVc}. We  set for simplicity the integration constants $t_1, t_2, t_3$ to zero  because they are not relevant as $t\rightarrow \infty$.

With the time variable $\tau=\ln (t)$, and defining the new variables $X(\tau)=x(\tau)-x_0 e^{\tau}$,  $\mathcal{Y}(\tau)= 6 \varepsilon V_0 e^{2 \tau} Y^{r}$ with $r=-\frac{3 w_m+1}{w_m-1}$ and  $Z(\tau)=z(\tau)-z_0 e^{\tau}$ the equations \eqref{eqVa}, \eqref{eqVb}, \eqref{eqVc} become
\begin{align*}
& \mathcal{Y}''(\tau )=\frac{4 w_m \mathcal{Y}'(\tau )^2}{3 w_m \mathcal{Y}(\tau )+\mathcal{Y}(\tau )}+\frac{(5-w_m) \mathcal{Y}'(\tau )}{3 w_m+1}-\frac{2 (w_m+3) \mathcal{Y}(\tau )}{3 w_m+1} \nonumber \\
& - Y_1 (w_m-1)^2 (w_m+1) (3 w_m+1) e^{-\frac{2 \tau  (w_m-1)}{3 w_m+1}}\mathcal{Y}(\tau )^{\frac{7 w_m+1}{3 w_m+1}},\\
& X''(\tau)-X'(\tau)-\frac{1}{3} \left(  w_{m}^{2}-1\right)  \beta \varepsilon \mathcal{Y}(\tau)  =0,   \quad  Z''(\tau)-Z'(\tau)+\frac{\left(  w_{m}+1\right)  }{w_{m}-1} \varepsilon \mathcal{Y}(\tau)   =0,  
\end{align*}
where $Y_1= \beta ^2 2^{\frac{1-w_m}{3 w_m+1}} 81^{-\frac{w_m}{3 w_m+1}}
  V_0^{\frac{1-w_m}{3 w_m+1}} \varepsilon ^{\frac{1-w_m}{3 w_m+1}}$. The solution \eqref{solVa}, \eqref{solVb}, \eqref{solVc} becomes
\begin{align}
& X_c(\tau)=\frac{\varepsilon Y_0 e^{\frac{w_m-1}{2 w_m}\tau }}{\beta  (w_m-1)^2},   \quad  \mathcal{Y}_c(\tau)= -\frac{3  Y_0 e^{\frac{\tau  (w_m-1)}{2 w_m}}}{4 \beta ^2 (w_m-1)^2
   w_m^2},  \quad  Z_c(\tau)=-\frac{3 \varepsilon Y_0 e^{\frac{w_m-1}{2 w_m}\tau}}{\beta ^2 (w_m-1)^4}.  
\end{align}
Defining the dimensionless variables 
\begin{equation}
    \phi_1= \frac{X}{X_c}, \quad \phi_2= \frac{\mathcal{Y}}{\mathcal{Y}_c}, \quad \phi_3= \frac{Z}{Z_c},
\end{equation}
we obtain the dynamical system
\begin{subequations}
\begin{align}
   & \phi_1'=  \Phi_{1}, \; 
   \Phi_1'=   \frac{ \Phi_{1}}{w_m}+\left(\frac{1}{4}-\frac{1}{4 w_m^2}\right)
    \phi_{1}+\frac{1}{4} \left(\frac{1}{w_m^2}-1\right)  \phi_{2}, \label{systemphi5a}\\
   & \phi_2'=  \Phi_{2}, \; 
   \Phi_2'= \frac{ \Phi_{2}}{w_m}  +\frac{4  \Phi_{2}^2
   w_m}{3 w_m  \phi_{2}+ \phi_{2}}+\frac{(w_m+1) (3 w_m+1)  \phi_{2}^{\frac{7 w_m+1}{3 w_m+1}}}{4 w_m^2}-\frac{(w_m+1) (3 w_m+1)  \phi_{2}}{4 w_m^2}, \label{systemphi5b}\\
   & \phi_3'=  \Phi_{3}, \; \Phi_3'= \frac{\Phi_{3}}{w_m}+\frac{1}{4}
   \left(\frac{1}{w_m^2}-1\right)  \phi_{2}+\left(\frac{1}{4}-\frac{1}{4 w_m^2}\right)  \phi_{3}. \label{systemphi5c} 
\end{align}
\end{subequations}
Now we analyze the stability of the fixed point 
$P:= (\phi_1, \Phi_1, \phi_2, \Phi_2, \phi_3, \Phi_3)= (1,0,1,0,1,0)$.
The Jacobian matrix of the full system is 
\begin{equation}
\label{jacobian5}
J:=    \left(
\begin{array}{cccccc}
 0 & 1 & 0 & 0 & 0 & 0 \\
 \frac{1}{4}-\frac{1}{4 w_m^2} & \frac{1}{w_m} & \frac{1}{4} \left(\frac{1}{w_m^2}-1\right) & 0 & 0 & 0 \\
 0 & 0 & 0 & 1 & 0 & 0 \\
 0 & 0 & \frac{(w_m+1) (3 w_m+1) (7 w_m+1)  \phi_{2}^{\frac{10
   w_m+2}{3 w_m+1}}-16  \Phi_{2}^2 w_m^3-(w_m+1) (3
   w_m  \phi_{2}+ \phi_{2})^2}{4  \phi_{2}^2 w_m^2 (3 w_m+1)}
   & \frac{8  \Phi_{2} w_m}{3 w_m  \phi_{2}+ \phi_{2}}+\frac{1}{w_m} & 0 & 0 \\
 0 & 0 & 0 & 0 & 0 & 1 \\
 0 & 0 & \frac{1}{4} \left(\frac{1}{w_m^2}-1\right) & 0 & \frac{1}{4}-\frac{1}{4 w_m^2} &
   \frac{1}{w_m} \\
\end{array}
\right).
\end{equation}
Evaluating $J$  at the fixed point $P$  the eigenvalues  $\left\{-1,-\frac{w_m-1}{2 w_m},-\frac{w_m-1}{2 w_m},\frac{w_m+1}{2 w_m},\frac{w_m+1}{2 w_m},\frac{ w_m+1}{w_m}\right\}$ are obtained. For $-1<w_m <0$ $P$ is a sink, or a saddle for $w_m>0$.

\begin{figure*}
    \centering
    \subfigure[\label{fig:PVwm-1f4} Projections in the space ($\phi_{1}$, $\phi_{2}$, $\phi_{3}$) (left) and ($\Phi_{1}$, $\Phi_{2}$, $\Phi_{3}$) (right) for $\omega_{m}=-\frac{1}{4}$. The point $P$ is asymptotically stable.]{\includegraphics[scale = 0.29]{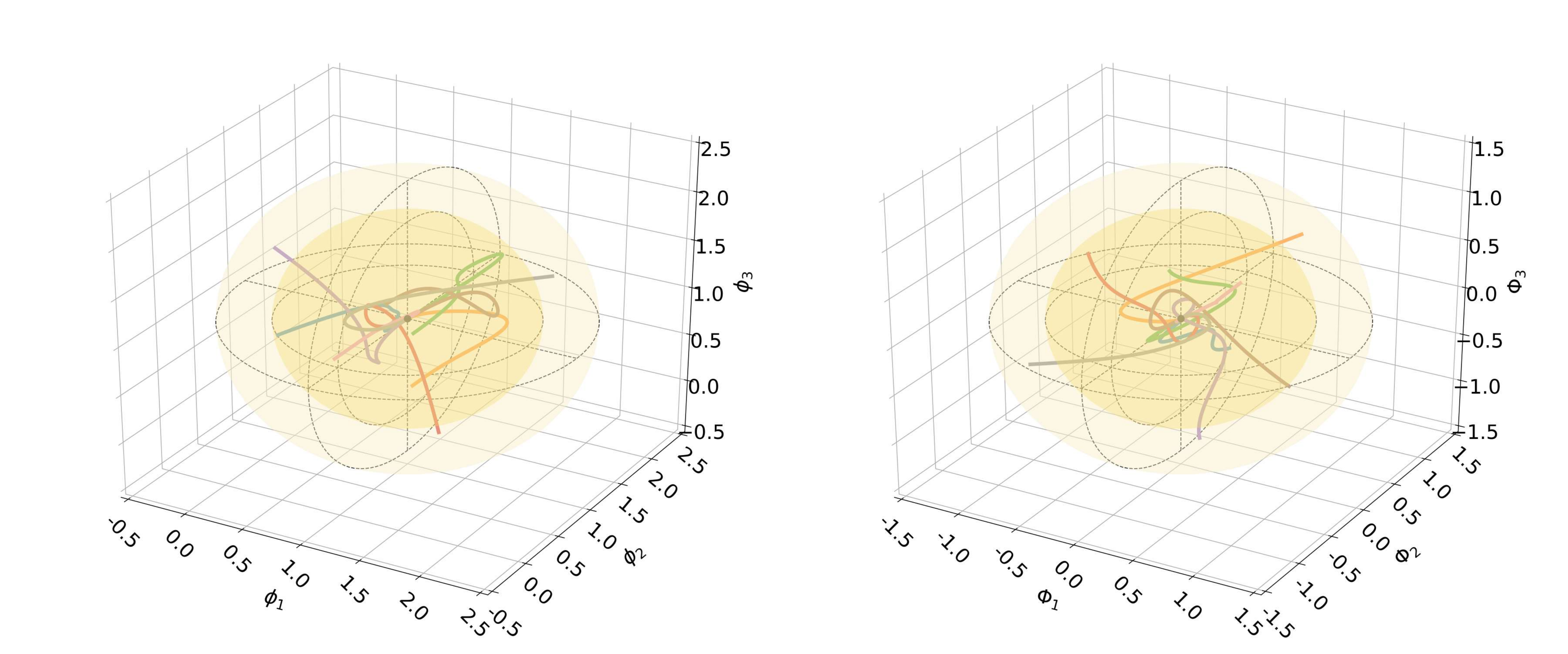}}
    \subfigure[\label{fig:PVwm-1f7} Projections in the space ($\phi_{1}$, $\phi_{2}$, $\phi_{3}$) (left) and ($\Phi_{1}$, $\Phi_{2}$, $\Phi_{3}$) (right) for $\omega_{m}=-\frac{1}{7}$. The point $P$ is asymptotically stable.]{\includegraphics[scale = 0.29]{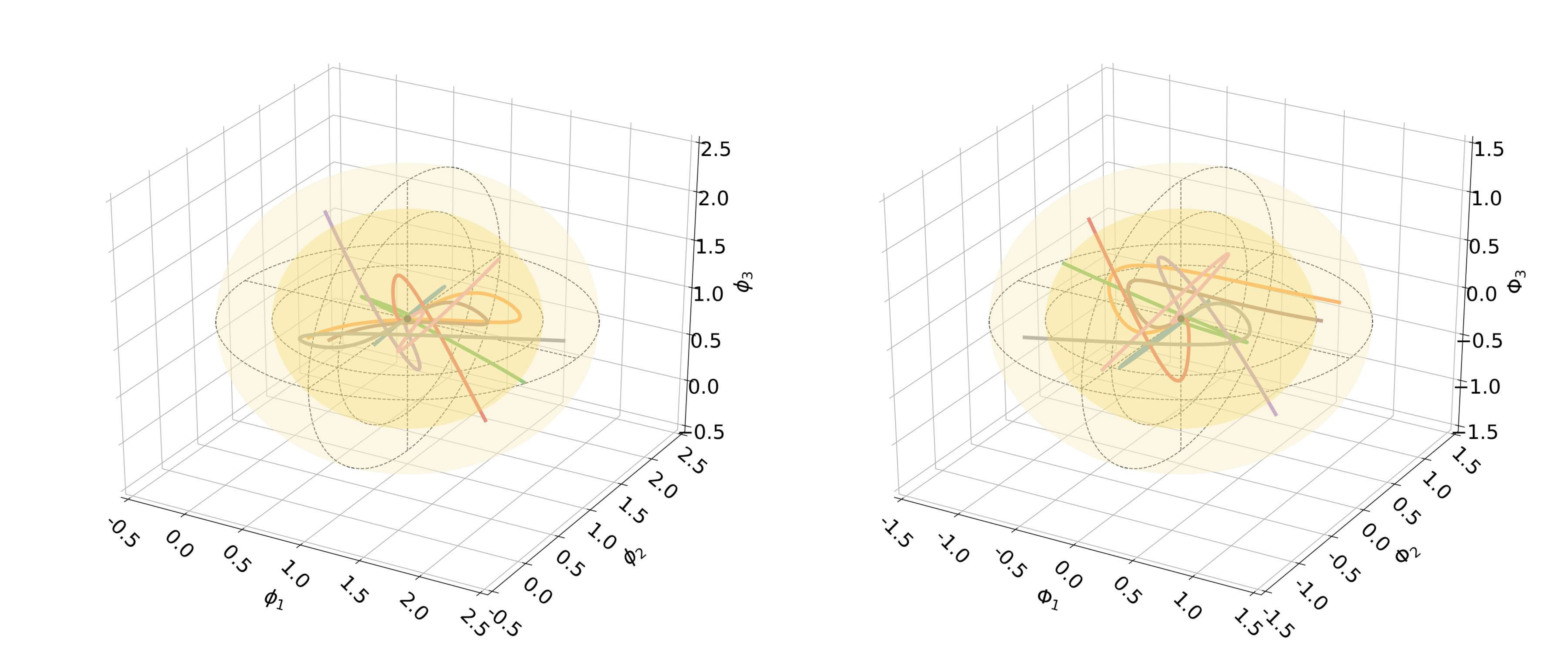}}
    \subfigure[\label{fig:PVwm1f3} Projections in the space ($\phi_{1}$, $\phi_{2}$, $\phi_{3}$) (left) and ($\Phi_{1}$, $\Phi_{2}$, $\Phi_{3}$) (right) for $\omega_{m}=\frac{1}{3}$. The point $P$ is a saddle.]{\includegraphics[scale = 0.29]{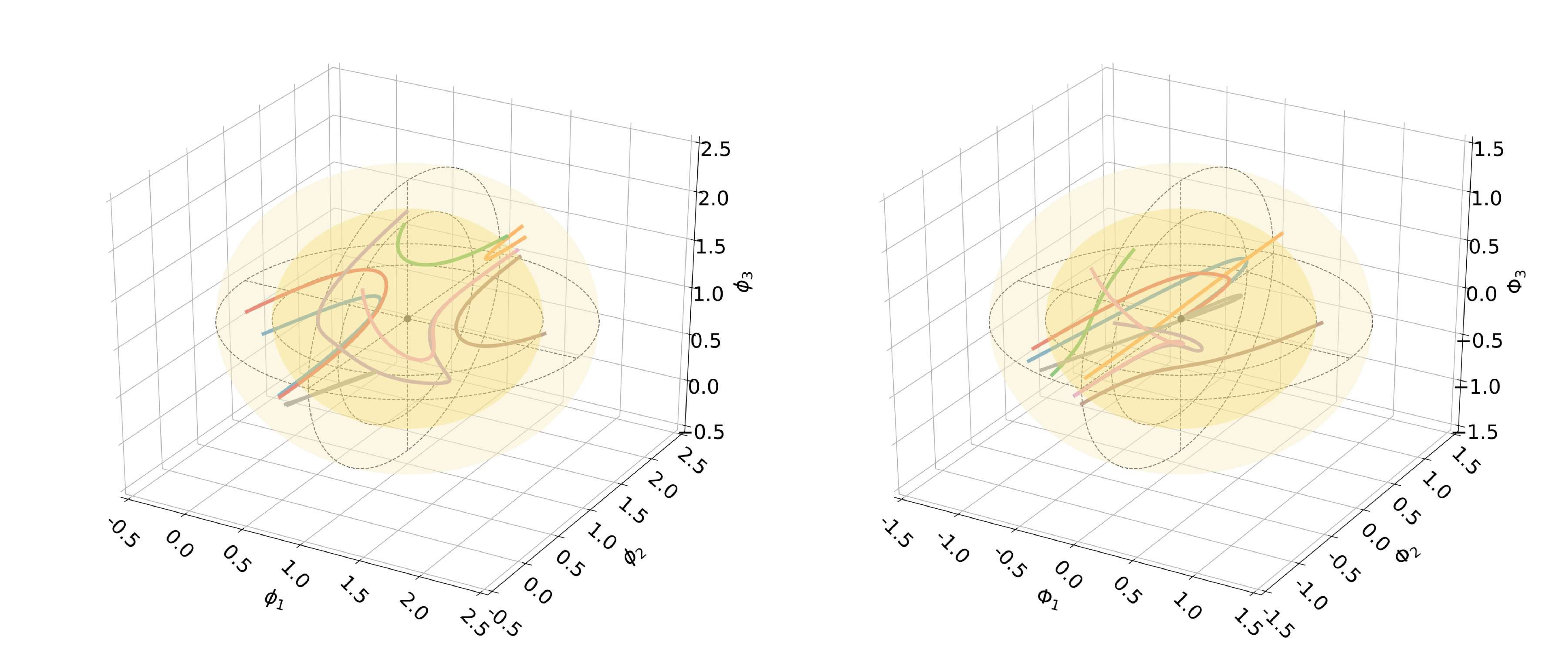}}
    \caption{Some solutions of the system \eqref{systemphi5a}-\eqref{systemphi5c} for the potential $V_{V}(\phi,\psi)$ when $\omega_{m}= -\frac{1}{4}$, $-\frac{1}{7}$ and $\frac{1}{3}$. In these figures we have represented the spheres $(\phi_{1}-1)^{2}+(\phi_{2}-1)^{2}+(\phi_{3}-1)^{2}=r^2$ and $\Phi_{1}^{2}+\Phi_{2}^{2}+\Phi_{3}^{2}=r^2$,  $r\in\{1,\sqrt{2}\}$.}
    \label{fig:PV}
\end{figure*}

In figures \ref{fig:PVwm-1f4}-\ref{fig:PVwm1f3} some solutions of the system \eqref{systemphi5a}-\eqref{systemphi5c} are represented for the potential $V_{V}(\phi,\psi)$ when $\omega_{m}=-\frac{1}{4}$, $-\frac{1}{7}$ and $\frac{1}{3}$, respectively. More specific, in each sub-figure the solutions are projected in the space ($\phi_{1}$, $\phi_{2}$, $\phi_{3}$) (left) and ($\Phi_{1}$, $\Phi_{2}$, $\Phi_{3}$) (right), where additionally we have represented the spheres $(\phi_{1}-1)^{2}+(\phi_{2}-1)^{2}+(\phi_{3}-1)^{2}=r^2$ and $\Phi_{1}^{2}+\Phi_{2}^{2}+\Phi_{3}^{2}=r^2$ with $r\in\{1,\sqrt{2}\}$. From these figures, it is confirmed that the point $(1,0,1,0,1,0)$ is a saddle when $\omega_{m}>0$ while for $-1<\omega_{m}<0$ is an attractor of the system, i. e., it  is asymptotically stable.

\subsection{Scalar field potential $V_{VI}\left(  \phi,\psi\right)  $}

In this section we study the stability of the powerlaw solution \eqref{dc.90}, \eqref{dc.91}, \eqref{dc.92} of the system \eqref{dc.82}, \eqref{dc.83}, \eqref{dc.84}. We  set for simplicity the integration constants $t_1, t_2, t_3$ to zero  because they are not relevant as $t\rightarrow \infty$.

With the time variable $\tau=\ln (t)$, and defining the new variables $\mathcal{U}(\tau)= 12 \beta  V_0 e^{2 \tau} U^{r}$ with $r=-\frac{3 w_m+1}{w_m-1}$, $X(\tau)=x(\tau)-x_0 e^{\tau}$,   and  $Z(\tau)=z(\tau)-z_0 e^{\tau}$ the equations \eqref{dc.82}, \eqref{dc.83}, \eqref{dc.84} become
\begin{align*}
& \mathcal{U}''(\tau )= \frac{4 w_m \mathcal{U}'(\tau )^2}{(1+3 w_m)\mathcal{U}(\tau )}+\frac{(5-w_m) \mathcal{U}'(\tau )}{3 w_m+1} + (w_m+1) (3 w_m+1)  e^{-\frac{2 \tau  (w_m-1)}{3 w_m+1}} \mathcal{U}(\tau )^{\frac{7 w_m+1}{3 w_m+1}}
   -\frac{2 (w_m+3) \mathcal{U}(\tau )}{3 w_m+1},\\
& X''(\tau)-X'(\tau)- \gamma\left(  w_{m}^{2}-1\right)  \mathcal{U}(\tau)    =0, \quad Z''(\tau)- Z'(\tau)+\frac{\left(  w_{m}+1\right)  }{w_{m}-1}\mathcal{U}(\tau)    =0,
\end{align*}
where 
$v_1= 2^{\frac{2 (3 w_m-1)}{3 w_m+1}} w_m^{\frac{2
   (w_m-1)}{3 w_m+1}} U_0^{-\frac{4 w_m}{3 w_m+1}} \gamma ^{\frac{4 w_m}{3 w_m+1}}
  \left(8-3 \beta  \gamma  (w_m-1)^2 \epsilon \right)^{\frac{4 w_m}{3 w_m+1}}$.
   
   The powerlaw solution \eqref{dc.90}, \eqref{dc.91}, \eqref{dc.92} transforms to 
\begin{align}
\label{sol6a}
& \mathcal{U}(\tau)=\frac{U_0 e^{\frac{\tau  (w_m-1)}{2 w_m}}}{8 \gamma  w_m^2 \left(8 - 3 \beta  \gamma  (w_m-1)^2 \varepsilon\right)},  \quad
X(\tau)= -\frac{U_0 e^{\frac{\tau  (w_m-1)}{2 w_m}}}{2(8-3 \beta  \gamma  (w_m-1)^2
   \varepsilon)}, \nonumber \\
   & Z(\tau)= \frac{U_0 e^{\frac{\tau  (w_m-1)}{2 w_m}}}{2
   \gamma  (w_m-1)^2 \left(8- 3 \beta  \gamma  (w_m-1)^2 \varepsilon\right)}.
\end{align}
Defining the dimensionless variables 
\begin{equation}
    \phi_1= \frac{X}{X_c}, \quad \phi_2= \frac{\mathcal{U}}{\mathcal{U}_c}, \quad \phi_3= \frac{Z}{Z_c}.
\end{equation}
we obtain the dynamical system \eqref{systemphi5a}, \eqref{systemphi5b}, \eqref{systemphi5c}.
Hence, as before, the fixed point 
$P:= (\phi_1, \Phi_1, \phi_2, \Phi_2, \phi_3, \Phi_3)= (1,0,1,0,1,0)$ is a sink for $-1<w_m <0$ or a saddle for $w_m>0$ as shown in figure \ref{fig:PV}.

\section{Conclusions}

\label{sec6}

We have considered a cosmological model consisting of two-scalar fields minimally
coupled to gravity and an ideal gas. The two scalar fields interact in the
kinetic and in the potential term. In particular the kinematics of the two
scalar fields lie on a space of constant curvature. This kind of scalar field
models are known also as Chiral models. The Chiral model is the main mechanism
for the description of the hyperbolic inflation. 

In the case of a spatially flat FLRW universe, the field equations form an
autonomous dynamical system which is described by a point-like Lagrangian. The
Lagrangian function depends on an unknown potential function which drives
the dynamics of the two scalar fields. In this work we focused on the
integrability properties for the field equations and we applied geometrical
selection rules in order to constraint the functional forms of the potential.

We applied the theory of symmetries of differential equations  to
constraint the scalar field potential such that  field equations admit conservation laws. Specifically, we investigate the existence of variational
symmetries where the corresponding conservation laws are constructed with the
use of Noether's theorem. Because the point-like dynamical system is defined
on a three dimensional space with dependent variables the scale factor and two
scalar fields $\left\{  a,\phi,\psi\right\}  $, and because the system is
autonomous, the Hamiltonian function is one conservation law. Thus, two
additional conservation laws should be found such that the system is
Liouville integrable. 

Therefore, we performed a classification of the variational symmetries for the
point-like Lagrangian which describes the field equations and we found six
potential functions $V\left(  \phi,\psi\right)  $ where the field equations
are Liouville integrable. For these six potential functions we define the
normal coordinates, which is used to construct the analytic solution for the
field equations. The free parameters of the cosmological model
are constrained in order to describe analytic and exact solutions for the scale factor which describe the hyperbolic inflation era. 

Finally, the stability properties of exact closed-form solutions were investigated using a dynamical systems formulation and numerical tools for these potential functions. In particular, for potentials $V_{I}\left(  \phi,\psi\right)  $ and $V_{II}\left(  \phi,\psi\right)  $ the scaling solution is a saddle for $-1<w_m<1, w_m\neq 0$ and stable, but not asymptotically stable for $w_m=0$; and for potential $V_{III}\left(  \phi,\psi\right)  $ the scaling solution is a saddle when $w_{m}>0$ while for $-1<w_{m}<0$ the scaling solution is stable, but not asymptotically stable. 
For potentials $V_{IV}\left(  \phi,\psi\right)  $, $V_{VI}\left(  \phi,\psi\right)  $ and $V_{VI}\left(  \phi,\psi\right)  $ the scaling solutions are saddle when $w_{m}>0$ while for $-1<w_{m}<0$ are asymptotically stable. 

\begin{acknowledgments}
This work is based on the research supported in part by the National Research Foundation of South Africa (Grant Numbers 131604). The research of AG was funded by Agencia Nacional de Investigaci\'{o}n
y Desarrollo - ANID through the program FONDECYT Iniciaci\'{o}n grant no. 1200293. The research of AP, GL and EG was funded by Agencia Nacional de Investigaci\'{o}n
y Desarrollo - ANID through the program FONDECYT Iniciaci\'{o}n grant no.
11180126. Additionally, GL was funded by Vicerrector\'{\i}a de
Investigaci\'{o}n y Desarrollo Tecnol\'{o}gico at Universidad Catolica del Norte. Ellen de los M. Fern\'andez Flores  is acknowledged for proofreading this manuscript and for improving the English. 
\end{acknowledgments}

\end{document}